\documentclass[11pt]{article}
\pdfoutput=1

\usepackage{jheppub}
\usepackage{amsmath,epsf,amssymb,latexsym,amsthm,setspace,bbm,array,pifont,enumerate}
\usepackage[table]{xcolor}
\usepackage{bbm}
\usepackage{mathtools}
\usepackage{nicematrix}
\usepackage{arydshln}
\usepackage{float}
\usepackage{mathrsfs}
\usepackage{upgreek}

\newcommand\zb{\overline{z}}
\newcommand\thetab{\overline{\theta}}
\def\rank{\operatorname{rank}}
\def\diag{\operatorname{diag}}
\def\rep#1{{{\boldsymbol{#1}}}}
\def\brep#1{{{\overline{\boldsymbol{#1}}}}}
\def\C{{\mathbb C}}
\def\P{{\mathbb P}}
\def\R{{\mathbb R}}
\def\Z{{\mathbb Z}}
\def\SO{\operatorname{SO}}
\def\SL{\operatorname{SL}}
\def\GL{\operatorname{GL}}
\def\SU{\operatorname{SU}}
\def\GU{\operatorname{U{}}}
\def\Spin{\operatorname{Spin}}
\def\GE{\operatorname{E}}
\def\so{\operatorname{\mathfrak{so}}}
\def\su{\operatorname{\mathfrak{su}}}
\def\Lu{\operatorname{\mathfrak{u}}}
\def\Le{\operatorname{\mathfrak{e}}}
\def\Lg{\operatorname{\mathfrak{g}}}
\def\la{\langle}
\def\ra{\rangle}
\def\cA{{\cal A}}
\def\cE{{\cal E}}
\def\cF{{\cal F}}
\def\cG{{\cal G}}
\def\cH{{\cal H}}
\def\cJ{{\cal J}}
\def\cK{{\cal K}}
\def\cL{{\cal L}}
\def\cN{{\cal N}}
\def\cO{{\cal O}}
\def\cS{{\cal S}}
\def\cT{{\cal T}}
\def\cW{{\cal W}}
\def\ii{\mathrm{i}}
\def\dd{\mathrm{d}}
\def\ee{\mathrm{e}}
\def\sleft{\text{\normalfont\tiny{L}}}
\def\sright{\text{\normalfont\tiny{R}}}
\def\LG{\text{\normalfont\tiny{LG}}}
\def\KM{\text{\normalfont\tiny{KM}}}
\def\LGsmall{\text{\normalfont\scalebox{.7}{\tiny{LG}}}}
\def\KMsmall{\text{\normalfont\scalebox{.7}{\tiny{KM}}}}
\def\balpha{{\boldsymbol{\alpha}}}
\def\blambda{{\boldsymbol{\lambda}}}
\def\bL{{\boldsymbol{L}}}
\def\bw{{\boldsymbol{w}}}
\def\tk{\mathrm{k}}
\def\tK{\mathrm{K}}
\def\tN{\mathrm{N}}
\def\tr{\mathrm{r}}
\def\tR{\mathrm{R}}
\newcommand{\colunderline}[1]{{\textcolor{pink}{\underline{\textcolor{black}{#1}}}}}
\newcommand{\GLSMblock}[2]{{ \textcolor{pink}{\overbracket{\textcolor{black}{ #2}}^{#1}} }}

\title{Calabi--Yau linear models with abelian bundles}
\author{Yann Proto\,}
\affiliation{Bethe Center for Theoretical Physics, Universit\"at Bonn, Bonn, Germany}
\emailAdd{yproto@uni-bonn.de}

\abstract{We introduce a construction of (0,2) gauged linear sigma models whose large radius phases realize Calabi--Yau sigma models with line bundles. We study a family of such models admitting Landau--Ginzburg orbifold phases, described as asymmetric orbifolds of (2,2) Gepner models and free left-moving fermions. These orbifolds extend the Gepner construction to a broader class of (0,2) SCFTs, and reveal unexpected features, including the emergence of free fermions in twisted sectors and substantial enhancements of the left-moving current algebra. The gauged linear sigma model connects the Calabi--Yau geometry to these rational points, where the volume modulus becomes charged under the enhanced gauge symmetry and freezes at the string scale.}

\begin{document}

\maketitle

\section{Introduction}
Supersymmetric sigma models with Calabi--Yau (CY) target spaces have played a central role in string compactifications since their early days, owing to their relation to two-dimensional superconformal theories. The (2,2) non-linear sigma model (NLSM) on a CY $d$-fold $Y$ is expected to flow to a superconformal field theory (SCFT) with central charges $c_\sleft=c_\sright=3d$ and integral $\GU(1)_\sleft\times\GU(1)_\sright$ charges---the RG flow deforms the K\"ahler metric on $Y$, while preserving the complexified K\"ahler class~\cite{Nemeschansky:1986yx}. The moduli space of (2,2) preserving deformations is locally parametrized  by (a,c) and (c,c) marginal deformations, corresponding near a large radius limit to complexified K\"ahler and complex structure deformations, and at special points in moduli space, the SCFT may admit an exact description as a (2,2) Gepner model~\cite{Gepner:1987vz}. However, non-chiral models capture only a restricted corner in the string landscape. In contrast, (0,2) theories offer access to a much richer locus through compactifications of the heterotic string, at the cost of significant technical complications. Several of these concern the holomorphic vector bundle $\cE\to Y$ supporting left-moving fermions in a (0,2) NLSM: supersymmetry imposes stability conditions on $\cE$ that are often difficult to verify. While considerable progress has been made for deformations of (2,2) models, in which $\cE$ is a deformation of the tangent bundle $T_Y$,  constructing models with more general bundles remains essential to understanding the dynamics of ``generic'' (0,2) theories.

Explorations of the (0,2) landscape have greatly benefited from the gauged linear sigma model (GLSM) approach~\cite{Witten:1993yc}. In this framework, the NLSM on $Y$ emerges as the low energy limit of a supersymmetric abelian gauge theory, with appropriately tuned parameters. For (0,2) models, the bundle $\cE$ is usually realized in the GLSM as a monad bundle, defined as the cohomology of a three-term complex on $Y$~\cite{Horrocks:1964,Barth:1978,Okonek:1980}. Several methods have been developed to establish the stability of certain classes of monads~\cite{Anderson:2007nc,Anderson:2016byt,Brambilla:2009}, notably through generalizations of Hoppe's criterion~\cite{Hoppe:1984,Jardim:2017}. In this note we pursue a different approach, and propose a linear model construction for CY compactifications with abelian bundles, in which $\cE$ is a direct sum of line bundles over $Y$.

Abelian bundles can, in principle, be readily realized in a (0,2) GLSM by introducing charged fermi multiplets without superpotential couplings. At the IR fixed point, these give rise to left-moving fermions minimally coupled to $\Lu(1)$ connections on line bundles over $Y$. However, this simple picture is obstructed by an awkward issue: the GLSM may suffer from gauge anomalies that cannot, in general, be canceled by the charged fermi multiplets alone. For certain models, this obstacle can be circumvented using a mechanism developed in the study of heterotic flux vacua in recent work with D. Isra\"el and I. Melnikov~\cite{Israel:2026bco}. The idea is to introduce additional charged superfields that describe left-moving fermions supported on a holomorphically trivial line bundle $\cO\to Y$, realized as a monad. In the IR limit, these degrees of freedom are liberated from the NLSM, emerging as free, decoupled left-moving fermions. In the UV gauge theory, however, the charged matter associated with such monadic $\cO$ factors remains interacting and contributes to the gauge anomalies. With suitably chosen charges, these contributions can be used to ensure anomaly cancellation.

We illustrate the construction with several examples, focusing in the main text on a class of CY $d$-folds realized as branched covers of $(\P^1)^d$, with further examples relegated to the appendix. This class is particularly appealing because the corresponding GLSMs admit a Landau--Ginzburg orbifold (LGO) phase, in which the abelian gauge symmetry is broken to a finite subgroup. In all our examples there is a locus in the GLSM parameter space where the LGO phase is described by a chiral orbifold of a (2,2) Gepner model and free left-moving fermions. Although the orbifold action on the Gepner sector is anomalous, its anomaly is canceled by the accompanying transformation of the fermions, yielding a well-defined (0,2) orbifold SCFT.

These explicit examples allow us to test the expectations from the large radius analysis by computing the spacetime massless spectrum in the LGO phase. In particular, the LGO models share a noteworthy feature: the emergence of free left-moving fermions, which arise as twisted ground states in the orbifold SCFT. Notably, the $\Lu(1)_\sleft$ symmetry of the parent Gepner model is enhanced to a considerably larger $\so(r)_\sleft$ current algebra, generated by bilinears of the $r$ emergent free fermions. We explain the emergence of these fermions using a spectral flow approach, tracing it to a special structure shared by the orbifold models. While remarkable from the LGO perspective, the appearance of these fermions is expected from the large radius description: they originate from the GLSM degrees of freedom associated with trivial bundle factors, which flow to free decoupled fermions in the IR. Their presence in the LGO spectrum provides compelling evidence that the two phases are connected.

We find substantial gauge symmetry enhancements at the Gepner point. For instance, one of the GLSMs presented below describes, in its large radius phase, a K3 surface with a rank-$12$ abelian bundle supported on a single anti-self-dual class. Compactification of the $\Spin(32)/\Z_2$ heterotic string on this background yields a six-dimensional (1,0) supergravity theory whose gauge group $G_{\text{\tiny{LR}}}$ is the commutant of the embedded $\GU(1)$ structure group of the bundle.\footnote{More precisely, a $\GU(1)$ factor is lifted through the  St\"uckelberg mechanism.} At the Gepner point, however, the gauge symmetry is enhanced to $G_{\text{\tiny{LGO}}}$, with
\begin{align}
G_{\text{\tiny{LR}}} = \mathrm{Spin}(8) \times \SU(12)
\qquad\subset\qquad
G_{\text{\tiny{LGO}}} = \mathrm{Spin}(8) \times \SU(12)\times\SO(4)~.
\end{align}
The first two factors in $G_{\text{\tiny{LGO}}}$ are linearly realized by bilinears of the heterotic fermions---including those emerging in twisted sectors. The enhanced $\SO(4)$ factor is of a very different nature: remarkably, \emph{all} hypermultiplets in the six-dimensional supergravity are charged under this symmetry. This includes, in particular, the universal hypermultiplet associated at large radius with the CY volume: it is frozen at the string scale by the enhanced $\SO(4)$ symmetry. This effect closely resembles the transitions discussed in~\cite{Baykara:2023plc} for six-dimensional supergravities without neutral hypermultiplets.

The heterotic CFTs describing these LGO models are also rather exotic. For the example above, the internal (0,4) SCFT, with central charges $(c_\sleft,c_\sright)=(20,6)$, has left-moving degrees of freedom purely described by non-abelian chiral currents. Indeed, the spacetime gauge symmetry $G_{\text{\tiny{LGO}}}$ is realized on the worldsheet by a left-moving Kac--Moody (KM) algebra, with the enhanced $\SO(4)$ factor arising from an $\su(2)\oplus\su(2)$ current algebra whose two factors are both at level $10$.
Not only are such high KM levels uncommon in perturbative heterotic constructions, but they also result in a total Sugawara central charge $c_\sleft=4+11+2\cdot\frac{5}{2}=20$ which completely saturates the left-moving central charge of the internal SCFT. Hence, the left-moving sector is entirely described by the current algebra for $G_{\text{\tiny{LGO}}}$, and the SCFT can be expressed as a non-diagonal pairing of left-moving KM primaries with right-moving primaries of the  $c_\sright=6$ superconformal algebra (SCA) of the Gepner model. The existence of such models is intriguing from the perspective of modular invariance. We find several examples with similar enhancements by higher level KM currents, including a model associated with a CY threefold compactification, which we study in detail. We expect the asymmetric Gepner orbifold construction uncovered through the present GLSM explorations to yield a broad class of new string vacua that could populate exotic corners of the string landscape. We therefore sketch out the basic framework for constructing these (0,2) SCFTs, independently of any GLSM realization.

The structure of this note is as follows. In section~\ref{sec:linebundleGLSMs}, we outline the GLSM construction for abelian bundles and present the examples that will illustrate the analysis in subsequent sections. We turn in section~\ref{sec:K3instantons} to K3 compactifications with abelian instantons, and introduce the corresponding GLSMs. Section~\ref{sec:GepnerSCFT} provides a self-contained account of the Gepner orbifold construction, develops the tools needed to analyze the CFT spectrum, and concludes with the analysis of the K3 models. In section~\ref{sec:CY}, we study in detail a GLSM describing a CY threefold compactification with abelian bundle, including the spacetime massless spectrum in its LGO phase. We end with a discussion and further directions in section~\ref{sec:discussion}. The appendices contain some technical details, together with further GLSM examples collected in appendix~\ref{app:additionalexamples}.

\section*{Acknowledgements}
It is a pleasure to thank G. Lockhart for useful discussions, and R. Singhal for helpful  explanations of constructions of stable bundles over K3. I am especially grateful to I. Melnikov and D. Isra\"el for discussions and collaboration in the early stages of this project, as well as for valuable comments on the manuscript. I also thank LPTHE for hospitality during the course of this work, as well as the organizers and participants of the ``GLSM26'' conference at MITP for the stimulating atmosphere. This work has received funding from the European Research Council (ERC) under the Horizon Europe (grant agreement No. 101078365) research and innovation program.

\section{Line bundles in linear sigma models}
\label{sec:linebundleGLSMs}
We begin by introducing a general construction of (0,2) GLSMs for heterotic compactifications with abelian gauge bundles. Throughout, we work on a Euclidean worldsheet with (0,2) superspace coordinates $(z;\zb,\theta,\thetab)$, and follow the superspace conventions of~\cite{Melnikov:2019tpl}.

\subsection{Set up}
Consider first a heterotic compactification on a  CY $d$-fold $Y$, with a holomorphic vector bundle $\cE\to Y$ encoding the left-moving fermion couplings. We realize $Y$ as a smooth hypersurface in a toric variety, and briefly review the relevant features of this construction. The toric variety $V$ is specified by a $(d+1)$-dimensional reflexive polytope $\Delta$, with dual $\Delta^\circ$. Its geometry can be described using the homogeneous coordinate ring~\cite{Cox:2000}: each ray $\rho\in\Delta^\circ$ determines a homogeneous coordinate, acted upon by the abelian group $G_\C=(\C^\ast)^r$.\footnote{In general, $G_\C$ may also contain a discrete factor, which is absent in the examples considered here.} We denote the corresponding charges by $Q^\alpha_\rho$, with $\alpha=1,\dots,r$, and assemble them into  column vectors $Q_\rho$. The homogeneous coordinates span $\C^{d+r+1}$, and the toric variety is obtained as a quotient $V=(\C^{d+r+1} -F)/G_\C$, where the exceptional locus $F$ is determined by a suitable triangulation of $\Delta^\circ$. The hypersurface
\begin{align}
Y=\{J_0=0\}\quad \subset V
\end{align}
is defined by a generic anticanonical section $J_0\in H^0(V,\cO_V(-K_V))$, which can be represented by a polynomial in the homogeneous coordinates of charge $\sum_\rho Q_\rho$. The toric divisors of $V$ restrict to divisor classes $\eta_\alpha\in H^2(Y,\Z)$, for $\alpha=1,\dots,r$. These define a natural collection of line bundles on $Y$: we denote by $\cO(Q)$ the line bundle with first Chern class $c_1(\cO(Q))=Q^\alpha\eta_\alpha$.

The holomorphic bundle $\cE\to Y$ may likewise admit a toric description---natural in the GLSM framework---as a monad bundle. In the simplest situation, $\cE$ is a kernel bundle, defined by a short exact sequence of the form
\begin{align}
0~\longrightarrow~\cE~\longrightarrow~\mathop{\oplus}_{A=1}^{\text{r}+1} \cO(Q_A)~\mathop{\longrightarrow}^{\boldsymbol{F}|_Y}~\cO\left(\Sigma_{A=1}^{\tr+1}Q_A\right)~\longrightarrow~0~.
\end{align}
The map is determined by $\boldsymbol{F}=(F_1,\dots,F_{\tr+1})$, where each $F_A$ is a section of $\cO_V(\Sigma_{B\neq A}Q_B)$. Equivalently, the $F_A$ are homogeneous polynomials of the corresponding charges under the action of $G_\C$; their restrictions to $Y$ define the map in the sequence. The bundle condition requires that the polynomials $F_A$ have no common zero on the hypersurface. This condition ensures that the map has constant rank along $Y$, so the sequence defines a vector bundle rather than a sheaf. For generic choices of the polynomials $F_A$, the resulting rank-$\tr$ bundle $\cE$ is irreducible and has $\SU(\tr)$ structure group---in particular, $c_1(\cE)$ vanishes.\footnote{In a general heterotic compactification, $c_1(\cE)$ need not vanish; cancellation of a global anomaly requires only that it be even in $H^2(Y,\Z)$~\cite{Witten:1985mj,Freed:1986zx,Distler:1986wm}. We restrict to bundles with $\SU(\tr)$ structure group, which admit a more convenient realization in the GLSM and, in particular, preserve a non-anomalous global $\GU(1)_\sleft$ symmetry.}

The (0,2) linear model realizing this geometry has $\GU(1)^r$ gauge group. There is for each homogeneous coordinate a chiral superfield with gauge charges $Q_\rho$ corresponding to the toric action, and a fermi superfield $\Gamma_0$ of charge $Q_{\Gamma_0}=-\sum_\rho Q_{\Phi_\rho}$. In addition, the bundle is realized using a collection of fermi multiplets $\Gamma_{A=1,\dots,\text{r}+1}$  carrying charges $Q_A$, and an extra chiral superfield $\Phi_0$, with charge $Q_{\Phi_0}=-\sum_A Q_A$. The GLSM has superpotential
\begin{align}
\cW = \Gamma_0 J_0+\Phi_0 \sum_{A=1}^{\tr+1}\Gamma_A F_A~,
\label{eq:GLSMsuperpotential}
\end{align}
where $J_0$ and $F_A$ are now understood as being valued in the superfields $\Phi_\rho$. The geometry may then be realized as the large radius phase of the GLSM, specified by a choice of Fayet--Iliopoulos (FI) parameters.
Because of the bundle condition on the polynomials $F_A$, the field $\Phi_0$ vanishes in the LR phase, and the remaining chiral superfields $\Phi_\rho$ describe the toric ambient space.

The GLSM with superpotential~\eqref{eq:GLSMsuperpotential} admits a $\GU(1)_\sleft\times\GU(1)_\sright$ global symmetry, under which the superfields carry the following charges.
\begin{align*}
\arraycolsep=6pt
\begin{array}{c}
 \\
q_\sleft \\
q_\sright \\
\end{array}
\;
\GLSMblock{Y}{
\begin{array}{cccc}
\Gamma_0 & \Phi_{\rho_1} & \dots & \Phi_{\rho_{d+r+1}} \\
0 & 0 & \dots & 0 \\
+1 & 0 & \dots & 0 \\
\end{array}
}
\;
\GLSMblock{\cE}{
\begin{array}{cccccccccccc}
\Phi_0 & \Gamma_1 &\dots & \Gamma_{\tr+1} \\
+1 & -1 & \dots & -1 \\
+1 & 0 & \dots & 0 \\
\end{array}
}
\begin{array}{c}
\theta \\
0 \\
+1 \\
\end{array}
\end{align*}
The superpotential~\eqref{eq:GLSMsuperpotential} takes the most general form compatible with this symmetry. In particular, the $\GU(1)_\sleft$ symmetry constrains it to be at most linear in the superfield $\Phi_0$. The global symmetry is non-anomalous, and we assume that it flows to a $\Lu(1)_\sleft\oplus\Lu(1)_\sright$ current algebra with
\begin{align}
J_\sleft(z)J_\sleft(0) &\sim \frac{\text{r}}{z^2}~,&
J_\sright(\zb)J_\sright(0) &\sim \frac{d}{\zb^2}~,
\end{align}
where the currents assign integer charges $q_\sleft$ and $q_\sright$ to local operators in the SCFT. In the IR, the $\GU(1)_\sleft\times\GU(1)_\sright$ symmetry acts on the massless left- and right-moving fermions, respectively. The current algebra levels are determined by 't Hooft anomaly matching in the GLSM:
\begin{align}
&\GU(1)_\sleft\cdot\GU(1)_\sleft~:&
\tr &=(\tr+1)\times(-1)^2-(+1)^2~,\nonumber\\
&\GU(1)_\sright\cdot\GU(1)_\sright~:&
-d&=(+1)^2+r\times(-1)^2-(d+r+1)\times(-1)^2~,
\end{align}
with no mixed $\GU(1)_\sleft\cdot\GU(1)_\sright$ 't Hooft anomaly. The current $J_\sright$ is identified with the R-current of the right-moving SCA, whose two-point function determines the central charge $c_\sright=3d$. The left-moving central charge then follows from the gravitational anomaly $c_\sleft-c_\sright$, computed in the UV, giving $c_\sleft=2d+\tr$.

Our primary guide in constructing consistent (0,2) GLSMs with tractable IR dynamics is the geometry of the large radius phase. This geometric perspective motivates a splitting of the chiral and fermi superfields into two sets, associated respectively with ``geometric'' and ``bundle'' degrees of freedom:
\begin{align}
\mathscr{C}_Y &=\{ \Phi_{\rho_1} , \dots , \Phi_{\rho_{d+r+1}} , \Gamma_0 \}~,&
\mathscr{C}_\cE &=\{ \Phi_0 , \Gamma_1 ,\dots , \Gamma_{\tr+1} \}~.
\end{align}
This distinction is indicated by the colored brackets in the charge table above. The fields in $\mathscr{C}_Y$ contain the massless scalars parameterizing $Y$ and the massless right-moving fermions valued in $T_Y$, while those in $\mathscr{C}_\cE$ contain the massless left-moving fermions valued in $\cE$. In the GLSM, this splitting only reflects a bias from our large radius understanding: nothing singles out a superfield as intrinsically ``geometric'' or ``bundle'', and the same fields may play different roles in other phases. Nevertheless, we will often use this distinction, as it provides a convenient shorthand for specifying the global symmetries of the model. In particular, ``bundle'' superfields  carry a $\GU(1)_\sleft$ symmetry with charges as specified above, and the superpotential is constrained to respect this symmetry.

The distinction between $\mathscr{C}_Y$ and $\mathscr{C}_\cE$ is also useful for analyzing gauge anomalies. The GLSM gauge symmetries must be non-anomalous, imposing the condition
\begin{align}
\sum_{\Gamma}Q_\Gamma Q_\Gamma^{\text{t}}-\sum_{\Phi} Q_\Phi Q_\Phi^{\text{t}} =0~,
\end{align}
where the sum runs over all chiral and fermi superfields of the theory, with their charges treated as column vectors. We can express this condition as $\cA_Y=-\cA_\cE$, where the two $r\times r$ anomaly matrices are
\begin{align}
\cA_Y &= \sum_{\Gamma\in\mathscr{C}_Y}Q_\Gamma Q_\Gamma^{\text{t}}-\sum_{\Phi\in\mathscr{C}_Y} Q_\Phi Q_\Phi^{\text{t}}~,&
\cA_\cE &= \sum_{\Gamma\in\mathscr{C}_\cE}Q_\Gamma Q_\Gamma^{\text{t}}-\sum_{\Phi\in\mathscr{C}_\cE} Q_\Phi Q_\Phi^{\text{t}}~.
\end{align}
In the LR phase, the GLSM vector multiplets give rise to divisors $\eta_\alpha\in H^2(Y,\Z)$, in terms of which the second Chern classes of $T_Y$ and $\cE$ are written as
\begin{align}
c_2(T_Y) &= \tfrac{1}{2}\cA^{\alpha\beta}_Y\eta_\alpha\wedge\eta_\beta~,&
c_2(\cE) &= -\tfrac{1}{2}\cA^{\alpha\beta}_\cE \eta_\alpha\wedge\eta_\beta~.
\end{align}
Thus, cancellation of the GLSM gauge anomalies implies in the large radius limit the heterotic Bianchi identity $c_2(T_Y)-c_2(\cE)=0$, as required for cancellation of sigma model anomalies.\footnote{The converse, however, need not hold. For example, on an elliptic curve $C$, standard (0,2) GLSM realizations of $C$ have a non-vanishing anomaly matrix $\cA_C$ even though $c_2(T_C)=0$.}

For ease of presentation, we have restricted our discussion to hypersurfaces and a single irreducible bundle factor, but both assumptions can be readily relaxed in the GLSM. Complete intersections can be realized by replacing $\Gamma_0$ with several fermi superfields, whose superpotential couplings specify the defining polynomials. Likewise, one may consider reducible holomorphic bundles $\cE=\oplus_s\cE_s$. Each irreducible factor $\cE_s$ is described by its own set of superfields for the corresponding monad, and carries an independent global $\GU(1)$ symmetry. In the IR, these symmetries give rise to a $\oplus_s\Lu(1)_\sleft^s$ current algebra, with currents $J_\sleft^s$ satisfying
\begin{align}
J_\sleft^s(z) J_\sleft^{s'}(0) \sim\frac{\text{r}_s\delta_{ss'}}{z^2}~,
\end{align}
where $\text{r}_s = \operatorname{rank}(\cE_s)$.
The $\Lu(1)_\sleft$ subalgebra defined by $J_\sleft = \sum_s J_\sleft^s$ has level $\tr=\sum_s\tr_s$. In a heterotic construction, this current implements the chiral GSO projection on the left-moving fermions.

\subsection{Line bundle data}
The main challenge in the GLSM construction outlined above is the choice of holomorphic bundle $\cE\to Y$: not every choice gives rise to a consistent IR SCFT, or at least one admitting a smooth large radius description. The consistency conditions in the LR phase are well understood: $\cE$ must admit an $\su(\tr)$ connection solving the HYM equations. The existence of a HYM connection on $Y$ is in turn related to a topological condition known as slope stability. Given a polarization $\upeta\in H^{1,1}(Y)\cap H^2(Y,\Z)$, the slope of a rank-$\tr$ bundle $\cE$ is defined as
\begin{align}
\mu(\cE) = \frac{1}{\tr}\int_Y c_1(\cE)\wedge \upeta^{d-1} ~.
\end{align}
The bundle $\cE$ is stable with respect to $\upeta$ if every proper subsheaf $\cF\subset\cE$ satisfies $\mu(\cF)<\mu(\cE)$, and $\cE$ admits a HYM connection if and only if it is a direct sum of stable bundles of vanishing slope~\cite{Donaldson:1985zz,Uhlenbeck:1986ntc}. Verifying stability of a given monad bundle is however notoriously difficult, and has proven a long-standing obstacle in the construction of heterotic (0,2) NSLMs and GLSMs. 

The stability conditions become much simpler when $\cE$ is an abelian bundle, namely a direct sum 
\begin{align}
\cE &=\mathop{\oplus}_{A=1}^{\text{r}} \cL_A
\end{align}
of holomorphic line bundles $\cL_A\to Y$. An abelian bundle admits a HYM connection if the first Chern classes $c_1(\cL_A)$, which specify its topology, are primitive (1,1) forms on $Y$. Fortunately, such bundles can also be realized, at least in principle, in a GLSM whose large radius phase is $Y$. Each line bundle summand is encoded by a charged fermi superfield $\Lambda_A$, with no superpotential coupling. The resulting bundle has
\begin{align}
 c_1(\cL_A) = Q^\alpha_{\Lambda_A}\eta_\alpha~.
\end{align}
The first Chern classes are (1,1) forms and become primitive on a sublocus of the FI parameter space.\footnote{In the LR limit, the primitivity condition imposes the constraint $\rho^{\alpha_1}\dots\rho^{\alpha_{d-1}}Q^{\alpha_d}_\Lambda \int_Y\eta_{\alpha_1}\wedge\dots\wedge\eta_{\alpha_d} = 0$ on the K\"ahler cone for each charged fermi multiplet $\Lambda$ encoding a line bundle. A worldsheet mechanism enforces this constraint in the GLSM: the phase rotation associated with $\Lambda$ is anomalous and induces an equivalence in the FI parameter space, effectively collapsing the phase diagram onto the locus compatible with the HYM equations. We refer to~\cite{Israel:2026bco} for details.} The GLSM admits an independent $\GU(1)$ global symmetry for each fermi multiplet $\Lambda_A$. We assume that the diagonal symmetry $\GU(1)_\sleft$, under which all $\Lambda_A$ have charge $q_\sleft=-1$ and all other fields are neutral, is non-anomalous: this is equivalent to the condition $c_1(\cE)=0$.  This $\GU(1)_\sleft$ symmetry ensures that the fermi multiplets $\Lambda_A$ do not appear in the superpotential.

In practice, gauge anomalies can obstruct the realization of a given abelian bundle in a GLSM. The contribution of the fermi multiplets encoding the bundle to the gauge anomalies is described by the matrix
\begin{align}
\cA_\cE &= \sum_{A=1}^{\text{r}} Q_{\Lambda_A}Q_{\Lambda_A}^{\text{t}} ~.
\end{align}
These contributions are often insufficient to cancel the gauge anomalies $\cA_Y$ associated with the CY hypersurface, because of the restricted form of $\cA_\cE$: for instance, all diagonal entries of $\cA_\cE$ are strictly positive. However, this obstruction can be circumvented by a trick proposed in~\cite{Israel:2026bco}, which consists of introducing extra interacting fermions coupled to the trivial line bundle $\cO\to Y$. Although these fields do not affect the IR theory, where they provide decoupled free fermions, they can contribute non-trivially to gauge anomalies in the UV. One way to realize such a trivial bundle is via a monad construction: the GLSM is supplemented by two fermi multiplets $\Gamma_1$, $\Gamma_2$ and a chiral multiplet $\Phi_0$ of charge $Q_{\Phi_0}=-(Q_{\Gamma_1}+Q_{\Gamma_2})$. The latter enters linearly in the superpotential:
\begin{align}
\cW \supset \Phi_0(\Gamma_1F_1+\Gamma_2F_2)~.
\end{align}
This data encodes the structure sheaf $\cO\to Y$, with global nowhere-vanishing section $(F_2,-F_1)$. The fields $\Phi_0$, $\Gamma_1$, $\Gamma_2$ contribute to the gauge anomalies as follows:
\begin{align}
\cA_{\cO} =-Q_{\Gamma_1} Q_{\Gamma_2}^{\text{t}}-Q_{\Gamma_2}Q_{\Gamma_1}^{\text{t}}~.
\end{align}
For the models considered in this work, this additional freedom is sufficient to construct anomaly-free GLSMs with abelian gauge bundles.

\subsection{CY \texorpdfstring{$d$}{d}-folds branched over \texorpdfstring{$(\P^1)^d$}{(P1)d}}
\label{subsec:threefamilies}
We now turn to a class of (0,2) GLSMs that will be the focus of the rest of this note. We illustrate the key ideas in the simplest setting, that of elliptic curves, before presenting the general construction for CY $d$-folds.

\subsubsection*{Three GLSMs for $T^2$}
Consider the three following descriptions of an elliptic curve:
\begin{enumerate}
\item $C_1$ is a degree $4$ hypersurface in $\P_{[1:1:2]}^2$;
\item  $C_2$ is a degree $3$ hypersurface in $\P^2$;
\item $C_3$ is a complete intersection of two quadrics in $\P^3$.
\end{enumerate}
In each case, the curve admits a GLSM description with $\GU(1)$ gauge group, but the naive (0,2) superfield content encoding its geometry suffers from a gauge anomaly. In all three models, the anomaly can be canceled by adding extra superfields that realize a ``monadic'' trivial bundle. The curve $C_1$ is realized by a GLSM with the following charges.
\begin{center}
$\GLSMblock{C}{
\begin{array}{cccc}
\Phi_1&\Phi_2&\Phi_3&\Gamma_3 \\
1&1&2&-4
\end{array}
}$
\qquad
$\GLSMblock{\cO}{
\begin{array}{cccc}
\Phi_4&\Gamma_1&\Gamma_2\\
-6&1&5
\end{array}
}$
\end{center}
Similarly, $C_2$ is realized by a GLSM with the following charges.
\begin{center}
$\GLSMblock{C}{
\begin{array}{cccc}
\Phi_1&\Phi_2&\Phi_3&\Gamma_3 \\
1&1&1&-3
\end{array}
}$
\qquad
$\GLSMblock{\cO}{
\begin{array}{cccc}
\Phi_4&\Gamma_1&\Gamma_2\\
-4&1&3
\end{array}
}$
\end{center}
Finally, the GLSM description of $C_3$ has the following charges.
\begin{center}
$\GLSMblock{C}{
\begin{array}{cccccc}
\Phi_1&\Phi_2&\Phi_3&\Phi_4&\Gamma_3&\Gamma_4\\
1&1&1&1&-2&-2
\end{array}
}$
\qquad
$\GLSMblock{\cO}{
\begin{array}{cccc}
\Phi_5&\Gamma_1&\Gamma_2\\
-3&1&2
\end{array}
}$
\end{center}
Each model has a global $\GU(1)_\sleft\times\GU(1)_\sright$ symmetry. Applying c-extremization to the GLSM field content yields the expected values $c_\sleft=c_\sright=3$ for the central charges of the IR theory.

All three GLSMs admit a LGO phase. Consider first the model for $C_1$. For negative values of the FI parameter, the chiral field $\Phi_4$ acquires a vacuum expectation value, breaking the gauge group to $\Z_6$. The remaining fields are described by a Landau--Ginzburg (LG) theory orbifolded by this residual symmetry, which acts on the superfields as
\begin{center}
$\begin{array}{ccccccc}
&\Phi_1&\Phi_2&\Phi_3&\Gamma_1&\Gamma_2&\Gamma_3 \\
g & \zeta & \zeta & \zeta^2 & \zeta & \zeta^5 & \zeta^2
\end{array}$
\end{center}
where $\zeta=\ee^{\frac{2\pi\ii}{6}}$. Although the orbifold action is chiral, the matter content itself is not: the chiral and fermi superfields can be assembled into (2,2) chiral multiplets $X_i=(\Phi_i,\Gamma_i)$. Moreover, the superpotential can be tuned to the (2,2) supersymmetric locus\footnote{In the UV linear model, this corresponds to a choice of superpotential $\cW_1=\Phi_4(\Gamma_1\Phi_1^5+\Gamma_2\Phi_2)+\Gamma_3\Phi_3^2$.}
\begin{align}
W_1 = \tfrac{1}{6}X_1^6+\tfrac{1}{2}X_2^2+\tfrac{1}{3}X_3^3~.
\end{align}
A similar treatment can be given for the other two models. Each has a locus in its moduli space described by a (2,2) LG model with superpotential
\begin{align}
W_2 &= \tfrac{1}{4}X_1^4+\tfrac{1}{2}X_2^2+\tfrac{1}{4}X_3^4~,&
W_3 &= \tfrac{1}{3}X_1^3+\tfrac{1}{2}X_2^2+\tfrac{1}{3}X_3^3+\tfrac{1}{3}X_4^3~,
\end{align}
orbifolded respectively by a $\Z_4$ and a $\Z_3$ symmetry. All three LGO phases correspond to Gepner models. A (2,2) chiral multiplet $X$ with LG superpotential $W=\frac{1}{k}X^k$ flows in the IR to a (2,2) minimal model with central charge $c=3-6/k$. In each of the three LGO models, the multiplet $X_2$ becomes massive and decouples. The orbifold then acts homogeneously on the remaining (2,2) multiplets, yielding the three $c=3$ Gepner models associated with $T^2$.

\subsubsection*{Generalization}
The three GLSMs introduced above share a common geometric structure. In each case, the chiral superfields $\Phi_1$ and $\Phi_2$ define homogeneous coordinates on a $\P^1$, because they cannot vanish simultaneously on the elliptic curve $C$. The hypersurface equation then determines $\Phi_3$ (or $\Phi_3$ and $\Phi_4$ in the third model) in terms of these coordinates. Thus, $C$ is naturally realized as a branched cover of $\P^1$. More precisely:
\begin{enumerate}
\item $C_1$ is a double cover of $\P^1$ branched at $4$ points;
\item $C_2$ is a triple cover of $\P^1$ branched at $3$ points;
\item $C_3$ is a biquadratic cover of $\P^1$ branched at $4$ points.
\end{enumerate}
Each of these constructions admits a natural generalization to a family of CY $d$-folds $Y$, realized as branched covers of $(\P^1)^d$:
\begin{enumerate}
\item $Y_1$ is a double cover, branched along a hypersurface of multidegree $(4,\dots,4)$;
\item $Y_2$ is a triple cover, branched along a hypersurface of multidegree $(3,\dots,3)$;
\item $Y_3$ is a biquadratic cover of $(\P^1)^d$, branched in two hypersurfaces of multidegree $(2,\dots,2)$.
\end{enumerate}
For generic choices of the defining equations, each resulting $Y$ is smooth. These constructions admit a toric realization: the ambient toric varieties are associated to lattice polytopes $\Delta_1$, $\Delta_2$, $\Delta_3$, whose dual polytopes have vertices given by the columns of the following matrices.
\begin{align}
\operatorname{Vert}_{\Delta_1^\circ}&=\begin{pmatrix}
1 & -1 & 0 & 0 & \cdots & 0 & 0 & 0\\
0 & 0 & 1 & -1 & \cdots & 0 & 0 & 0\\
\vdots & \vdots & \vdots & \vdots & \ddots & \vdots & \vdots & \vdots\\
0 & 0 & 0 & 0 & \cdots & 1 & -1 & 0\\
0 & -2 & 0 & -2 & \cdots & 0 & -2 & 1
\end{pmatrix}~,&
\operatorname{Vert}_{\Delta_2^\circ}&=\begin{pmatrix}
1 & -1 & 0 & 0 & \cdots & 0 & 0 & 0\\
0 & 0 & 1 & -1 & \cdots & 0 & 0 & 0\\
\vdots & \vdots & \vdots & \vdots & \ddots & \vdots & \vdots & \vdots\\
0 & 0 & 0 & 0 & \cdots & 1 & -1 & 0\\
0 & -1 & 0 & -1 & \cdots & 0 & -1 & 1
\end{pmatrix}~,
\nonumber\\
\operatorname{Vert}_{\Delta_3^\circ}&=\begin{pmatrix}
1 & -1 & 0 & 0 & \cdots & 0 & 0 & 0 & 0\\
0 & 0 & 1 & -1 & \cdots & 0 & 0 & 0 & 0\\
\vdots & \vdots & \vdots & \vdots & \ddots & \vdots & \vdots & \vdots & \vdots\\
0 & 0 & 0 & 0 & \cdots & 1 & -1 & 0 & 0\\
0 & -1 & 0 & -1 & \cdots & 0 & -1 & 1 & 0\\
0 & -1 & 0 & -1 & \cdots & 0 & -1 & 0 & 1
\end{pmatrix}~.
\end{align}
The ambient toric variety $V$ is singular for $d>1$, but the CY hypersurface (or complete intersection) $Y$ does not cross the singularities of the ambient space. 

The toric description naturally leads to a GLSM realization. As we will see in the next sections, each $Y$ supports abelian instantons obtained by pulling back line bundles from $(\P^1)^d$ to the covering space. In the GLSM realization, these cancel all mixed gauge anomalies, while the remaining diagonal anomalies are canceled by monadic $\cO$ factors. The resulting GLSMs also possess a LG phase, in which the chiral multiplets associated with the trivial bundle factors acquire a vacuum expectation value. These particularly simple examples will be the main protagonists of this note, although the construction is more general; further CY linear models with abelian bundles are provided in appendix~\ref{app:additionalexamples}.

\section{Abelian instantons on K3}
\label{sec:K3instantons}
In this section, we discuss the simplest non-trivial realization of our construction: a K3 surface with an abelian bundle fibered over a single curve. Compactifying the heterotic string on such a background yields a six-dimensional (1,0) supergravity theory with one tensor multiplet and a rank-$15$ gauge group. We first introduce the general framework for abelian bundles with a single K3 class and review the derivation of the massless spacetime spectrum. We then turn to our three central examples, presenting the corresponding GLSMs with abelian bundle configurations and analyzing their phases.

\subsection{Free fermion realization}
A compactification of the heterotic string on a K3 surface $M$ requires a non-trivial gauge bundle $\cE\to M$, with structure group contained in $\GE_8\times\GE_8$ or $\Spin(32)/\Z_2$. The heterotic Bianchi identity imposes the topological constraint
\begin{align}
\tfrac{1}{2}p_1(\cE)=\tfrac{1}{2}p_1(T_M)~.
\end{align}
Since $H^4(M,\Z)=\Z$ and $p_1(T_M)=-48$, this requires a configuration of instanton number twenty-four. In addition, the connection $\cA$ on $\cE$ must satisfy the zero-slope HYM equations.

In order to describe abelian bundles, we first introduce some notation for the relevant weight lattices. We represent the root lattice of $\so(2\tr)$ as
\begin{align}
\Lambda_{\so(2\tr)} &= \left\lbrace \bL = (l_1,\dots,l_\tr)\in\Z^\tr~,~\bL\cdot\bw\in\Z \right\rbrace~,
\end{align}
where $\bw=\frac{1}{2}(1,1,\dots,1)$ is the highest weight of the spinor representation, and we use $\cdot$ to denote the Euclidean inner product. For $\tr=8s$, it has an even self-dual overlattice, which we denote by $\Lambda_{8s}$, obtained by gluing the spinor weight:
\begin{align}
\Lambda_{8s}=\Lambda_{\so(16s)} \cup (\Lambda_{\so(16s)}+\bw)~.
\end{align}
The first lattice in this family, $\Lambda_8$, coincides with the $\Le_8$ root lattice. Let $\balpha_I$, $I=1,\dots,16$, denote a choice of simple roots of either $\Lambda_8\oplus\Lambda_8$ or of $\Lambda_{\so(32)}$, and let $H_I$ be the corresponding Cartan generators of the respective Lie algebra. For an abelian bundle $\cE$, the connection $\cA$ can be expanded as
\begin{align}
\cA = A\, x^I H_I~,
\end{align}
where $A$ is a $\Lu(1)$ connection one-form on K3 with curvature $F=\dd A$ normalized such that
\begin{align}
\upomega = [\tfrac{1}{2\pi}F] \in H^2(M,\Z)~.
\end{align}
The coefficients $x^I$ are subject to quantization conditions. Indeed, the bundle $\cE$ induces an associated vector bundle $\cE_{\rep{r}}$, with transition functions valued in $\GL(\dim\rep{r})$, for each representation $\rep{r}$ of $\Le_8\oplus\Le_8$ or $\so(32)$. The induced connection $\cA_{\rep{r}}$ is diagonal, with each weight $\blambda$ carrying charge $q(\blambda)=x^I\balpha_I\cdot\blambda$. For $\cA$ to define a $\GE_8\times\GE_8$ or $\Spin(32)$ connection, all charges $q(\blambda)$ must be integral. Equivalently,
\begin{align}
\bL = x^I\balpha_I
\end{align}
must belong to the root lattice $\Lambda_8\oplus\Lambda_8$ or $\Lambda_{\so(32)}$. This condition is relaxed for a $\Spin(32)/\Z_2$ connection: charge integrality is required only for weights invariant under the $\Z_2$ action $\exp(2\pi\ii\blambda\cdot\bw)$. Consequently, $\bL$ must take values in the lattice $\Lambda_{16}$.

Hence, the abelian instanton is specified by a K3 class $\upomega$ and a vector $\bL\in\Lambda$ valued in the appropriate rank-$16$ even self-dual lattice ($\Lambda_8\oplus\Lambda_8$ or $\Lambda_{16}$ depending on the heterotic gauge group). The heterotic Bianchi identity reduces to
\begin{align}
(\bL\cdot\bL)(\upomega.\upomega) = -48~,
\label{eq:heteroticBianchiK3}
\end{align}
where we denote by~.~the pairing on $H^2(M,\Z)$. There are finitely many solutions for $\bL$. Moreover, Weyl transformations act on $\bL$ by lattice isometries, so it suffices to classify solutions of~\eqref{eq:heteroticBianchiK3} up to automorphisms of $\Lambda$. The class $\upomega$ is also constrained. Without loss of generality, we can take it to be primitive in the  K3 lattice, rescaling $\bL$ accordingly. Since $\upomega$ is an anti-self-dual class on a smooth K3 surface, it satisfies $\upomega.\upomega\leq-4$~\cite{Barth:2004,Aspinwall:1996mn}. Consequently, only a few values of $\bL\cdot\bL$ are possible.

In the heterotic CFT, the abelian bundle $\cE$ determines the couplings between the sixteen left-moving Weyl fermions and the scalar coordinate fields of the (0,2) NLSM. We assume that the gauge connection admits a regular embedding in $\so(16)\oplus\so(16)$ or $\so(32)$, meaning that $\bL$ belongs to the sublattice $\Lambda_{\so(16)}\oplus\Lambda_{\so(16)}$ or  $\Lambda_{\so(32)}$, respectively. This allows us to describe the configuration in terms of weakly-coupled free fermions.\footnote{Arbitrary principal bundles can be described using fibered WZW models; see~\cite{Distler:2007av}.} Consider the $\Spin(32)/\Z_2$ string. An abelian bundle with $\bL = (l_1,\dots,l_{16})\in\Lambda_{\so(32)}$ may be represented by sixteen Weyl fermions transforming as sections of the holomorphic bundle\footnote{We use the standard notation $\cL^l=\cL^{\otimes l}$, with $\cL^{-l}=(\cL^\ast)^{\otimes l}$.}
\begin{align}
\cE = \cL^{l_1}\oplus\dots\oplus\cL^{l_{16}}~,
\end{align}
where $\cL$ is a line bundle with $c_1(\cL)=\upomega$. The condition $\bL\in\Lambda_{\so(32)}$ ensures that $c_1(\cE)$ is even in $H^2(M,\Z)$, as required for the absence of global anomalies~\cite{Witten:1985mj,Freed:1986zx,Distler:1986wm}. The $\so(32)$ Weyl group acts by redefinitions on the Weyl fermions. Modulo these symmetries, there are thirteen inequivalent choices for $\bL$:
\begin{align}
\bL\cdot\bL &=2~:&
&(1^2,0^{14})~,
\nonumber\\
\bL\cdot\bL &=4~:&
&(1^4,0^{12})~,&
&(2,0^{15})~,
\nonumber\\
\bL\cdot\bL &=6~:&
&(1^6,0^{10})~,&
&(2,1^2,0^{13})~,
\nonumber\\
\bL\cdot\bL &=8~:&
&(1^8,0^8)~,&
&(2,1^4,0^{11})~,&
&(2^2,0^{14})~,
\nonumber\\
\bL\cdot\bL &=12~:&
&(1^{12},0^4)~,&
&(2,1^8,0^7)~,&
&(2^2,1^4,0^{11})~,&
&(2^3,0^{13})~,&
&(3,1^3,0^{12})~,
\label{eq:spin32instantonconfigurations}
\end{align}
where $l^a$ denotes $a$ repeated entries equal to $l$. Each choice corresponds to a class $\upomega$ with self-intersection determined by the Bianchi identity~\eqref{eq:heteroticBianchiK3}. 

For concreteness, we work with the $\Spin(32)/\Z_2$ heterotic string throughout the rest of this work. This is a matter of convenience; the $\GE_8\times\GE_8$ string can be treated similarly. In that case, the abelian configuration is specified by a lattice vector $\bL = (l_1,\dots,l_8)\oplus(l_9,\dots,l_{16})$, represented by a bundle $\cE=\cE_1\oplus\cE_2$, with
\begin{align}
\cE_1 &= \cL^{l_1}\oplus\dots\oplus\cL^{l_8}~,&
\cE_2 &= \cL^{l_9}\oplus\dots\oplus\cL^{l_{16}}~.
\end{align}
The generalization to several curves is also straightforward. Denoting by $\upomega^I$, $I=1,\dots,D$, the corresponding Chern classes, the abelian bundle takes the form $\cE=\oplus_{A=1}^{16} \cL_A$, where $c_1(\cL_A)=\sum_{I=1}^D l_{AI}\upomega^I$. The integers $l_{AI}$ may be assembled into lattice vectors $\bL_I$, and the Bianchi identity reads
\begin{align}
(\boldsymbol{L}_I\cdot\bL_J) (\upomega^I.\upomega^J) =-48~.
\end{align}

\subsection{Massless spectrum}
\label{subsec:K3spectrum}
Heterotic compactifications on a K3 surface $M$ with abelian instantons yield six-dimensional (1,0) supergravity theories with one tensor multiplet. Their gauge symmetry and charged hypermultiplet spectrum can be determined using a Born--Oppenheimer approximation~\cite{Distler:1987ee}. The left-moving current algebra, that characterizes the six-dimensional gauge symmetry, is generated by holomorphic left-moving fermion bilinears. These correspond to roots $\balpha$ of $\so(32)$ satisfying $\balpha\cdot\bL=0$, which generate a $\Lg\oplus \Lu(1)\subset\so(32)$ subalgebra. Under this subalgebra, the $496$ roots of $\Lambda_{\so(32)}$ decompose as
\begin{align}
\rep{496} = \oplus_q \rep{r}_q~,
\end{align}
where $\rep{r}$ denotes a representation of $\Lg$, and $q=\balpha\cdot\bL$ its $\Lu(1)$ charge. Each charged representation $\rep{r}_q$ appearing in this branching contributes a number $n_{\rep{r}_q}$ of hypermultiplets, with multiplicity determined by an index formula and suitable vanishing theorems:
\begin{align}
n_{\rep{r}_q}=h^1(M,\cL^{q}) = -\tfrac{1}{2}q^2\upomega.\upomega-2~.
\end{align}
In addition, there are $20$ neutral hypermultiplets associated to the K3 geometry; their scalar components parametrize deformations of the metric and B-field on $M$.

This description requires an important correction: in the presence of a line bundle, the abelian gauge field is lifted through a  St\"uckelberg mechanism.\footnote{A lucid account of this mechanism for abelian bundles over K3 can be found in~\cite{Melnikov:2012cv}.} On the worldsheet, the quaternionic dimension of the CFT moduli space is reduced by one. Three real constraints arise from the HYM condition on the line bundle connection. The fourth one is due to the trivialization of the component of the B-field along $\upomega$, so that its moduli are valued in $H^2(M,\R)/\la\upomega\ra$. Hence, one full hypermultiplet pairs up with the $\Lu(1)$ vector multiplet into a massive vector multiplet.

This mechanism is also reflected directly in the gauge anomalies of the six-dimensional supergravity. Consider, for instance, a compactification on $M$ with an abelian instanton $\bL=(1^{\text{r}},0^{16-\text{r}})$, with $\tr$ even. Naively, the gauge symmetry is $\so(32-2\text{r})\oplus\su(\text{r})\oplus\Lu(1)$, with hypermultiplets in the representation
\begin{align}
\mathscr{H} = (\rep{f}_{\so(32-2\text{r})},\rep{f}_{\su(\text{r})})_{+1}^{\oplus (\frac{24}{\text{r}}-2)}\oplus(\rep{1},\Lambda^2\rep{f}_{\su(\text{r})})_{+2}^{\oplus(\frac{96}{\text{r}}-2)}\oplus(\rep{1},\rep{1})_{0}^{\oplus20}~,
\label{eq:hypersK3instanton}
\end{align}
where $\rep{f}_{\mathfrak{h}}$ denotes the defining representation of the Lie algebra $\mathfrak{h}$. This gives a supergravity model with one tensor multiplet, $N_{\text{v}}=496-63\text{r} +3\text{r}^2$ vector multiplets, and $N_{\text{h}}=740-63\text{r} +3\text{r}^2$ hypermultiplets. However, the spectrum fails the usual anomaly cancelation conditions, reviewed in appendix~\ref{app:6danomalies}. Indeed, although  $N_{\text{h}}-N_{\text{v}}=244$---as required by the gravitational anomaly constraint---the anomaly polynomial $I_8$ does not factorize but instead takes the form
\begin{align}
I_8 = \tfrac{1}{2}X_4\cdot X_4 + F_{\Lu(1)}X_6~,
\end{align}
where $X_4$ and $X_6$ are polynomials in the Riemann and gauge curvatures.\footnote{This partial factorization of $I_8$ was studied in~\cite{Honecker:2006dt,Honecker:2006qz} in the context of heterotic compactifications with line bundles. Explicit expressions for $X_4$ and $X_6$ are provided in appendix~\ref{app:6danomalies}.} Anomaly cancellation then involves a gauge-variant coupling $C X_6$, where $C$ is an axionic scalar that shifts linearly under gauge transformations. This axion originates from the reduction of the ten-dimensional heterotic B-field over the two-cycle dual to $\upomega$. Its kinetic term reduces to a mass term for the $\Lu(1)$ vector field upon fixing the unitary gauge $C=0$.

After accounting for the Stückelberg mechanism, the heterotic compactification yields a gauge algebra $\Lg$ of rank $15$, and only $19$ neutral hypermultiplets associated to deformations of $M$. The result generalizes to abelian configurations supported on $D$ independent classes in $H^2(M,\Z)$: the St\"uckelberg axions give masses to $D$ abelian gauge bosons, leading to a gauge algebra with $\rank(\Lg) = 16-D$ and a CFT moduli space of quaternionic dimension $20-D$. The algebra $\Lg$ may, however, contain additional non-anomalous $\Lu(1)$ factors.

\subsection{GLSM description}
\label{subsec:K3GLSMs}
We now turn to the GLSM embedding of the abelian instanton configurations. For an abelian bundle $\cE$ fibered over a single curve, the class $\upomega$ is realized in the GLSM as a linear combination
\begin{align}
\upomega = w^\alpha\eta_\alpha
\end{align}
of the integral (1,1) forms $\eta_\alpha$ inherited from the ambient toric variety, with coefficients $w^\alpha\in\Z$ collected into a vector $w$. Its self-intersection is $\upomega.\upomega=w^{\text{t}}Dw$, where $D_{\alpha\beta}=\eta_\alpha.\eta_\beta$ is the intersection pairing on $\operatorname{Pic}(M)$. The class $\upomega$ is anti-self-dual (ASD) on the locus in FI parameter space defined by the linear constraint $w^{\text{t}}D\rho=0$. Given $\upomega$, the abelian bundle bundle $\cE$ is then specified by a lattice vector $\bL$ compatible with the Bianchi identity~\eqref{eq:heteroticBianchiK3}. This vector determines the charges of fermi multiplets $\Lambda_A$, namely $Q_{\Lambda_A}=l_A w$. Their contribution to the gauge anomaly is encoded in the rank-$1$ matrix
\begin{align}
\cA_\cE = (\bL\cdot\bL)ww^{\text{t}}~.
\end{align}
The remaining superfields of the (0,2) GLSM characterize, in the large radius phase, the K3 geometry $M$, together with trivial monadic factors for $\cO\to M$  necessary to cancel the gauge anomalies.

Let us illustrate the construction with the three examples of section~\ref{subsec:threefamilies}, corresponding to heterotic compactifications on K3 surfaces $M_1$, $M_2$, and $M_3$ defined respectively as double, triple, and biquadratic coverings of $\P^1\times\P^1$.\footnote{These K3 geometries were previously used to construct linear models for heterotic flux vacua in~\cite{Israel:2026bco}.} For generic defining polynomials, each surface has a rank-$2$ Picard group generated by $\eta_1,\eta_2\in H^2(M,\Z)$, and intersection pairing
\begin{align}
D =\begin{pmatrix}
0&k\\
k&0\\
\end{pmatrix}~,
\end{align}
where $k=2,3,4$ is the degree of the cover. For definiteness, we choose $\upomega=\eta_1-\eta_2$, which is anti-self-dual for equal FI parameters. We consider the configuration $\bL=(1^{\text{r}},0^{16-\text{r}})$, where $\tr=24/k$, realized by fermi multiplets $\Lambda_{A=1,\dots,\text{r}}$ of appropriate charges.\footnote{The lattice vector encoding the fermi charges is more precisely $(1^{\frac{\tr}{2}},-1^{\frac{\tr}{2}},0^{16-\tr})$. It is related to $\bL$ by a lattice isometry, corresponding to a redefinition $\Lambda'=\overline{\Lambda}$ of some of the fermi superfields. This ensures that the non-anomalous $\GU(1)_\sleft$ symmetry associated with the abelian bundle assigns the same charge to all $\Lambda_A$.} 

The first GLSM is associated to a compactification on $M_1$ equipped with a rank-$12$ abelian bundle $\cE_1$ supported on a $(-4)$-curve. The superfield content and gauge charges are listed in the table below.
\begin{table}[h!]
\centering
$\GLSMblock{T_M}{
\begin{array}{cccccc}
\Phi_{1,2} & \Phi_{3,4} & \Phi_5 & \Gamma_5 \\
1 & 0 & 2 & -4 \\
0 & 1 & 2 & -4 \\
\end{array}
}$
\quad
$\GLSMblock{\cO}{
\begin{array}{ccc}
\Phi_6 & \Gamma_1 & \Gamma_2 \\
-12 & 1 & 11 \\
0 & 0 & 0 \\
\end{array}
}$
\quad
$\GLSMblock{\cO}{
\begin{array}{ccc}
\Phi_7 & \Gamma_3 & \Gamma_4 \\
0 & 0 & 0 \\
-12 & 1 & 11 \\
\end{array}
}$
\quad
$\GLSMblock{\cE}{
\begin{array}{cc}
\Lambda_{1,\dots,6} & \Lambda_{7,\dots,12} \\
1 & -1 \\
-1 & 1 \\
\end{array}
}$
\caption{Field content and gauge charges of the GLSM for $\cE_1\to M_1$}
\label{tab:K3modelM1}
\end{table}

\noindent
The second model describes a compactification on $M_2$ with a rank-$8$ abelian bundle $\cE_2$ supported on a $(-6)$-curve, and has the following superfield content.
\begin{table}[h!]
\centering
$\GLSMblock{T_M}{
\begin{array}{cccccc}
\Phi_{1,2} & \Phi_{3,4} & \Phi_5 & \Gamma_5 \\
1 & 0 & 1 & -3 \\
0 & 1 & 1 & -3 \\
\end{array}
}$
\quad
$\GLSMblock{\cO}{
\begin{array}{ccc}
\Phi_6 & \Gamma_1 & \Gamma_2 \\
-8 & 1 & 7 \\
0 & 0 & 0 \\
\end{array}
}$
\quad
$\GLSMblock{\cO}{
\begin{array}{ccc}
\Phi_7 & \Gamma_3 & \Gamma_4 \\
0 & 0 & 0 \\
-8 & 1 & 7 \\
\end{array}
}$
\quad
$\GLSMblock{\cE}{
\begin{array}{cc}
\Lambda_{1,\dots,4} & \Lambda_{5,\dots,8} \\
1 & -1 \\
-1 & 1 \\
\end{array}
}$
\caption{Field content and gauge charges of the GLSM for $\cE_2\to M_2$}
\label{tab:K3modelM2}
\end{table}

\noindent
The third model describes a compactification on $M_3$ with a rank-$6$ abelian bundle $\cE_3$ supported on a $(-8)$-curve, its superfields are listed in the table below.
\begin{table}[h!]
\centering
$\GLSMblock{T_M}{
\begin{array}{cccccc}
\Phi_{1,2} & \Phi_{3,4} & \Phi_{5,6} & \Gamma_{5,6} \\
1 & 0 & 1 & -2 \\
0 & 1 & 1 & -2 \\
\end{array}
}$
\quad
$\GLSMblock{\cO}{
\begin{array}{ccc}
\Phi_7 & \Gamma_1 & \Gamma_2 \\
-6 & 1 & 5 \\
0 & 0 & 0 \\
\end{array}
}$
\quad
$\GLSMblock{\cO}{
\begin{array}{ccc}
\Phi_8 & \Gamma_3 & \Gamma_4 \\
0 & 0 & 0 \\
-6 & 1 & 5 \\
\end{array}
}$
\quad
$\GLSMblock{\cE}{
\begin{array}{cc}
\Lambda_{1,\dots,3} & \Lambda_{4,\dots,6} \\
1 & -1 \\
-1 & 1 \\
\end{array}
}$
\caption{Field content and gauge charges of the GLSM for $\cE_3\to M_3$}
\label{tab:K3modelM3}
\end{table}

\noindent
In each case, the GLSM admits a geometric phase where both FI parameters are positive. In the large radius limit we expect the IR dynamics to be accurately described by the (0,2) NLSM on $M$ with left-moving fermions supported on the bundle $\cE\to M$. Each model also contains two trivial $\cO\to M$ factors. For instance, one can show that, for the model specified in table~\ref{tab:K3modelM1}, the most general superpotential compatible with the global symmetries can be brought by holomorphic field redefinitions to the form
\begin{align}
\cW = \Phi_6(\Gamma_1\Phi_1^{11}+\Gamma_2\Phi_2)+\Phi_7(\Gamma_3\Phi_3^{11}+\Gamma_4\Phi_4)+\Gamma_5 (\Phi_5^2 - f)~,
\label{eq:superpotentialK3model}
\end{align}
where $f$ is a polynomial in $\Phi_{1,2}$ and $\Phi_{3,4}$ of degree four in each pair. The fermi superfields $\Lambda_{1,\dots,6}$ are excluded from the superpotential by imposing a global $\GU(1)$ symmetry under which they have unit charge and all other fields are neutral. In the large radius phase, the chiral multiplets $\Phi_{1,\dots,4}$ parametrize $\P^1\times\P^1$, while $\Phi_5$ is the coordinate on the double cover, branched along the locus defined by $\{f=0\}\subset(\P^1)^2$. The two monads specified by the superpotential are manifestly trivial, with holomorphic sections given by $(\Phi_2,-\Phi_1^{11})$ and $(\Phi_4,-\Phi_3^{11})$, respectively. 

Compactifying the $\Spin(32)/\Z_2$ heterotic string on these background yields six-dimensional (1,0) theories, and applying the methods reviewed in the previous section, it is straightforward to derive their massless spectra. The resulting gauge groups $G$ and hypermultiplet representations $\mathscr{H}$ are
\begin{align}
G_1&=\Spin(8)\times\SU(12)~,&
\mathscr{H}_1 &=(\rep{1},\rep{66})^{\oplus6}\oplus(\rep{1},\rep{1})^{\oplus19}~,
\nonumber\\
G_2&=\Spin(16)\times\SU(8)~,&
\mathscr{H}_2 &=(\rep{16},\rep{8}) \oplus (\rep{1},\rep{28})^{\oplus 10}\oplus (\rep{1},\rep{1})^{\oplus 19}~,
\nonumber\\
G_3&=\Spin(20)\times\SU(6)~,&
\mathscr{H}_3 &=(\rep{20},\rep{6})^{\oplus2}\oplus (\rep{1},\rep{15})^{\oplus14}\oplus(\rep{1},\rep{1})^{\oplus 19}~.
\label{eq:largeradiusK3spectrum}
\end{align}
The $19\times 4$ neutral real scalars parametrize the metric and B-field deformations on $M$. A particularly appealing feature of the three models---and our main reason to focus on them---is that they all admit a LGO phase. This phase occurs when both FI parameters are negative: in that case the last two chiral superfields acquire vacuum expectation values, breaking the gauge group to a finite subgroup $\Gamma$. Because the fermi multiplets $\Lambda_A$ have no superpotential couplings, they flow to holomorphic Weyl fermions. The remaining superfield content can be assembled into (2,2) supermultiplets $X_i=(\Phi_i,\Gamma_i)$. Moreover, there is a locus in moduli space where the LG superpotential $\cW$ is (2,2) supersymmetric,\footnote{For the first model, this amounts to setting the polynomial $f$ in the superpotential~\eqref{eq:superpotentialK3model} to zero.} and derives from the (2,2) superpotential
\begin{align}
W_1 &= \tfrac{1}{12}(X_1^{12}+X_3^{12})+\tfrac{1}{3}X_5^3 ~,&
\Gamma_1 &=\Z_{12}\times\Z_{12}~,
\nonumber\\
W_2 &= \tfrac{1}{8}(X_1^8 + X_3^8) + \tfrac{1}{4}X_5^4~,&
\Gamma_2 &=\Z_8\times\Z_8~,
\nonumber\\
W_3 &= \tfrac{1}{6}(X_1^6 + X_3^6) + \tfrac{1}{3}(X_5^3 + X_6^3)~,&
\Gamma_3 &=\Z_6\times\Z_6~.
\label{eq:K3modelsLGO}
\end{align}
We also indicate the orbifold group $\Gamma$; it acts asymmetrically on the LG degrees of freedom and the chiral fermions. The LGO models~\eqref{eq:K3modelsLGO} are rational: each (2,2) supermultiplet flows to a (2,2) minimal model, so the IR fixed point is an asymmetric orbifold of a collection of minimal models and free fermions. Moreover, the subgroup of $\Gamma$ that acts trivially on the left-moving fermions coincides with the non-chiral orbifold used in the Gepner construction~\cite{Gepner:1987vz} to project onto integer R-charges: the theory can therefore be viewed as a chiral orbifold of a $c=6$ Gepner model, with a combined action on free left-moving fermions. The resulting SCFT is exactly solvable, allowing us to determine its complete set of primaries, and, in particular, extract the spacetime massless spectrum. This will provide several checks of our supergravity expectations, including the decoupling of the free fermions associated with the trivial bundle factors, as well as the large radius massless spectrum. 

The study of these LGO models is the subject of section~\ref{sec:GepnerSCFT}. We will see that they indeed describe K3 compactifications with an abelian instanton, albeit in a stringy small-volume regime. Before proceeding with the analysis, we pause to make a few comments.

\subsubsection*{General instanton configurations}
The branched cover geometries introduced in section~\ref{subsec:threefamilies} allow to construct many more K3 models with abelian instantons.  There are essentially two ways to generalize the construction: by modifying either the bundle configuration $\bL$ or the class $\upomega$ specifying the principal fibration. Consider, for instance, the GLSM defined in table~\ref{tab:K3modelM3}, which realizes the abelian configuration $\bL=(1^6,0^{10})$. By changing the fermi superfield content, one can realize other configurations with $\bL\cdot\bL=6$. One example is $\bL=(2,1^2,0^{13})$, realized by three fermi multiplets $\Lambda_{1,2,3}$ with the following charges.
\begin{center}
$\GLSMblock{\cE}{
\begin{array}{ccc}
\Lambda_1&\Lambda_2&\Lambda_3 \\
2 & -1 & -1 \\
-2 & 1 & 1 \\
\end{array}
}$
\end{center}
The resulting linear model has similar properties: it still admits a LGO phase with the same $\Z_6\times\Z_6$ orbifold group. The LGO phase only differs in the orbifold action on the left-moving fermions, leading, in particular, to a different spacetime gauge group.

Second, keeping the geometry fixed, we can also vary the charge vector $w$, which specifies the ASD curve over which the bundle is fibered. There are several choices compatible with the Bianchi identity~\eqref{eq:heteroticBianchiK3}, and for each one, a suitable choice of trivial monadic bundles yields an anomaly free (0,2) GLSM with an abelian bundle configuration. By varying $\bL$ and $w$, we can thus realize \emph{all} the bundle configurations in~\eqref{eq:spin32instantonconfigurations} describing K3 compactifications of the $\Spin(32)/\Z_2$ heterotic string with a single abelian instanton. We defer the details of these GLSMs to appendix~\ref{app:generalU1instanton}. They share the essential features of the three models discussed above and can be analyzed in the same way. In particular, they all admit an LGO phase described by an asymmetric orbifold of an LG model and left moving fermions. Moreover, at a special locus in parameter space, the LG degrees of freedom have (2,2) supersymmetry, and the theory can be viewed as a chiral orbifold of a (4,4) K3 Gepner model and of free fermions.

Finally, many further models with abelian bundles can be constructed using different realizations of the K3 surface $M$, beyond the branched covers of $\P^1\times\P^1$ considered here. In the appendices, we present additional constructions in which $M$ is realized as a branched cover of a Hirzebruch surface (appendix~\ref{app:Hirzebruchcover}) or as a hypersurface in $\P^1\times\P^1\times\P^1$ (appendix~\ref{app:CYhypersurfaces}).

\subsubsection*{Relation to heterotic flux vacua}
The GLSMs for abelian instantons on K3 discussed in this section are closely related to another class of (0,2) theories, known as torsional linear sigma models (TLSMs). Introduced in~\cite{Adams:2006kb}, these linear models describe heterotic flux vacua in their large radius phase: the geometry $X$ is a principal $T^D$ fibration over K3, and heterotic compactification on $X$ yields a $(6-D)$-dimensional theory preserving eight supercharges. The torus fibration is characterized by ASD classes $\upomega^I$, $I=1,\dots,D$, and supports non-trivial $H$-flux. In the TLSM, these classes are encoded by charges $w^I$ associated with shift multiplets, namely (0,2) chiral multiplets transforming inhomogeneously under the gauge symmetries.

The structure of TLSMs closely parallels that of the GLSMs for K3 with line bundles. An abelian bundle $\cE$ fibered over $D$ independent classes $\upomega^I$, with configuration vectors $\bL_I$, is encoded by fermi multiplets $\Lambda_A$ with charges $Q_{\Lambda_A}^\alpha=l_{AI}w^{I\alpha}$. Their contribution to the gauge anomaly, arising from the anomalous path-integral measure, is
\begin{align}
\cA_\cE &=(\boldsymbol{L}_I\cdot\bL_J) w^I(w^J)^{\text{t}}~.
\end{align}
This is precisely the contribution of shift multiplets describing a principal $T^D$ fibration with metric $\cG_{IJ}=\frac12\bL_I\cdot\bL_J$ (measured in units of $\alpha'$). The three GLSMs discussed in this section, each with a single $\Lu(1)$ instanton, can thus be used to construct TLSMs describing five-dimensional flux backgrounds, as discussed in~\cite{Israel:2026bco}. In appendix~\ref{app:CYhypersurfaces}, we also present a K3 model with two independent ASD classes, which can be used to construct a six-dimensional flux background. The resulting complex geometry has $\SU(3)$ structure but is non-K\"ahler.

\section{Fermion liberation in (0,2) LGO models}
\label{sec:GepnerSCFT}

In the previous section, we introduced (0,2) GLSMs with a large radius K3 phase and a $\Lu(1)$ instanton. A suitable limit in the FI parameter space of the GLSMs leads to a stringy regime where the K3 volume is of $\alpha'$ size, and the corresponding SCFTs admit an exact description in terms of the LGO models~\eqref{eq:K3modelsLGO}. In this section, we set out the notation and introduce the tools to study these models. In the first two subsections, we give a general description of the orbifolds appearing in these LGO phases, which combine a chiral action on a LG model with an action on left-moving fermions. Using spectral flow, we explain in subsection~\ref{subsec:fermionliberation} the appearance of free fermions in twisted sectors of the orbifold. We then apply this framework in subsection~\ref{subsec:stringyK3}  to the K3 models, and derive the massless spectra of the resulting six-dimensional (1,0) supergravity theories.

This section is somewhat more technical than the rest of the note, drawing on methods from KM algebras and (2,2) Gepner models. Readers who do not wish to delve into these details  may proceed directly to the results in section~\ref{subsec:stringyK3}.

\subsection{Orbifolds and modular orbits}
Consider a two-dimensional CFT with a symmetry $g$ generating a finite cyclic group $\Gamma$. Gauging this symmetry amounts to introducing twisted sectors and projecting onto $\Gamma$-invariant states. This procedure is not always consistent: the symmetry may carry an anomaly, classified by $H^3(\Gamma,\GU(1))$, which obstructs gauging. A necessary consistency condition is modular invariance of the orbifold partition function, which takes the form
\begin{align}
Z_{\text{orb}}(\tau,\overline{\tau}) = \frac{1}{|\Gamma|}\sum_{a,b\in\Gamma} Z^{(a,b)}(\tau,\overline{\tau})~.
\label{eq:modularorbits}
\end{align}
Here $Z^{(a,b)}$ denotes the trace over the $g^a$-twisted Hilbert space with an insertion of $g^b$. In practice, these partial traces are often constructed using modular orbits, without explicitly identifying the twisted Hilbert spaces: one computes the untwisted traces and generates the remaining ones through $\SL(2,\Z)$ transformations. Modular invariance alone, however, does not guarantee consistency. In particular, the resulting partition function need not have non-negative integer multiplicities, as required for a Hilbert space interpretation---a recent discussion of these subtleties may be found in~\cite{Robbins:2019zdb}.

A key simplification arises when $\Gamma$ is a subgroup of a global $\GU(1)$ symmetry realized on the worldsheet as a current algebra. Suppose the current is left-moving, with generator $J_\sleft$, normalized so that all local operators carry integer charges $q_\sleft$. Its OPE is
\begin{align}
J_\sleft(z)J_\sleft(0) \sim \frac{2\text{k}}{z^2}~,
\end{align}
where $\text{k}$ is a positive integer. We can then introduce the flavored partition function of the original CFT,
\begin{align}
Z_{\text{flav}}(\tau,\bar{\tau},z)= \operatorname{tr}_{\cH}\!\left[ \text{q}^{L_0-c_\sleft/24}\overline{\text{q}}^{\overline{L}_0-c_\sright/24} \ee^{2\pi\ii z J_{\sleft 0}} \right]
~,
\end{align}
which obeys the universal modular transformation law~\cite{Dijkgraaf:1987vp,Dijkgraaf:1996xw,Kraus:2006wn,Benjamin:2016fhe,Dyer:2017rul}
\begin{align}
Z_{\text{flav}}(-1/\tau,-1/\bar{\tau},z/\tau)= \ee^{2\pi\ii\text{k}\frac{z^2}{\tau}} Z_{\text{flav}}(\tau,\bar{\tau},z)~,
\end{align}
independently of the details of the theory. The generator of the finite subgroup $\Gamma\subset\GU(1)$ can be written as
\begin{align}
g = \exp(2\pi\ii \mu J_{\sleft 0})~,
\end{align}
with $\mu\in\frac{1}{|\Gamma|}\Z$. Acting with $\SL(2,\Z)$ on the untwisted traces with insertions of $g^b$ generates the partial traces
\begin{align}
Z^{(a,b)}(\tau,\overline{\tau}) = \ee^{2\pi\ii\text{k}\mu^2 a( a\tau + b )} Z_{\text{flav}}(\tau,\overline{\tau},(a\tau+b)\mu)~,
\label{eq:partialtraces}
\end{align}
which transform as
\begin{align}
Z^{(a,b)}(\tau+1,\overline{\tau}+1)&= Z^{(a,b+a)}(\tau,\overline{\tau})~,&
Z^{(a,b)}(-1/\tau,-1/\overline{\tau})&= Z^{(b,-a)}(\tau,\overline{\tau})~.
\end{align}
The indices $a,b$ are initially integers, but consistency of the orbifold partition function~\eqref{eq:modularorbits} requires them to be periodic modulo \(|\Gamma|\). The condition $Z^{(a,b+|\Gamma|)}=Z^{(a+|\Gamma|,b)}=Z^{(a,b)}$ is equivalent to
\begin{align}
\ee^{2\pi\ii\text{k}\mu^2|\Gamma|}=1~.
\end{align}
The phase on the left encodes the class in $H^3(\Gamma,\GU(1))\simeq\Gamma$ whose nontriviality obstructs gauging $\Gamma$.

The twisted Hilbert spaces can be constructed explicitly by spectral flow.\footnote{See~\cite{Melnikov:2019tpl} for a pedagogical account of this construction.} Bosonizing the left-moving current in terms of a free chiral boson, one can introduce twist fields $\Sigma_a(z)$, realized as vertex operators of this boson. Their OPEs with charged operators produce the required monodromies. The states $|\Sigma_a\rangle$ correspond to the twisted vacua, and spectral flow provides an isomorphism between the twisted Hilbert spaces $\cH_a$ and the untwisted Hilbert space $\cH$. In particular, an operator $\cO$ acting in the $a$-th twisted sector can be mapped to an operator $\cO^{(a)}$ in the untwisted sector, such that
\begin{align}
\cO|\Sigma_a\ra = \cO^{(a)}|0\ra~.
\end{align}
The spectral flow isomorphism respects the current and Virasoro algebras, with twisted generators
\begin{align}
L^{(a)}_m&= L_m+\mu a J_{\sleft m}+ \text{k}\mu^2 a^2 \delta_{m,0} ~,&
J^{(a)}_{\sleft m}&=J_{\sleft m}+2\text{k}\mu a\delta_{m,0}~.
\end{align}
In particular, the partial traces~\eqref{eq:partialtraces} can be expressed as
\begin{align}
Z^{(a,b)}(\tau,\overline{\tau}) = \operatorname{tr}_{\cH_a}\!\left[ \text{q}^{ L_0-c_\sleft/24}\overline{\text{q}}^{\overline{L}_0-c_\sright/24} \ee^{2\pi\ii b\mu( J_{\sleft 0} - \text{k}\mu a )} \right]
~,
\end{align}
confirming their Hilbert space interpretation. Note that the orbifold projection in the $a$-th twisted sector retains states satisfying $\mu q_\sleft-\text{k}\mu^2 a \in\Z$.

The generalization to a product of cyclic groups $\Gamma$, each embedded in a (possibly distinct) left-moving $\Lu(1)$ symmetry, is straightforward. The twist fields $\Sigma_a$ are now labeled by multi-indices $a=(a_1,\dots,a_r)$ specifying the twisted sectors, and are constructed by bosonizing the corresponding currents. Unlike the cyclic case, there may be discrete torsion, classified by \(H^2(\Gamma,\GU(1))\), which modifies the orbifold projection in the twisted sectors. The spectral flow construction also extends to superconformal theories. In what follows, we use twist fields to construct the NS Hilbert space of the orbifold (0,2) SCFT, imposing, in particular, integrality of R-charges after the orbifold projection. The Ramond sector may then be obtained by spectral flow with respect to the right-moving R-symmetry current.

Since the twist field approach gives an explicit construction of the twisted Hilbert spaces $\cH_a$, spectral flow also provides a convenient way to decompose the action of an orbifold on a product theory. The LGO models considered here all have the same structure: $\Gamma$ acts on the product of a (2,2) LG model and the sixteen heterotic Weyl fermions. Each generator $g_\alpha$ decomposes as $g_\alpha=g_\alpha^{\LG} g_\alpha^{\KM}$, where $g_\alpha^{\LG} $ is embedded in a $\Lu(1)$ symmetry of the LG model, while $g_\alpha^{\KM}$ belongs to a $\Lu(1)$ subalgebra of the KM symmetry of the fermion system. More explicitly,
\begin{align}
g_\alpha = \exp\left(2\pi\ii \sum_{i=1}^{N_{\LGsmall}}\mu^\alpha_i J^i_{\sleft0} + 2\pi\ii \sum_{j=1}^{N_{\KMsmall}} \nu^\alpha_j K^j_{\sleft0} \right)~,
\end{align}
where $J_\sleft^i$ and $K_\sleft^j$ are left-moving $\Lu(1)$ currents of the LG model and the heterotic fermions, respectively---we will introduce the notation in detail in the following subsection. The twisted Hilbert spaces $\cH_a^{\LG}$ and $\cH_a^{\KM}$ can then be constructed separately for the two factors, using their respective twist fields. In particular, the partial traces are well-defined for each factor, even though the corresponding orbifold actions are individually anomalous. This is reflected in their modular transformations,
\begin{align}
Z^{(a,b)}_{\LG}(-1/\tau,-1/\bar{\tau}) &= \ee^{2\pi\ii a_\alpha b_\beta A^{\alpha\beta}_{\LG}} Z^{(b,-a)}_{\LG}(\tau,\bar{\tau})~,&
Z^{(a,b)}_{\KM}(-1/\tau) &= \ee^{2\pi\ii a_\alpha b_\beta A^{\alpha\beta}_{\KM}} Z^{(b,-a)}_{\KM}(\tau)~,
\label{eq:anomalyLGKM}
\end{align}
where the matrices $A_{\LG}$ and $A_{\KM}$ have rational entries that give rise to non-trivial phases. If their sum $A_{\LG}+A_{\KM}$ has integer entries, these phases cancel between the two factors, rendering the combined orbifold consistent. The resulting SCFT has (NS,NS) partition function
\begin{align}
Z = \frac{1}{|\Gamma|}\sum_{a,b\in\Gamma} Z^{(a,b)}_{\LG} Z^{(a,b)}_{\KM}~.
\label{eq:orbifoldpartitionfunctionLGKM}
\end{align}

As a final general remark, because the SCFTs considered below are rational, their partition functions determine not only the state degeneracies at each Virasoro level, but also the decomposition of the spectrum into representations of the left- and right-moving chiral algebras. Although we will not explicitly construct the full orbifold CFT---we will not attempt a complete characterization of OPEs among twisted sector operators---we can nevertheless determine the complete spectrum of primary representations by identifying the corresponding characters from the partition function.

\subsection{(0,2) Gepner orbifolds}
\label{subsec:LGOconstruction}
The construction of twisted Hilbert spaces via spectral flow provides a useful perspective on the LGO models introduced in section~\ref{subsec:K3GLSMs}. The three models~\eqref{eq:K3modelsLGO} share a common structure: their (0,2) orbifold SCFTs arise from a combined orbifold action that acts chirally on a (2,2) LG model with central charge $c=3d$ and on free left-moving fermions. To interpret these theories as heterotic string vacua, we fermionize the $c_\sleft=16$ chiral bosons associated with $\Spin(32)/\Z_2$ into $16$ left-moving Weyl fermions. We then regard the LG model as accounting for $d$ of the heterotic left-moving fermions, while the remaining $16-d$ fermions generate an $\so(32-2d)$ current algebra. The left-moving fermions $\Lambda_A$ of the (0,2) GLSM are embedded into this heterotic fermion system. The orbifold group takes the form
\begin{align}
\Gamma = (\Z_{\text{N}})^r~.
\end{align}
We restrict to orbifold groups of this form for definiteness, but the methods presented here also apply to more general abelian groups $\Gamma$.

\subsubsection*{LG model}
Consider a (2,2) LG model with $N_{\LG}$  chiral multiplets $X_i$ and a Fermat-type superpotential
\begin{align}
W= \sum_{i=1}^{N_{\LGsmall}} \tfrac{1}{k_i}X_i^{k_i}~.
\end{align}
We set $\alpha_i=1/k_i$, with the levels $k_i$ satisfying
\begin{align}
\sum_{i=1}^{N_{\LGsmall}}(1-2\alpha_i)=d~.
\end{align}
Decomposing each multiplet as $X_i=(\Phi_i,\Gamma_i)$, with $\Phi_i$ a (0,2) chiral multiplet and $\Gamma_i$ a fermi multiplet, makes a $\GU(1)_\sleft^{N_{\LGsmall}}\times\GU(1)_\sright$ global symmetry of this superpotential manifest. The corresponding charges are listed below.
\begin{center}
\arraycolsep=8pt
$\begin{array}{cccc}
&\Phi_i & \Gamma_i & \theta \\
q_\sleft^j & \alpha_i\delta_{ij} & (\alpha_i-1)\delta_{ij} & 0 \\
q_\sright & \alpha_i & \alpha_i & 1 \\
\end{array}$
\end{center}
The LG model flows in the IR to a (2,2) SCFT with central charge $c_\sleft=c_\sright=3d$. The global symmetries give rise to a $\Lu(1)_\sleft^{\oplus N_{\LGsmall}}\oplus\Lu(1)_\sright$ current algebra. The two-point functions of the currents take the form
\begin{align}
\la J_\sleft^i(z)J_\sleft^j(0) \ra &= \frac{M_{\LG}^{ij}}{z^2}~,&
\la J_\sright(\zb)J_\sright(0) \ra &= \frac{c_\sright/3}{\zb^2}~,
\end{align}
where $M_{\LG}$ is the positive definite matrix
\begin{align}
M_{\LG}^{ij}&=\delta^{ij}(1-2\alpha_i)~.
\end{align}
Each multiplet $X_i$ flows to a (2,2) minimal model, whose primary representations  are described by the coset $(\su(2)_{k_i-2}\oplus\Lu(1)_2)/\Lu(1)_{k_i}$. The partition function in the NS sector takes the form
\begin{align}
Z_{\LG}^{(0,0)} &=\frac{1}{2^{N_{\LGsmall}}}\sum_{\boldsymbol{l},\boldsymbol{m}} 
\chi^{\boldsymbol{l}}_{\boldsymbol{m}} \overline{\chi}^{\boldsymbol{l}}_{\boldsymbol{m}}~,
\end{align}
where $\chi^{\boldsymbol{l}}_{\boldsymbol{m}}$ are products of minimal model characters.\footnote{Our conventions for (2,2) minimal models follow~\cite{Israel:2026bco} and are reviewed in appendix~\ref{app:minimalmodels}.}

Consider now an orbifold by $\Gamma=(\Z_{\text{N}})^r$, with generators $g_\alpha^\LG$ acting as
\begin{align}
\Phi_i&\to \ee^{2\pi\ii \mu^\alpha_i \alpha_i}\Phi_i~,&
\Gamma_i&\to \ee^{2\pi\ii \mu^\alpha_i (\alpha_i-1)}\Gamma_i~,
\end{align}
where $\mu^\alpha_i$ are rational numbers satisfying $\mu^\alpha_i\alpha_i\text{N}\in\Z$. We assemble them into an $r\times N_{\LG}$ matrix $\boldsymbol{\mu}$, whose columns are the charge vectors $\mu_i$.  The orbifold group $\Gamma$ is embedded in the global symmetry and, at the IR fixed point, it can be represented by the operator
\begin{align}
g^{\LG}_\alpha = \exp\left(2\pi\ii\sum_{i=1}^{N_{\LGsmall}}\mu^\alpha_i J_{\sleft0}^i\right)~.
\end{align}
The twisted Hilbert spaces can be constructed explicitly using twist fields $\Sigma^\LG_a$ built from the currents. The resulting partial traces $Z^{(a,b)}_\LG$ in the NS sector have the modular transformation properties given in~\eqref{eq:anomalyLGKM}, with anomalous phases determined by the matrix 
\begin{align}
A_{\LG} &= \boldsymbol{\mu}M_{\LG}\boldsymbol{\mu}^{\text{t}}~.
\label{eq:anomalyLG}
\end{align}
When $\mu^\alpha_i\in\tfrac{1}{2}\Z$, these twisted NS partition functions may be expressed in terms of coset characters as
\begin{align}
Z_{\LG}^{(a,0)} &=\frac{1}{2^{N_{\LGsmall}}}\sum_{\boldsymbol{l},\boldsymbol{m},\boldsymbol{\varepsilon}} 
\chi^{\boldsymbol{l}}_{\boldsymbol{m}+2a\cdot\boldsymbol{\mu}}{}^{(2\boldsymbol{\varepsilon}+2a\cdot\boldsymbol{\mu})} \overline{\chi}^{\boldsymbol{l}}_{\boldsymbol{m}}~,
\label{eq:twistedLGpartitionfunction}
\end{align}
where $a\cdot\boldsymbol{\mu}=a^{\text{t}}\boldsymbol{\mu}$. See appendix~\ref{app:minimalmodels} for the notation and further details.

\subsubsection*{Chiral fermions}
The LG model is supplemented by a collection of $16-d$ free left-moving Weyl fermions $\lambda_A(z)$, with two-point functions
\begin{align}
\la \lambda_A(z)\overline{\lambda}_B(0)\ra = \frac{\delta_{AB}}{z}~.
\end{align}
Their bilinears generate a $\so(32-2d)$ KM algebra at level $1$. We decompose this symmetry as
\begin{align}
\so(32-2d)\quad\supset\quad \so(2\text{r}_0)\oplus \Lu(\text{r}_1)\oplus \dots\oplus \Lu(\text{r}_{N_{\KMsmall}})~,
\label{eq:chiralfermionembedding}
\end{align}
where $\text{r}_1,\dots,\text{r}_{N_\KMsmall}$ are even integers and $\text{r}_0=16-d-\sum_{i=1}^{N_{\KMsmall}}\text{r}_i$. Accordingly, we relabel the fermions as $\lambda^i_s$, with $i=0,1,\dots,N_{\KM}$ and $s=1,\dots,\text{r}_i$. Each $\Lu(\text{r}_i)=\su(\text{r}_i)\oplus\Lu(1)$ algebra is generated by the fermion bilinears  $:\lambda^i_s\overline{\lambda}^i_{s'}:$, with the $\Lu(1)$ factor associated to the current
\begin{align}
K_\sleft^i &= \sum_{s=1}^{\text{r}_i} :\lambda^i_s\overline{\lambda}^i_s:~.
\end{align}
These chiral currents have two-point functions
\begin{align}
\la K_\sleft^i  (z)K_\sleft^j(0) \ra &= \frac{M_{\KM}^{ij} }{z^2}~,&
M_{\KM}^{ij} &= \delta^{ij}\text{r}_i~.
\end{align}
Thus, each $\Lu(1)$ current algebra has level $\tr_i/2$---see appendix~\ref{app:conformalembeddings} for our KM algebra conventions. The partition function of the chiral fermion system may be decomposed into characters of the KM algebra~\eqref{eq:chiralfermionembedding}. In the NS sector, it takes the form
\begin{align}
Z_{\KM}^{(0,0)} &= \chi^{\so(32-2d)}_{\rep{1}} + \chi^{\so(32-2d)}_{\rep{v}} 
\nonumber\\
&=( \chi^{\so(2\text{r}_0)}_{\rep{1}} + \chi^{\so(2\text{r}_0)}_{\rep{v}} ) \prod_{i=1}^{N_\KMsmall}\sum_{m_i\in\Z_{\text{r}_i}}\chi^{\su(\text{r}_i)}_{\rep{r}_{m_i}}\chi^{\Lu(1)_{\tr_i/2}}_{m_i}~.
\end{align}
Here $\chi^{\Lg}_{\rep{r}}$ denotes a character of the level-$1$ $\Lg$ current algebra, labeled by a representation $\rep{r}$ of the Lie algebra $\Lg$, while $\chi^{\Lu(1)_\tk}_m$ are characters for the $\Lu(1)$ current algebra at level $\tk$, labeled by $m\in\Z_{2\tk}$. Note that~\eqref{eq:chiralfermionembedding} is a conformal embedding.

The orbifold group $\Gamma=(\Z_{\text{N}})^r$ leaves the fermions generating the $\so(2\tr_0)$ symmetry invariant. Its generators $g_\alpha^\KM$ act on the remaining fermions as
\begin{align}
\lambda^i_s \to \ee^{2\pi\ii \nu^\alpha_i }\lambda^i_s~,
\end{align}
where $\nu^\alpha_i$ are rational numbers satisfying $\nu^\alpha_i\text{N}\in\Z$. We assemble them into an $r\times N_{\KM}$ matrix $\boldsymbol{\nu}$, whose columns are the charge vectors $\nu_i$.  The symmetry generators may be represented by the operators
\begin{align}
g^{\KM}_\alpha &= \exp\left(2\pi\ii \sum_{i=1}^{N_\KMsmall} \nu^\alpha_i K_{\sleft0}^i\right)~.
\end{align}
As in the LG sector, the twisted Hilbert spaces can be constructed explicitly using twist fields $\Sigma^\KM_a$, and the resulting NS-sector partial traces $Z^{(a,b)}_\KM$ obey the modular transformation law~\eqref{eq:anomalyLGKM}. The anomalous phases are determined by the matrix 
\begin{align}
A_{\KM} &= \boldsymbol{\nu}M_{\KM}\boldsymbol{\nu}^{\text{t}}~.
\label{eq:anomalyKM}
\end{align}

The twisted Hilbert spaces $\cH^{(a,b)}_\KM$ can be decomposed into primary representations of the KM algebra~\eqref{eq:chiralfermionembedding}. The only subtlety concerns the $\Lu(1)$ factor: the action of $\Gamma$ need not preserve the primary representation of the level-$\frac{\tr_i}{2}$ algebras. We therefore make the decomposition $\Lu(1)_{\tr_i/r}\supset \Lu(1)_{\tR_i}$, with $\tR_i=\tr_i\tN^2/2$, so that $\Gamma$ acts homogeneously within each representation; see appendix~\ref{app:fermionorbifold} for details. With this refinement, the twisted NS partition functions  take the form
\begin{align}
Z^{(a,0)} =( \chi^{\so(2\text{r}_0)}_{\rep{1}} + \chi^{\so(2\text{r}_0)}_{\rep{v}} ) \prod_{i=1}^{N_\KMsmall}\sum_{m_i\in\Z_{\text{r}_i}}\sum_{l_i\in\Z_\tN} \chi^{\su(\text{r}_i)}_{\rep{r}_{m_i}} \chi^{\Lu(1)_{\tR_i}}_{\tN(m_i+\text{r}_i l_i+\text{r}_i\nu_i\cdot a )} ~.
\end{align}

\subsubsection*{Chiral GSO projection and (0,2) supersymmetry}
Assuming the chiral orbifold by $\Gamma$ is non-anomalous, additional bells and whistles are required for the orbifold to define a consistent heterotic background. The orbifold theory must possess a right-moving $c_\sright=3d$ superconformal symmetry, and the heterotic left-moving fermions are subject to a chiral GSO projection.

The superconformal symmetry is inherited from the direct sum of the minimal model right-moving SCAs, each with $c_\sright=3(1-2\alpha_i)$. In a (2,2) model, the Gepner construction involves orbifolding the LG model by the symmetry
\begin{align}
g_{\text{gep}} = \exp\left(2\pi\ii J_{\sleft0}\right)~,
\end{align}
where 
\begin{align}
J_\sleft = \sum_{i=1}^{N_{\LGsmall}} J_\sleft^i
\end{align}
is the current that becomes the left-moving R-current of the (2,2) orbifold SCFT. This projection ensures integral left- and right-moving R-charges $q_\sleft$ and $q_\sright$. We denote by $\Gamma_{\text{gep}}$ the cyclic subgroup generated by $g_{\text{gep}}$, and we assume that $\Gamma_{\text{gep}}\subset\Gamma$. Since the states of the original LG model satisfy $q_\sleft-q_\sright\in\Z$, the orbifold projection ensures integral right-moving R-charges in the orbifold (0,2) SCFT. 

This also gives a useful perspective on the orbifold symmetry. The group $\Gamma$ belongs to a short exact sequence
\begin{align}
0 \to \Gamma_{\text{gep}}\to\Gamma\to\Gamma_{\text{chir}}\to 0~,
\end{align}
where $\Gamma_{\text{chir}}=\Gamma/\Gamma_{\text{gep}}$. We can therefore view the orbifold by $\Gamma=\Gamma_{\text{gep}}\rtimes\Gamma_{\text{chir}}$ in two steps. First, orbifolding by $\Gamma_{\text{gep}}$ yields a $c_\sleft=c_\sright=3d$ Gepner model with (2,2) supersymmetry, describing a sigma model on a CY $d$-fold at a special point in moduli space. A further orbifold by $\Gamma_{\text{chir}}$ acts chirally on the Gepner model, with a combined action on the left-moving heterotic fermions. The resulting (0,2) SCFT has an integral R-charge spectrum.

Finally, the chiral GSO projection on the left-moving fermions is implemented by the $\Z_2$ symmetry
\begin{align}
\ee^{\ii\pi F_\sleft} = \ee^{\ii\pi J_{\sleft0}}  \ee^{\ii\pi K_{\sleft0}} ~,
\end{align}
where the current $K_\sleft$ is given by
\begin{align}
K_\sleft = \sum_{A=1}^{16-d}:\lambda_A\overline{\lambda}_A:~.
\end{align}
The operator $\ee^{\ii\pi K_{\sleft0}}$ may also be identified with a center symmetry of the algebra~\eqref{eq:chiralfermionembedding}. The projection onto $\ee^{\ii\pi F_\sleft} = 1$ may be thought of as a $\Z_2$ orbifold, whose twisted sector corresponds to the left-moving Ramond sector. From this perspective, the orbifold group is enlarged to $\Gamma\rtimes\Z_2$, and NS and R boundary conditions are introduced only on the right-moving side.

\subsection{Spectral flow and free fermions}
\label{subsec:fermionliberation}
With this framework in place, we can now examine the orbifold (0,2) SCFT. The orbifold group $\Gamma=(\Z_\tN)^r$ is generated by $g_\alpha=g_\alpha^\LG g_\alpha^\KM$, where $g_\alpha^\LG$ acts on a (2,2) LG model with central charge $c=3d$, while $g_\alpha^\KM$ acts on $16-d$ free left-moving complex fermions. The two orbifold actions are individually anomalous, with modular transformation laws~\eqref{eq:anomalyLGKM} encoded in the matrices $A_\LG$ and $A_\KM$. However, the combined action is anomaly free for the models studied here, which arise as the LGO phases of (0,2) GLSMs. In each case, we will find that
\begin{align}
A_{\LG}^{\alpha\beta}+A_{\KM}^{\alpha\beta}=\delta^{\alpha\beta}~.
\label{eq:totalanomaly}
\end{align}
This ensures that the NS partition function~\eqref{eq:orbifoldpartitionfunctionLGKM} of the orbifold SCFT is well defined.

The relation~\eqref{eq:totalanomaly} has further important consequences. Let us write the orbifold generators as $g_\alpha = \exp(2\pi\ii \cJ^\alpha_{\sleft0})$, where $\cJ^\alpha_{\sleft0}$ are the zero modes of the currents
\begin{align}
\cJ^\alpha_\sleft = \sum_{i=1}^{N_\LG} \mu^\alpha_i J^i_\sleft+\sum_{j=1}^{N_\KM}\nu^\alpha_j K^j_\sleft~.
\label{eq:JLcurrents}
\end{align}
Using~\eqref{eq:anomalyLG} and~\eqref{eq:anomalyKM}, we readily find their two-point functions:
\begin{align}
\la\cJ^\alpha_\sleft(z)\cJ^\beta_\sleft(0) \ra =\frac{\delta^{\alpha\beta}}{z^2}~.
\end{align}
The orbifold twist fields $\Sigma_a = \Sigma^\LG_a\Sigma^\KM_a$ can then be expressed directly in terms of these currents. To see this, we bosonize the currents as
\begin{align}
\cJ^\alpha_\sleft=\ii\partial\varphi^\alpha~,    
\end{align}
where $\varphi^\alpha$ are free chiral bosons with $\la\varphi^\alpha(z)\varphi^\beta(0)\ra=-\delta^{\alpha\beta}\ln z$. Their vertex operators $\ee^{\ii\xi\cdot\varphi}$, with $\xi\in\R^r$, obey the OPE\footnote{Here and in what follows, normal ordering of operators is understood.}
\begin{align}
\ee^{\ii\xi_1\varphi(z)} \ee^{\ii\xi_2\varphi(0)} \sim z^{\xi_1\cdot\xi_2} \ee^{\ii\xi_1\varphi(z)+\ii\xi_2\varphi(0)}~.
\label{eq:vertexoperatorsOPE}
\end{align}
These vertex operators induce nontrivial monodromies in charged fields and can therefore be used to construct the twisted Hilbert spaces. In particular, the twist fields take the form
\begin{align}
\Sigma_a = \ee^{\ii a\cdot\varphi}~,
\end{align}
where the labels $a_\alpha\in\Z_{\text{N}}$ are chosen in the range $-\text{N}/2<a_\alpha\leq\text{N}/2$. Their conformal dimensions are
\begin{align}
h_\sleft(\Sigma_a) &= \tfrac{1}{2}(a_1^2+\dots+a_r^2)~,&
h_\sright(\Sigma_a) &=0~.
\end{align}
Thus, the twist fields give rise to a distinguished set of holomorphic operators in the orbifold SCFT. Among these twisted ground states, those of low conformal dimension, listed in table~\ref{tab:twistfields}, play a special role. We find $2r$ twisted ground states with spin $\frac{1}{2}$, and it is tempting to interpret these as free left-moving Majorana--Weyl fermions. We also find $2r(r-1)$ spin-$1$ currents, suggesting an enhancement of the $\Lu(1)^{\oplus r}$ algebra generated by the $\cJ^\alpha_\sleft$ to $\so(2r)$.
\begin{table}[h!]
\centering
\arraycolsep=6pt
\renewcommand{\arraystretch}{1.2}
$\begin{array}{c|ccccc}
\Sigma &  \ee^{\ii\varphi^\alpha} & \ee^{-\ii\varphi^\alpha} & \ee^{\ii(\varphi^\alpha+\varphi^\beta)}  & \ee^{-\ii(\varphi^\alpha+\varphi^\beta)}  & \ee^{\ii(\varphi^\alpha-\varphi^\beta)} \\ \hline
h_\sleft(\Sigma) & \frac{1}{2} & \frac{1}{2} & 1 & 1 & 1 \\
\#(\Sigma) & r & r & \frac{1}{2}r(r-1) & \frac{1}{2}r(r-1) & r(r-1)
\end{array}$
\caption{Twist fields $\Sigma$ of small conformal weight}
\label{tab:twistfields}
\end{table}

Both expectations are indeed verified. From the twist fields with $h_\sleft=\frac12$, we can construct anticommuting fields
\begin{align}
\Sigma^\alpha &= :\ee^{\ii \varphi^\alpha}: c_\alpha~,&
\overline{\Sigma}^\alpha &= :\ee^{-\ii \varphi^\alpha}: c_\alpha ~,
\end{align}
which satisfy the free-fermion OPEs
\begin{align}
\Sigma^\alpha(z)\overline{\Sigma}^\beta(0) \sim \frac{\delta^{\alpha\beta}}{z}~.
\end{align}
The cocycle factors $c_\alpha$ ensure that the fermions anticommute.\footnote{These cocycles, familiar in string theory textbooks, can be written explicitly as $c_\alpha=\exp(\ii\pi\sum_{\beta=1}^{\alpha-1}\cJ_{\sleft0}^\beta)$~\cite{Kostelecky:1986xg}.} The spin-one currents are precisely the bilinears of these fermions. Using the OPEs~\eqref{eq:vertexoperatorsOPE}, one can verify explicitly that they generate an $\so(2r)$ current algebra at level $1$, whose Cartan subalgebra is spanned by the currents $\cJ^\alpha_\sleft$.

\subsection{Stringy K3 SCFTs}
\label{subsec:stringyK3}
We now return to the (0,2) GLSMs introduced in section~\ref{subsec:K3GLSMs}, associated with K3 compactifications with a $\Lu(1)$ instanton. For the three models considered there, we identified a locus in the GLSM parameter space for which the IR theory is an orbifold of the type described above. Let us focus on the first of the three LGO models~\eqref{eq:K3modelsLGO}. Its orbifold group $\Gamma=\Z_{12}\times\Z_{12}$ is embedded in a $\GU(1)\times\GU(1)$ global symmetry of the LG model and the chiral fermion system. The action on the fermions breaks the $\so(28)$ symmetry to $\so(4)\oplus\su(12)\oplus\Lu(1)$. The $\Lu(1)$ factor is generated by the current $K_\sleft$ entering the construction of the orbifold generators, while the $\so(4)\oplus\su(12)$ symmetry remains unaffected by the orbifold. In the notation introduced above, the charges and levels take the form
\begin{align}
\boldsymbol{\mu} &=\begin{pmatrix}
1&0&\frac{1}{2} \\
0&1&\frac{1}{2}
\end{pmatrix}~,&
\boldsymbol{\nu} &=\begin{pmatrix}
\frac{1}{12} \\
-\frac{1}{12}
\end{pmatrix}~,&
M_\LG &=\diag(
\tfrac{5}{6},\tfrac{5}{6},\tfrac{1}{3})~,&
M_\KM &=12~,
\end{align}
so that the relation~\eqref{eq:totalanomaly} is satisfied, with
\begin{align}
A_\LG &= \begin{pmatrix}
\frac{11}{12} & \frac{1}{12} \\
\frac{1}{12} & \frac{11}{12} \\
\end{pmatrix}~,&
A_\KM &= \begin{pmatrix}
\frac{1}{12} & -\frac{1}{12} \\
-\frac{1}{12} & \frac{1}{12} \\
\end{pmatrix}~.
\end{align}
Consequently, the twisted ground states of the $\Z_{12}\times\Z_{12}$ orbifold include four free left-moving Majorana--Weyl fermions, together with currents for the associated $\so(4)$ symmetry. Since we have an explicit description of the (0,4) SCFT spectrum in terms of Gepner and KM characters, we can verify these claims directly. A numerical analysis confirms the expectations from spectral flow: the NS partition function factorizes into a contribution from the free fermion sector and a remaining interacting part.\footnote{We verified this factorization by expanding the NS partition function $Z(\tau,\overline{\tau})$ as a series in $\text{q}$ and $\overline{\text{q}}$, refined by the $\Lu(1)$ symmetries, and checking the factorization to the order considered.} The symmetry generated by the fermion bilinears recombines into $\so(8)\oplus\su(12)$.

However, the story does not end here: we find extra symmetry enhancements in the orbifold SCFT. To see this, we introduce a new basis of currents,
\begin{align}
\cK_\sleft^I= \sum_{i=1}^{N_\LGsmall} \rho^I_i J_\sleft^i+\sum_{i=1}^{N_\KMsmall} \sigma^I_i K_\sleft^i~,
\label{eq:newcurrentbasis}
\end{align}
where
\begin{align}
\boldsymbol{\rho} &=
\begin{pmatrix}
1&1&1\\
1&-1&0\\
2&0&-5 \\
0&2&-5
\end{pmatrix}~,&
\boldsymbol{\sigma} &=
\begin{pmatrix}
0\\
\frac{1}{6}\\
-\frac{5}{6} \\
\frac{5}{6}
\end{pmatrix}~.
\end{align}
One can verify that the currents $\cK^I_\sleft$ assign integer charges to all states in the orbifold SCFT. Furthermore, their two-point functions take the form
\begin{align}
\la \cK_\sleft^I(z)\cK_\sleft^J(0) \ra = \frac{2\text{k}^I\delta^{IJ}}{z^2}~,
\end{align}
with integer levels $\text{k}^I$, collected into the vector $\text{\textbf{k}}$ given by
\begin{align}
\text{\textbf{k}} = (1,1,10,10)~.
\end{align}
Surprisingly, we find that each of these $\Lu(1)$ symmetries is enhanced to an $\su(2)$, with the additional currents arising in various twisted sectors of the $\Z_{12}\times\Z_{12}$ orbifolds. The first two $\su(2)$ factors account for the $\so(4)$ symmetry associated with the free fermions, and recombine into $\so(8)\supset\so(4)\oplus\so(4)$. The remaining two factors, however, correspond to an extra $\so(4)$ KM algebra at level $10$, which is unusually high for standard heterotic constructions.  The physical interpretation of this $\SO(4)$ symmetry is less clear: it is a genuinely stringy symmetry of the SCFT, that emerges only deep in the LGO phase of the (0,2) GLSM.

A similar analysis applies to the other two models in~\eqref{eq:K3modelsLGO}. The second model, with $\Z_8\times\Z_8$ orbifold group, is specified by the data
\begin{align}
\boldsymbol{\mu} &=\begin{pmatrix}
1&0&\frac{1}{2} \\
0&1&\frac{1}{2}
\end{pmatrix}~,&
\boldsymbol{\nu} &=\begin{pmatrix}
\frac{1}{8} \\
-\frac{1}{8}
\end{pmatrix}~,&
M_\LG &=\diag(
\tfrac{3}{4},\tfrac{3}{4},\tfrac{1}{2})~,&
M_\KM &=8~,
\nonumber\\
\boldsymbol{\rho} &=
\begin{pmatrix}
1&1&1\\
1&-1&0\\
2&0&-3\\
0&2&-3\\
\end{pmatrix}~,&
\boldsymbol{\sigma} &=
\begin{pmatrix}
0\\
\frac{1}{4}\\
-\frac{3}{4}\\
\frac{3}{4}\\
\end{pmatrix}~,&
\text{\textbf{k}} &= (1,1,6,6)~,
\end{align}
and exhibits a similar enhancement to $\so(4)\oplus\so(4)$. One factor appears at level $1$ and combines with the other fermion bilinears to reconstruct $\so(16)\supset\so(12)\oplus\so(4)$, while the other factor appears at level $6$. The third model, with $\Z_6\times\Z_6$ orbifold group, is specified by the data
\begin{align}
\boldsymbol{\mu} &=\begin{pmatrix}
1&0&\frac{1}{2}&\frac{1}{2} \\
0&1&\frac{1}{2}&\frac{1}{2}
\end{pmatrix}~,&
\boldsymbol{\nu} &=\begin{pmatrix}
\frac{1}{6} \\
-\frac{1}{6}
\end{pmatrix}~,&
M_\LG &=\diag(
\tfrac{2}{3},\tfrac{2}{3},\tfrac{1}{3},\tfrac{1}{3})~,&
M_\KM &=6~,\nonumber\\
\boldsymbol{\rho} &=
\begin{pmatrix}
1&1&1&1\\
1&-1&0&0\\
2&0&-2&-2\\
0&2&-2&-2\\
0&0&3&-3\\
\end{pmatrix}~,&
\boldsymbol{\sigma} &=
\begin{pmatrix}
0\\
\frac{1}{3}\\
-\frac{2}{3}\\
\frac{2}{3}\\
0\\
\end{pmatrix}~,&
\text{\textbf{k}} &= (1,1,4,4,3)~,
\end{align}
and exhibits an enhancement of the left-moving symmetry to $\so(4)\oplus\so(4)\oplus\Lu(1)$. Again, the first factor is generated by free fermions and combines with the remaining currents to reconstruct $\so(20)\supset\so(16)\oplus\so(4)$, while the second $\so(4)$ factor appears at level $6$.

Since the (0,2) orbifold SCFTs are rational, we can systematically determine their marginal deformations. Those preserving (0,2) supersymmetry correspond to right-moving chiral primary operators of charge $q_\sright=1$ and conformal weights $(h_\sleft,h_\sright)=(1,\frac{1}{2})$~\cite{Bertolini:2014ela}. By scanning the SCFT spectrum, we can identify these operators explicitly and determine their representations under the global symmetry $\Lg$ associated with the left-moving current algebra. In the compactified theory---a six-dimensional (1,0) supergravity with one tensor multiplet---the (0,2) marginal deformations determine the hypermultiplets and their representation $\mathscr{H}$ under the six-dimensional gauge symmetry $G$. The resulting massless spectra for the three models are given by\footnote{Although we do not pay particular attention to the global form of the gauge group here, we indicate the non-simply connected form $\SO(4)=\mathrm{Spin}(4)/\Z_2$ for the last factor. This global structure follows from the fact that the $\Z_2$ operator $\exp(\pi\ii( \cK_{\sleft0}^3- \cK_{\sleft0}^4))$ acts trivially in the SCFT. It is also reflected in the presence of a simple current of the $\so(4)$ level-$10$ current algebra in the spectrum; see~\cite{Lockhart:2026gvb} for details.}
\begin{align}
G_1&=\Spin(8)\times\SU(12)\times\SO(4)~,&
\mathscr{H}_1 &=(\rep{1},\rep{66},\rep{6})\oplus(\rep{1},\rep{1},\rep{25})~,
\nonumber\\
G_2&=\Spin(16)\times\SU(8)\times\SO(4)~,&
\mathscr{H}_2 &=(\rep{16},\rep{8},\rep{1}) \oplus (\rep{1},\rep{28},\rep{6}\oplus\rep{4})\oplus (\rep{1},\rep{1},\rep{16}\oplus\rep{9})~,
\nonumber\\
G_3&=\Spin(20)\times\SU(6)\times\SO(4)\times\GU(1)~,&
\mathscr{H}_3 &= (\rep{20},\rep{6},\rep{1}_{+1}\oplus\rep{1}_{-1}) \oplus (\rep{1},\rep{15},\rep{4}_{+1}\oplus\rep{4}_{-1}\oplus\rep{6}_0)
\nonumber\\
&&&\quad\oplus (\rep{1},\rep{1},\rep{9}_{+2}\oplus\rep{4}_{+3})^{\oplus 2}
~.
\label{eq:stringyK3spectrum}
\end{align}
In each case, the massless spectrum is anomaly free. In particular, the gravitational anomaly cancellation condition $N_{\text{h}}-N_{\text{v}}=244$ is satisfied. The remaining anomaly cancellation conditions are verified in appendix~\ref{app:6danomalies}.

The (1,0) supergravity theories defined by the massless spectra~\eqref{eq:stringyK3spectrum} are rather special: none of these models contains neutral hypermultiplets. This makes them particularly difficult to realize in F-theory, as they lack the universal hypermultiplet that would parametrize the volume of the base of the elliptically fibered CY threefold. On the other hand, the enhanced $\SO(4)$ symmetry at the LGO point can be Higgsed, leading back to the geometric models in~\eqref{eq:largeradiusK3spectrum}, obtained as heterotic compactifications on K3. In particular, this Higgsing gives rise to $19$ neutral hypermultiplets, which precisely parametrize the K3 moduli in the presence of a $\Lu(1)$ instanton. The (0,2) GLSMs presented here allow us to track the deformation between these two phases. This is an example of the type of transition discussed in~\cite{Baykara:2023plc}, in which the small-volume limit of the CY geometry leads to a special point where the gauge symmetry is enhanced and the volume modulus becomes charged.

On the worldsheet, the deformations responsible for Higgsing the enhanced gauge symmetry correspond to (0,2)-preserving marginal operators in twisted sectors of the LGO model. In the first model, for instance, the marginal states neutral under the large-radius gauge symmetry form the $\rep{25}=(\rep{5},\rep{5})$ representation of $\SO(4)$, distributed across several twisted sectors of the $\Z_{12}\times\Z_{12}$ orbifold. A single marginal state lies in the untwisted sector, with charges $(+4,+4)$ under the Cartan subalgebra of $\so(4)$. This state has a natural interpretation: it deforms the Gepner point into a (0,2) LG model with superpotential
\begin{align}
\cW_{\LG} = \Gamma_1 \Phi_1^{11} + \Gamma_3\Phi_3^{11}+\Gamma_5\Phi_5^2 + \alpha\Gamma_5\Phi_1^4\Phi_3^4~,
\end{align}
parameterized by a complex constant $\alpha$.\footnote{The most general superpotential also contains a term proportional to $\Gamma_5\Phi_1^2\Phi_3^2\Phi_5$, which can be removed by a holomorphic field redefinition of $\Phi_5$.} However, this untwisted deformation is obstructed on its own: because it carries a charge of definite sign with respect to a $\Lu(1)$ current, conformal perturbation theory develops a logarithmic divergence~\cite{Bertolini:2014ela}, corresponding to a D-term potential in spacetime. Thus, turning on $\alpha$ induces an RG flow that drives the theory back to the $\alpha=0$ fixed point, so that the LGO limit of the GLSM reduces to a single point. Similar considerations apply to the other models, in close analogy with obstructions previously observed in~\cite{Israel:2026bco} for the LG phases of TLSMs.

The heterotic SCFTs describing these compactifications are also fairly unusual: their left-moving chiral algebras includes KM currents at uncommonly high levels. Such higher KM levels can appear in asymmetric orbifolds~\cite{Font:1990uw}, either through permutation orbifolds of identical level-$1$ factors---as in the famous example of the CHL string~\cite{Chaudhuri:1995fk}---or through orbifolds involving Weyl automorphisms. It would be interesting to investigate whether the models considered here admit an asymmetric orbifold realization. In addition, the left-moving KM algebra $\mathscr{A}$ appearing in these models is unusually ``large''. More precisely, the left-moving Virasoro algebra of the internal (0,4) SCFT, which is at central charge $c_\sleft=20$, admits a Sugawara decomposition. For the three models considered here, the current algebras $\mathscr{A}$ and their corresponding Sugawara contributions $c_\sleft(\mathscr{A})$ to the left-moving central charge are respectively given by
\begin{align}
\mathscr{A}_1 &= \so(8)_1\oplus \su(12)_1 \oplus \su(2)_{10}^{\oplus 2}~,&
c_\sleft(\mathscr{A}_1) &=4+11+2\cdot\tfrac{5}{2}=20~,
\nonumber\\
\mathscr{A}_2 &= \so(16)_1\oplus \su(8)_1 \oplus \su(2)_6^{\oplus 2}~,&
c_\sleft(\mathscr{A}_2) &=8+7+2\cdot\tfrac{9}{4}=\tfrac{39}{2}~,
\nonumber\\
\mathscr{A}_3 &= \so(20)_1\oplus \su(6)_1 \oplus \su(2)_4^{\oplus 2}\oplus\Lu(1)_3~,&
c_\sleft(\mathscr{A}_3) &=10+5+2\cdot2+1=20~,
\end{align}
where the subscripts indicate the KM levels. In two of the three models, the KM algebra completely saturates the left-moving central charge. In other words, the left-moving $c_\sleft=20$ chiral algebra is generated entirely by the KM currents (possibly together with extensions by higher-spin chiral operators). In the second model, the Sugawara decomposition instead leaves a residual stress tensor with central charge $c_\sleft=\frac{1}{2}$, corresponding to the Ising model. Such two-dimensional (0,4) SCFTs, whose left-moving chiral algebras are almost entirely saturated by currents, were recently considered in~\cite{Lockhart:2025lea,Lockhart:2026xml} in the study of BPS strings coupled to six-dimensional supergravity theories without tensor multiplets. The models found here are close relatives of these theories, with one tensor multiplet.

\section{Abelian bundles on CY threefolds}
\label{sec:CY}
In this final section we present several examples of (0,2) GLSMs whose LR phases describe sigma models on CY threefolds equipped with abelian bundles. We then focus on one of these models and use the GLSM to  connect the geometry to a stringy regime, deep in the LGO phase, where the CY volume is frozen at a small value. We compare the massless spectra in the two phases and argue that the spacetime superpotential receives no instanton corrections.

\subsection{Examples with \texorpdfstring{$h^{1,1}=3$}{h1,1=3}}

Specializing to $d=3$, the construction of section~\ref{subsec:threefamilies} produces three smooth CY threefolds, realized as abelian branched coverings $\pi:Y\to(\P^1)^3$. All three have $h^{1,1}(Y)=3$, with K\"ahler cone spanned by classes $\eta_{\alpha=1,2,3}$ inherited from the $(\P^1)^3$ base, satisfying $(\eta_\alpha)^2=0$. The only non-vanishing triple intersection number is
\begin{align}
\int_Y \eta_1\eta_2\eta_3=k~,
\end{align}
where $k$ is the degree of the covering map $\pi$. The topological data of these CY $3$-folds are collected in table~\ref{tab:topologicalCYdata}.
\begin{table}[h!]
\centering
\arraycolsep=6pt
$\begin{array}{cccccc}
Y & h^{1,1}(Y) & h^{2,1}(Y) & \chi(Y) & c_2(T_Y).\eta_\alpha & k  \\ \hline
Y_1 & 3 & 115 & -224 & (24,24,24) & 2 \\
Y_2 & 3 & 81 & -156 & (24,24,24) & 3 \\
Y_3 & 3 & 59 & -112 & (24,24,24) & 4 \\
\end{array}$
\caption{Topological data for the CY branched coverings of $(\P^1)^3$}
\label{tab:topologicalCYdata}
\end{table}
For consistency of the (0,2) NLSM on $Y$, the holomorphic bundle $\cE\to Y$ must have the same second Chern class as the tangent bundle of $Y$, namely,
\begin{align}
c_2(T_Y) = \tfrac{24}{k}(\eta_1\eta_2+\eta_2\eta_3+\eta_3\eta_1)~.
\end{align}
For an abelian bundle $\cE=\oplus_{A=1}^\tr\cL_A$, each line bundle in the summand is specified by its first Chern class, which must be a primitive (1,1) form for spacetime supersymmetry. We restrict to the locus $\rho^\alpha=\rho$ in FI parameter space, which corresponds in the LR limit to a K\"ahler form proportional to the ample divisor class $\upeta=\eta_1+\eta_2+\eta_3$. This class defines a polarization of degree $6k$. For a line bundle $\cL$ with first Chern class $\upomega = l^\alpha\eta_\alpha$, primitivity of $\upomega$ is then equivalent to the vanishing of the slope $\mu(\cL)$, which is given by 
\begin{align}
\mu(\cL) = 2k(l^1+l^2+l^3)~.
\end{align}

We consider bundles of the form $\cE=\cE_1\oplus\cE_2\oplus\cE_3$, where each of the $\tr_i$ line bundle summands of the rank-$\tr_i$ subbundle $\cE_i$ has first Chern class proportional to the same primitive class $\upomega_i$. This defines an embedding of three line bundles $\cL_i$ into the abelian bundle with structure group $\Lu(\tr_1)\oplus\Lu(\tr_2)\oplus\Lu(\tr_3)$. Since $h^{1,1}(Y)=3$, there are only two linearly independent primitive (1,1) classes. Accordingly, although the structure group commutant contains three abelian factors arising from the decompositions $\Lu(\tr_i)=\su(\tr_i)\oplus\Lu(1)$, we expect two of the corresponding $\Lu(1)$ vector fields to be lifted through the St\"uckelberg mechanism in the four-dimensional compactification. The resulting rank-$14$ gauge group takes the form
\begin{align}
G = \Spin(2\tr_0)\times\SU(\tr_1)\times\SU(\tr_2)\times\SU(\tr_3)\times\GU(1)~,
\end{align}
where $\tr_0=16-\tr_1-\tr_2-\tr_3$.

The (0,2) GLSMs associated with these geometries have gauge group $\GU(1)^3$, and the abelian bundle configuration is realized by charged fermi multiplets $\Lambda_1,\dots,\Lambda_\tr$ with no superpotential couplings. These fields alone do not suffice to cancel the GLSM gauge anomalies. We therefore supplement the construction in each case with superfields for a rank-$3$ trivial bundle $\cO\oplus\cO\oplus\cO$, which contributes to the gauge anomaly in the UV but gives rise to free decoupled fermions in the IR. 

The first GLSM has a LR phase describing a sigma model on $Y_1$ equipped with a rank-$12$ abelian bundle. The resulting heterotic background yields a four-dimensional theory with gauge group $G_1 = \Spin(8)\times\SU(4)^3\times\GU(1)$. The superfield content and gauge charges are listed in the table below.
\begin{table}[H]
\centering
$\GLSMblock{T_Y}{
\begin{array}{ccccc}
\Phi_{1,2} & \Phi_{3,4} & \Phi_{5,6} & \Phi_7 & \Gamma_7 \\
1 & 0 & 0 & 2 & -4 \\
0 & 1 & 0 & 2 & -4 \\
0 & 0 & 1 & 2 & -4 \\
\end{array}
}$
\;
$\GLSMblock{\cO}{
\begin{array}{ccc}
\Phi_8 & \Gamma_1 & \Gamma_2 \\
-18 & 1 & 17 \\
0 & 0 & 0\\
0 & 0 & 0 \\
\end{array}
}$
\;
$\GLSMblock{\cO}{
\begin{array}{ccc}
\Phi_9 & \Gamma_3 & \Gamma_4 \\
0 & 0 & 0\\
-18 & 1 & 17 \\
0 & 0 & 0 \\
\end{array}
}$
\;
$\GLSMblock{\cO}{
\begin{array}{ccc}
\Phi_{10} & \Gamma_5 & \Gamma_6 \\
0 & 0 & 0\\
0 & 0 & 0 \\
-18 & 1 & 17 \\
\end{array}
}$
\;
$\GLSMblock{\cE}{
\begin{array}{ccc}
\Lambda_{1,\dots,4} & \Lambda_{5,\dots,8} & \Lambda_{9,\dots,12} \\
2 & -1 & -1 \\
-1 & 2 & -1 \\
-1 & -1 & 2 \\
\end{array}
}$
\caption{Field content and gauge charges of the GLSM for $Y_1$}
\label{tab:CYmodelY1}
\end{table}

The second model corresponds to a compactification on $Y_2$ with a rank-$6$ abelian bundle. The compactification at large radius results in a four-dimensional theory with gauge group $G_2 = \Spin(20)\times\SU(2)^3\times\GU(1)$. The GLSM has the following superfield content.
\begin{table}[H]
\centering
$\GLSMblock{T_Y}{
\begin{array}{ccccc}
\Phi_{1,2}& \Phi_{3,4} & \Phi_{5,6} & \Phi_7 & \Gamma_7 \\
1 & 0 & 0 & 1 & -3 \\
0 & 1 & 0 & 1 & -3 \\
0 & 1 & 1 & 1 & -3 \\
\end{array}
}$
\,
$\GLSMblock{\cO}{
\begin{array}{ccc}
\Phi_8 & \Gamma_1 & \Gamma_2 \\
-12 & 1 & 11 \\
0 & 0 & 0\\
0 & 0 & 0 \\
\end{array}
}$
\,
$\GLSMblock{\cO}{
\begin{array}{ccc}
\Phi_9 & \Gamma_3 & \Gamma_4 \\
0 & 0 & 0\\
-12 & 1 & 11 \\
0 & 0 & 0 \\
\end{array}
}$
\,
$\GLSMblock{\cO}{
\begin{array}{ccc}
\Phi_{10} & \Gamma_5 & \Gamma_6 \\
0 & 0 & 0\\
0 & 0 & 0 \\
-12 & 1 & 11 \\
\end{array}
}$
\,
$\GLSMblock{\cE}{
\begin{array}{cccccc}
\Lambda_1 & \Lambda_2 & \Lambda_3 & \Lambda_4 & \Lambda_5 & \Lambda_6 \\
2 & -2 & 2 & -2 & 0 & 0 \\
-2 & 2 & 0 & 0 & 2 & -2 \\
0 & 0 & -2 & 2 & -2 & 2 \\
\end{array}
}$
\caption{Field content and gauge charges of the GLSM for $Y_2$}
\label{tab:CYmodelY2}
\end{table}

Finally, the third model describes a CY threefold $Y_3$ with a rank-$6$ abelian bundle. The compactified four-dimensional theory has gauge group $G_3 = \Spin(20)\times\SU(2)^3\times\GU(1)$, identical to that of the second model. The GLSM has the following superfield content.
\begin{table}[H]
\centering
$\GLSMblock{T_Y}{
\begin{array}{ccccc}
\Phi_{1,2} & \Phi_{3,4} & \Phi_{5,6} & \Phi_{7,8} & \Gamma_{7,8} \\
1 & 0 & 0 & 1 & -2 \\
0 & 1 & 0 & 1 & -2 \\
0 & 0 & 1 & 1 & -2 \\
\end{array}
}$
\;
$\GLSMblock{\cO}{
\begin{array}{ccc}
\Phi_9 & \Gamma_1 & \Gamma_2 \\
-9 & 1 & 8 \\
0 & 0 & 0\\
0 & 0 & 0 \\
\end{array}
}$
\;
$\GLSMblock{\cO}{
\begin{array}{ccc}
\Phi_{10} & \Gamma_3 & \Gamma_4 \\
0 & 0 & 0\\
-9 & 1 & 8 \\
0 & 0 & 0 \\
\end{array}
}$
\;
$\GLSMblock{\cO}{
\begin{array}{ccc}
\Phi_{11} & \Gamma_5 & \Gamma_6 \\
0 & 0 & 0\\
0 & 0 & 0 \\
-9 & 1 & 8 \\
\end{array}
}$
\;
$\GLSMblock{\cE}{
\begin{array}{ccc}
\Lambda_{1,2} & \Lambda_{3,4} & \Lambda_{5,6} \\
2 & -1 & -1 \\
-1 & 2 & -1\\
-1 & -1 & 2 \\
\end{array}
}$
\caption{Field content and gauge charges of the GLSM for $Y_3$}
\label{tab:CYmodelY3}
\end{table}

Each GLSM has a LGO phase where the gauge group is broken to a finite subgroup of the form $\Gamma=(\Z_\tN)^3$  by the expectation values of the three chiral superfields associated with the monadic $\cO$ factors. Before orbifolding, the fermi multiplets $\Lambda_A$ describe free left-moving fermions, while the remaining superfields can be assembled into (2,2) multiplets $X_i=(\Phi_i,\Gamma_i)$. Moreover, there is a locus in the GLSM parameter space where the superpotential preserves this enhanced supersymmetry, and can be expressed as the superspace derivative of a (2,2) superpotential $W$. In each model, three of the $X_i$ multiplets become massive, which we take to be $X_2$, $X_4$ and $X_6$. The massless chiral multiplets form a (2,2) LG model with central charges $c_\sleft=c_\sright=9$. The corresponding superpotentials $W$ and orbifold groups $\Gamma$ are respectively
\begin{align}
W_1 &= \tfrac{1}{18}(X_1^{18}+X_3^{18}+X_5^{18})+\tfrac{1}{3}X_7^3~,&
\Gamma_1 &= \Z_{18}\times\Z_{18}\times\Z_{18}~,
\nonumber\\
W_2 &= \tfrac{1}{12}(X_1^{12}+X_3^{12}+X_5^{12})+\tfrac{1}{4}X_7^4~,&
\Gamma_2 &= \Z_{12}\times\Z_{12}\times\Z_{12}~,
\nonumber\\
W_3 &= \tfrac{1}{9}(X_1^9+X_3^9+X_5^9)+\tfrac{1}{3}(X_7^3+X_8^3)~,&
\Gamma_3 &= \Z_9\times\Z_9\times\Z_9~.
\label{eq:LGsuperoptentialsCY}
\end{align}
The orbifold acts chirally on the LG degrees of freedom; its action is specified in table~\ref{tab:CYorbifoldaction} using the notation of section~\ref{sec:GepnerSCFT}. The orbifold data satisfy the relation~\eqref{eq:totalanomaly}, which implies, in particular, the presence of six free left-moving Majorana--Weyl fermions among the twisted ground states. The diagonal $\Z_\tN$ subgroup generated by $g=g_1g_2g_3$ acts symmetrically on the LG model and leaves the fermions invariant. This non-chiral subgroup enforces integral right-moving R-charges in the orbifold (0,2) SCFT.
\begin{table}[h!]
\centering
\arraycolsep=8pt
\renewcommand{\arraystretch}{1.3}
$\begin{array}{c|c|c|c}
Y & Y_1 & Y_2 & Y_3 \\ \hline
M_\LG & \diag(\tfrac{8}{9},\tfrac{8}{9},\tfrac{8}{9},\tfrac{1}{3}) & \diag(\tfrac{5}{6},\tfrac{5}{6},\tfrac{5}{6},\tfrac{1}{2}) & \diag(\tfrac{7}{9},\tfrac{7}{9},\tfrac{7}{9},\tfrac{1}{3},\tfrac{1}{3}) \\
M_\KM & \diag(4,4,4) & \diag(2,2,2) & \diag(2,2,2) \\
\boldsymbol{\mu} & \arraycolsep=2pt\begin{pmatrix}
1&0&0&\frac{1}{3}\\
0&1&0&\frac{1}{3}\\
0&0&1&\frac{1}{3}\\
\end{pmatrix} & \arraycolsep=2pt\begin{pmatrix}
1&0&0&\frac{1}{3}\\
0&1&0&\frac{1}{3}\\
0&0&1&\frac{1}{3}\\
\end{pmatrix} & \arraycolsep=2pt\begin{pmatrix}
1&0&0&\frac{1}{3}&\frac{1}{3}\\
0&1&0&\frac{1}{3}&\frac{1}{3}\\
0&0&1&\frac{1}{3}&\frac{1}{3}\\
\end{pmatrix} \\

\boldsymbol{\nu} &  \arraycolsep=2pt\begin{pmatrix}
\frac{2}{18} & -\frac{1}{18} & -\frac{1}{18} \\
-\frac{1}{18} & \frac{2}{18} & -\frac{1}{18} \\
-\frac{1}{18} & -\frac{1}{18} & \frac{2}{18} \\
\end{pmatrix} & \arraycolsep=2pt\begin{pmatrix}
\frac{1}{6} & -\frac{1}{6} & 0\\
-\frac{1}{6} & 0 & \frac{1}{6} \\
0 & \frac{1}{6} & -\frac{1}{6}
\end{pmatrix}  &  \arraycolsep=2pt\begin{pmatrix}
\frac{2}{9} & -\frac{1}{9} & -\frac{1}{9} \\
-\frac{1}{9} & \frac{2}{9} & -\frac{1}{9} \\
-\frac{1}{9} & -\frac{1}{9} & \frac{2}{9} \\
\end{pmatrix} \\
\end{array}$
\caption{Orbifold action in the LGO phase for the three CY models}
\label{tab:CYorbifoldaction}
\end{table}

\subsection{Transition at small volume}
\label{eq:smallvolumetransition}
For the remainder of this note, we focus on the model associated with the threefold $Y_3$, the biquadratic cover of $(\P^1)^3$, specified by the field content in table~\ref{tab:CYmodelY3}. Of the three, this model has the smallest orbifold group, $\Gamma=(\Z_9)^3$. Moreover, as we will see, the chiral orbifold action on some of the LG degrees of freedom admits a straightforward realization as a shift orbifold of a compact boson. We determine the massless spectrum of the four-dimensional theory in both the LR and LGO phases. The spectra agree across the two phases, suggesting that worldsheet instantons do not destabilize the vacuum. We argue that this is indeed the case, and that the (0,2) GLSM interpolates continuously between the two descriptions. The LGO limit leads to a transition in which all matter fields become charged, including the would-be CY volume modulus, which is frozen at a small, fixed value.

\subsubsection*{LR phase}
For notational simplicity, we henceforth write $Y$ for $Y_3$. Compactification of the $\Spin(32)/\Z_2$ heterotic string on $Y$ yields a four-dimensional supergravity theory with $4$ supercharges. At large radius, the charged matter is counted by the dimension of certain bundle-valued cohomology groups. More precisely, the rank-$6$ abelian bundle realized by the GLSM is
\begin{align}
\cE = (\cL_1\oplus\cL_2\oplus\cL_3)^{\oplus 2}~,
\end{align}
where the line bundles $\cL_i$ have first Chern classes $\upomega_i$ given by
\begin{align}
\upomega_1 &= 2\eta_1-\eta_2-\eta_3~,&
\upomega_2 &= -\eta_1+2\eta_2-\eta_3~,&
\upomega_3 &= -\eta_1-\eta_2+2\eta_3~.
\end{align}
The charged chiral spectrum follows from the branching
\begin{align}
\so(32) &\supset \so(20)\oplus\su(2)^{\oplus 3}\oplus\Lu(1)^{\oplus 3} \nonumber\\
\rep{496} &= \oplus_Q \rep{R}_Q~,
\end{align}
where $Q=(q_1,q_2,q_3)$ denotes the $\GU(1)^3$ charges. The adjoint representation of the unbroken gauge algebra gives rise to the four-dimensional vector multiplets. Each additional representation $\rep{R}_Q$ appearing in this decomposition contributes $\cN(\rep{R}_Q)$ chiral matter multiplets transforming in $\rep{R}_Q$, with multiplicity
\begin{align}
\mathcal{N}(\rep{R}_Q) = h^1(Y,\cL_Q)~,
\label{eq:h1line}
\end{align}
where $\cL_Q$ is the holomorphic line bundle with first Chern class $c_1(\cL_Q)=\sum_i q_i\upomega_i$. Additional massless moduli arise from deformations of the geometry. Usually, these correspond to simultaneous holomorphic deformations of $Y$ and $\cE$, subject to the constraints encoded by the Atiyah sequence~\cite{Atiyah:1957,Anderson:2011ty,Anderson:2014xha,delaOssa:2014cia}. In our situation, since holomorphic line bundles on $Y$ have no continuous moduli, the relevant deformations are those of the K\"ahler and complex structures. The Atiyah map leaves all complex structure deformations unobstructed, while the zero-slope condition on the bundle lifts two K\"ahler moduli, one for each independent primitive (1,1) class that support the abelian bundle configurations. Altogether, the deformations of $Y$ give rise to
\begin{align}
h^{1,1}(Y)+h^{2,1}(Y)-2 = 60
\end{align}
gauge-neutral chiral multiplets; these are supplemented by the axio-dilaton chiral multiplet.

We expect two of the three $\GU(1)$ vector fields to acquire masses through the St\"uckelberg mechanism. The surviving abelian gauge symmetry acts on matter with charge $q=q_1+q_2+q_3$.
The resulting spacetime gauge group is
\begin{align}
G_{\text{\tiny{LR}}} = \Spin(20)\times\SU(2)^3\times\GU(1)~.
\label{eq:gaugegroupLR}
\end{align}
The line bundle-valued cohomology groups~\eqref{eq:h1line} can be readily computed  using the description of $Y$ as a branched covering  of $(\P^1)^3$, which reduces the calculation to simpler cohomology computations on the base. The method and the relevant cohomology dimensions are given in appendix~\ref{app:linebundlecohomology}. We obtain the following multiplicities for the charged chiral matter,
\begin{center}
\renewcommand{\arraystretch}{1.1}
$\begin{array}{cc}
\rep{R}_q& \cN(\rep{R}_q) \\ \hline 
\oplus_i(\rep{20},\rep{2}_i)_{-1} & 8 \\
\oplus_{i<j}(\rep{1},\rep{2}_i\otimes\rep{2}_j)_{+2} & 8 \\
\oplus_{i<j}(\rep{1},\rep{2}_i\otimes\rep{2}_j)_{0} & 16 \\
(\rep{1},\rep{1})_{-2} & 192 \\
\end{array}$
\end{center}
where $\rep{2}_i$ denotes the representation $\rep{2}$ embedded in the $i$-th factor of $\SU(2)^3$. The spacetime massless spectrum is symmetric under permutations of the three $\SU(2)$ factors.

\subsubsection*{LGO phase}
In the LGO phase, the theory is described by a chiral $(\Z_9)^3$ orbifold of a $c=9$ (2,2) LG model and of six free left-moving Weyl fermions. The orbifold action is specified by the data in the last column of table~\ref{tab:CYorbifoldaction}. The separate orbifold actions on the LG model and on the free fermions are inconsistent, with anomalous modular transformations encoded by the matrices
\begin{align}
A_\LG &=
\begin{pmatrix}
\frac{23}{27}&\frac{2}{27}&\frac{2}{27}\\
\frac{2}{27}&\frac{23}{27}&\frac{2}{27}\\
\frac{2}{27}&\frac{2}{27}&\frac{23}{27}\\
\end{pmatrix}~,&
A_\KM &=\begin{pmatrix}
\frac{4}{27}&-\frac{2}{27}&-\frac{2}{27}\\
-\frac{2}{27}&\frac{4}{27}&-\frac{2}{27}\\
-\frac{2}{27}&-\frac{2}{27}&\frac{4}{27}\\
\end{pmatrix}~.
\end{align}
The combined orbifold is anomaly free. There is a locus in parameter space where the LG superpotential preserves (2,2) supersymmetry and takes the form
\begin{align}
\cW = \Gamma_1\Phi_1^8 + \Gamma_3\Phi_3^8 + \Gamma_5\Phi_5^8 + \Gamma_7\Phi_7^2 + \Gamma_8\Phi_8^2~,
\label{eq:chiralZ9action}
\end{align}
which derives from the (2,2) superpotential $W_3$ in~\eqref{eq:LGsuperoptentialsCY} via the decomposition $X_i=(\Phi_i,\Gamma_i)$. We omit the massive superfields $X_{2,4,6}$ which do not affect the IR dynamics. The orbifold includes a $\Z_9$ action on $X_7$ of the form
\begin{align}
\Phi_7&~\to~\ee^{\frac{2\pi\ii}{9}}\Phi_7~,&
\Gamma_7&\to\ee^{-\frac{4\pi\ii}{9}}\Gamma_7~,
\end{align}
with a similar action on $(\Phi_8,\Gamma_8)$. In the IR, the $X_7$ multiplet flows to a (2,2) minimal model with $c_\sleft=c_\sright=1$, whose left- and right-moving primary representations are described by the coset $(\su(2)_1\oplus\Lu(1)_2)/\Lu(1)_3$.  Since the orbifold action does not preserve the individual coset representations, the partition function cannot be expressed directly in the form~\eqref{eq:twistedLGpartitionfunction} used for the previous K3 LG models. Fortunately, the orbifold admits a simpler description. The $X_7$ multiplet can equivalently be represented by a free compact boson at radius $r=\sqrt{6}$, and the $12$ coset primary representations correspond to representations of the $\Lu(1)_6$ current algebra. In this description, the orbifold action~\eqref{eq:chiralZ9action} becomes a simple shift orbifold and, although the shift does not preserve the $\Lu(1)_6$ KM algebra, it preserves a $\Lu(1)_{54}\subset\Lu(1)_6$ subalgebra, in terms of which the twisted NS partition functions can be written as
\begin{align}
Z^{(a,0)}_{X_7} = \sum_{m\in\Z_6}\sum_{\ell\in\Z_6}\chi^{\Lu(1)_{54}}_{18\ell-6m+6a\cdot\mu}\,\overline{\chi}^0_m~,
\end{align}
where $\mu=(\frac{1}{3},\frac{1}{3},\frac{1}{3})$. The derivation and notation are detailed in appendix~\ref{app:chiralbosonorbifold}. This allows us to decompose the LG contributions to the orbifold partition function into a finite set of known primary representations.

Turning to the heterotic fermions, the $c=9$ LG model is supplemented with $13$ free left-moving fermions, of which the orbifold preserves a $\so(14)\oplus\Lu(2)^{\oplus 3}\subset\so(26)$ symmetry. The decomposition $\Lu(2)=\su(2)\oplus\Lu(1)$  leads to $\Lu(1)$ factors at level $1$ (each equivalent to an $\su(2)$ level-$1$ KM algebra). The orbifold, however, preserves only a $\Lu(1)_{81}\subset\Lu(1)_1$ subalgebra. Using this decomposition, the twisted NS partition functions of the $c_\sleft=13$ chiral fermion system can be expressed as
\begin{align}
Z_\KM^{(a,0)} = (\chi^{\so(14)}_{\rep{1}}+\chi^{\so(14)}_{\rep{14}})\prod_{i=1,2,3}\sum_{l_i=0,1}\chi^{\su(2)^{i}}_{[l_i]}\sum_{\ell_i\in\Z_9}\chi^{\Lu(1)_{81}^{i}}_{9l_i+18\ell_i+18\nu_i\cdot a}~,
\end{align}
where $l=0,1$ labels the $\su(2)$ representations $\rep{1}=[0]$ and $\rep{2}=[1]$. Further details are given in appendix~\ref{app:fermionorbifold}.

\subsubsection*{Enhanced symmetry and frozen volume}
Using the description above, we can explicitly determine the spectrum of the (0,2) SCFT describing the LGO phase of the GLSM. We can then identify states of particular interest---those include the holomorphic left-moving currents, which generate the gauge symmetry of the four-dimensional heterotic compactification.
Before orbifolding, the LG model and the heterotic fermions have a left-moving current algebra $\Lu(1)^{\oplus 5}\oplus\so(26)$, arising from fermion bilinears and from the left-moving R-symmetry currents of the five (2,2) minimal model factors composing the LG theory. The orbifold acts non-trivially on some of the fermion bilinears, and the currents surviving the orbifold projection in the untwisted sector generate a current algebra $\mathscr{A}_\LG\oplus\mathscr{A}_\KM$, where
\begin{align}
\mathscr{A}_\LG&=\Lu(1)^{\oplus 5}~,&
\mathscr{A}_\KM&=\so(14)\oplus\su(2)^{\oplus 3}\oplus\Lu(1)^{\oplus 3}~.
\label{eq:untwistedcurrents}
\end{align}
However, the chiral symmetry of the orbifold SCFT is larger: extra currents appear in the twisted sectors and enhance the algebra~\eqref{eq:untwistedcurrents}. To make this enhancement manifest, we change basis for the eight abelian symmetries. We introduce currents $\cJ^{\alpha=1,2,3}_\sleft$ and $\cK^{I=1,\dots,5}_\sleft$  whose non-vanishing OPEs take the form
\begin{align}
\cJ^\alpha_\sleft(z)\cJ^\alpha_\sleft(0) &\sim \frac{1}{z^2}~,&
\cK^I_\sleft(z)\cK^I_\sleft(0) &\sim \frac{2\text{k}^I}{z^2}~,
\end{align}
with levels $\text{\textbf{k}} = (3,3,7,7,7)$. The currents $\cJ_\sleft^\alpha$ are defined as in~\eqref{eq:JLcurrents}. They can be used to express the orbifold generators as $g_\alpha=\exp(2\pi\ii\cJ^\alpha_{\sleft0})$ and to construct the twisted Hilbert spaces by spectral flow. As explained in section~\ref{subsec:fermionliberation}, we expect the corresponding twist fields to give rise to six left-moving fermions, together with twelve extra currents that enhance the symmetry generated by the $\cJ_\sleft^\alpha$ to an $\so(6)$ current algebra at level one. An explicit analysis of the twisted partition functions confirms this expectation. Furthermore, we find holomorphic $h_\sleft=1$ states transforming in the representation $(\rep{14},\rep{6})$ of the $\so(14)\oplus\so(6)$ algebra generated by the original free fermions and the spin-$\frac{1}{2}$ twist fields. These states complete the current algebra to $\so(20)$, reproducing the factor in~\eqref{eq:gaugegroupLR} predicted by the large radius analysis.

Beyond these enhancements, the orbifold SCFT exhibits extra symmetries that are not present in the large radius description. The five currents $\cK^I_\sleft$ are defined as in~\eqref{eq:newcurrentbasis}, with coefficients given by
\begin{align}
\boldsymbol{\rho} &=
\begin{pmatrix}
0&0&0&0&0\\
0&0&0&3&-3\\
2&0&0&-\frac{7}{3}&-\frac{7}{3}\\
0&2&0&-\frac{7}{3}&-\frac{7}{3}\\
0&0&2&-\frac{7}{3}&-\frac{7}{3}\\
\end{pmatrix}~,&
\boldsymbol{\sigma} &=
\begin{pmatrix}
1&1&1\\
0&0&0\\
-\frac{14}{9}&\frac{7}{9}&\frac{7}{9}\\
\frac{7}{9}&-\frac{14}{9}&\frac{7}{9}\\
\frac{7}{9}&\frac{7}{9}&-\frac{14}{9}\\
\end{pmatrix}~.
\end{align}
One can verify that these currents assign integer charges to all states in the orbifold SCFT. The current $\cK^1_\sleft$ corresponds to the $\GU(1)\subset\GU(2)^3$ factor that survives the St\"uckelberg mechanism in the LR phase and gives rise to the abelian factor of the gauge group~\eqref{eq:gaugegroupLR}. The remaining currents involve global symmetries of the LG model at the (2,2) locus and are therefore accidental symmetries specific to this locus. The current $\cK^2_\sleft$ generates an additional abelian factor. On the other hand, an explicit analysis of the orbifold partition function reveals further currents in the twisted sectors, which enhance each of the $\cK_\sleft^{3,4,5}$ currents to an $\su(2)$ current algebra at level $7$.\footnote{These three enhanced level-$7$ $\su(2)$ factors bring the Sugawara central charge to $c_\sleft=22$, saturating that of the internal SCFT.} All in all, the left-moving current algebra gives rise to the spacetime gauge symmetry
\begin{align}
G_{\text{\tiny{LGO}}} = \Spin(20) \times \SU(2)^3 \times \GU(1) \times \SU(2)^3 \times \GU(1)~.
\end{align}
Comparing this with the gauge symmetry $G_{\text{\tiny{LR}}}$ found in the large radius limit of the CY compactification, the extra factor of $\SU(2)^3\times\GU(1)$ factor is a genuinely stringy enhancement that emerges in the strongly coupled regime of the NLSM, when the target space is taken to string-size.

Finally, it is instructive to examine the space of first order deformations of the (0,2) SCFT, consisting of right-chiral primary states with weights $(h_\sleft,h_\sright)=(1,\frac{1}{2})$. In the four-dimensional compactification, these states give rise to chiral matter multiplets. We find a $1510$-dimensional space of infinitesimal deformations. The marginal operators can be organized into representations of the global $G_{\text{\tiny{LGO}}}$ symmetry. Remarkably, every marginal operator is charged. The complete chiral matter spectrum is given in table~\ref{tab:CYmasslessspectrumLGO}, where we list the representations $\rep{R}_{\text{\tiny{LGO}}}$ under the global symmetry of the SCFT, together with their branching under $G_{\text{\tiny{LGO}}}\supset G_{\text{\tiny{LR}}}$. The notation $\oplus_{\{i,j,k\}}$ denotes the direct sum over distinct indices $i,j,k$. 
\begin{table}[h!]
\centering
\renewcommand{\arraystretch}{1.1}
$\begin{array}{cc}
\rep{R}_{\text{\tiny{LGO}}} & \rep{R}_{\text{\tiny{LR}}} \\ \hline
\\[-10pt]

\left.\begin{array}{c}
\oplus_{i}(\rep{20},\rep{2}_i,\rep{2}_i)_{-1,+1} \\
\oplus_{i}(\rep{20},\rep{2}_i,\rep{2}_i)_{-1,-1}  \\
\oplus_{\{i,j,k\}}(\rep{20},\rep{2}_i,\rep{2}_j\otimes\rep{2}_k)_{-1,0}  \\
\end{array}\right\rbrace & \oplus_{i}(\rep{20},\rep{2}_i)_{+1}^{\oplus 8} \\
\\[-10pt]

\left.\begin{array}{c}
\oplus_{i<j}(\rep{1},\rep{2}_i\otimes\rep{2}_j,\rep{2}_i\otimes\rep{2}_j)_{+2,0} \\
\oplus_{\{i,j,k\}}(\rep{1},\rep{2}_i\otimes\rep{2}_j,\rep{2}_k)_{+2,+1} \\
\oplus_{\{i,j,k\}}(\rep{1},\rep{2}_i\otimes\rep{2}_j,\rep{2}_k)_{+2,-1} \\
\end{array}\right\rbrace & \oplus_{i<j}(\rep{1},\rep{2}_i\otimes\rep{2}_j)_{+2}^{\oplus 8} \\
\\[-10pt]

\left.\begin{array}{c}
\oplus_{i<j} (\rep{1},\rep{2}_i\otimes\rep{2}_j,\rep{4}_i\otimes\rep{2}_j)_{0,0} \\
\oplus_{i<j} (\rep{1},\rep{2}_i\otimes\rep{2}_j,\rep{2}_i\otimes\rep{4}_j)_{0,0} \\
\end{array}\right\rbrace & \oplus_{i<j}(\rep{1},\rep{2}_i\otimes\rep{2}_j)_{0}^{\oplus 16} \\
\\[-10pt]

\left.\begin{array}{c}
\oplus_{i}(\rep{1},\rep{1},\rep{5}_i)_{-2,0} \\
\oplus_{\{i,j,k\}}(\rep{1},\rep{1},\rep{4}_i\otimes\rep{2}_j\otimes\rep{2}_k)_{-2,+1} \\
\oplus_{\{i,j,k\}}(\rep{1},\rep{1},\rep{4}_i\otimes\rep{2}_j\otimes\rep{2}_k)_{-2,-1} \\
(\rep{1},\rep{1},\rep{3}_1\otimes\rep{3}_2\otimes\rep{3}_3)^{\oplus 3}_{-2,0}  \\
\end{array}\right\rbrace & (\rep{1},\rep{1})_{-2}^{\oplus 192} \\
\\[-10pt]

\left.\begin{array}{c}
(\rep{1},\rep{1},\rep{2}_1\otimes\rep{2}_2\otimes\rep{2}_3)_{0,+3} \\
(\rep{1},\rep{1},\rep{2}_1\otimes\rep{2}_2\otimes\rep{2}_3)_{0,-3} \\
(\rep{1},\rep{1},\rep{3}_1\otimes\rep{3}_2\otimes\rep{3}_3)_{0,+2} \\
(\rep{1},\rep{1},\rep{3}_1\otimes\rep{3}_2\otimes\rep{3}_3)_{0,-2} \\
\end{array}\right\rbrace & (\rep{1},\rep{1})_{0}^{\oplus 70} \\
\end{array}$
    
\caption{Spacetime massless spectrum in the LGO phase}
\label{tab:CYmasslessspectrumLGO}
\end{table}

The massless spectrum of the LGO model is fully consistent with the large radius description. Indeed, provided there exists marginal deformations that take the SCFT back to the LR locus, these would result in the Higgsing of the enhanced $\SU(2)^3\times\GU(1)$ symmetry emerging the LGO limit. Examining the states listed in table~\ref{tab:CYmasslessspectrumLGO}, we find that the Higgsed spectrum reproduces the large radius result. In particular, the $70$ $G_{\text{\tiny{LR}}}$-singlets give rise to $70-10=60$ neutral massless chiral multiplets, matching the number expected from the K\"ahler and complex deformations of the CY target $Y$. This agreement is non-trivial: a discrepancy could have signaled a destabilization of the vacuum by worldsheet instantons, whose contributions to the spacetime superpotential can generate masses for some of the gauge-singlet fields~\cite{Dine:1986zy}. The Gepner computation, however, reproduces all $60$ deformations of the (0,2) NLSM, suggesting that such non-perturbative corrections vanish. Indeed, our model belongs to a broad class of GLSMs for which the absence of destabilizing worldsheet instantons has been established~\cite{Silverstein:1995re,Beasley:2003fx,Bertolini:2014dna}.

To formulate the compactness assumptions of~\cite{Bertolini:2014dna} precisely, we must revisit two aspects of the GLSM that were omitted from our earlier discussion. The first concerns the inclusion of spectator fields: a chiral-fermi pair $(\Phi_{\text{sp}},\Gamma_{\text{sp}})$ with respective charges $+5$ and $-5$ under each $\GU(1)$ gauge factor, which supplements the field content listed in table~\ref{tab:CYmodelY3}. The spectators enter the superpotential only through the mass term $\cW\supset \Gamma_{\text{sp}}\Phi_{\text{sp}}$, so they decouple at low energies while ensuring that the total charge of chiral superfields vanishes~\cite{Distler:1993mk}. The second aspect concerns the secondary fan associated to $Y$. The toric charge vectors associated with $\Phi_1,\dots,\Phi_8$ generate the ``classical cone''
\begin{align}
K_{\text{cl}} = \left\lbrace (\rho^1,\rho^2,\rho^3)~|~\rho^\alpha\geq0\right\rbrace
\end{align}
in the FI parameter space. This $K_{\text{cl}}$ is further subdivided into six chambers of the form $K_V = \{\rho^\alpha\geq\rho^\beta\geq\rho^\gamma\geq0\}$, corresponding to toric varieties $V$ obtained from a distinct triangulation of the face fan generated by the toric rays. Each toric variety $V$ is singular, but the CY complete intersection $Y\subset V$ is smooth. Importantly, $Y$ remains non-singular across the walls $\rho^\alpha=\rho^\beta$, even though the ambient toric varieties develop more severe singularities there. The space $Y$ is in fact smooth throughout the interior of the cone $K_{\text{cl}}$, and is everywhere described as a quadratic branched cover of $(\P^1)^3$, with each $\rho^\alpha$ controlling the volume of the corresponding $\P^1$ factors. This reassuringly confirms that the locus $\rho^\alpha=\rho$ considered earlier, where the classes $\eta_\alpha-\eta_\beta$ are primitive, yields a smooth CY geometry despite lying on the boundary of the ambient toric K\"ahler cone.

With these preliminaries in place, we are now in a position to apply the vanishing theorem of~\cite{Bertolini:2014dna}. Its argument relies on the compactness of the gauge instanton moduli space in the GLSM with superpotential set to zero. Potential non-compact zero-modes could arise from the lowest components of the superfields $\Phi_9$, $\Phi_{10}$, $\Phi_{11}$ and $\Phi_{\text{sp}}$. The instanton moduli space is compact provided their gauge charges satisfy
\begin{align}
- Q_9,- Q_{10},- Q_{11}\;&\in K_V~,&
- Q_{\text{sp}}\;&\notin K_{\text{cl}}~,
\end{align}
for some subcone $K_V\subset K_{\text{cl}}$ that leads to a smooth $Y$. Since we can choose $K_V= K_{\text{cl}}$, these conditions are satisfied and the theorem applies to our GLSM. This establishes the absence of worldsheet instanton corrections to the spacetime superpotential. The same argument extends to all the other examples considered above.

These results provide substantial evidence for the existence of the putative (0,2) SCFT associated to the  sigma model on $Y$ with line bundles, and for the identification of some of its marginal deformations with the GLSM parameters. The linear model gives access to a strongly coupled regime of the NLSM, reached by taking the FI parameter $\rho$---identified at large radius with the overall volume of $Y$---to large negative values. In the LGO limit, the volume modulus becomes charged under the enhanced gauge symmetry, which freezes its value at the string scale.

\section{Discussion}
\label{sec:discussion}
In this note we constructed a number of (0,2) GLSMs that describe Calabi--Yau sigma models with abelian gauge bundles. The linear models admit LGO phases with solvable points.  The resulting (0,2) SCFTs generalize the Gepner construction through a chiral orbifold action on the heterotic left-moving fermions. These Gepner points correspond to small volume limits of the target space geometries, where stringy gauge symmetries emerge and all geometric moduli become charged.

Although our exploration is restricted to the simpler setting of line bundles rather than general stable holomorphic bundles allowed by spacetime supersymmetry, we expect these configurations to capture some of the essential features of a generic (0,2) model. Indeed, abelian instanton can be viewed as particular loci in the moduli space of non-abelian bundles, where the connection becomes reducible while the supergravity solution remains smooth. When unobstructed, deformations in moduli space can lead to more general bundle configurations---in spacetime, this motion is then realized by a super-Higgs mechanism in which the vector multiplets associated with the broken gauge symmetries become massive. It would be interesting to realize such a deformation explicitly in a GLSM. A promising route would be to draw on constructions of stable monad bundles on branched cover CY geometries~\cite{Festi:2025}, which may be amenable to a linear model construction.

It would also be useful to determine the scope of the present GLSM construction, particularly for CY threefold compactifications, and identify the four-dimensional gauge groups it can realize. Of particular interest are the ``heterotic line bundle models'' in which a rank-$5$ abelian bundle with $\GU(1)^4\subset\SU(5)$ structure group breaks the gauge symmetry to $\GE_8\times\SU(5)$. Suitable discrete quotients and Wilson lines can then yield phenomenologically interesting models~\cite{Anderson:2011ns}. Realizing such compactifications would require GLSMs with worldsheet abelian gauge groups of higher rank. Indeed, the CY $3$-fold $Y$ must support four linearly independent primitive (1,1) classes, requiring $h^{1,1}(Y)\geq 5$, and these classes must be accessible through the toric divisors of the GLSM. A possible avenue would be to investigate whether any of these heterotic line bundle models admit a GLSM embedding.\footnote{Some of the models studied in~\cite{Anderson:2011ns,Anderson:2012yf,Anderson:2013xka} satisfy a Bianchi identity of the form $c_2(T_Y)-c_2(\cE)=[C]$, where $C$ is an effective class in $H_2(Y,\Z)$ supporting five-branes. In the present work, this five-brane class vanishes.}

Finally, it would be worthwhile to investigate further the asymmetric orbifolds of Gepner models and free left-moving fermions uncovered here, and their relation to existing constructions~\cite{Blumenhagen:1995tt,Schellekens:1989wx,Israel:2013wwa,Blumenhagen:2016axv}. More broadly, these constructions  motivate the development of a systematic framework for asymmetric orbifolds in the Gepner setting, analogous to that of asymmetric toroidal orbifolds~\cite{Narain:1986qm}. The possibility of a relation between two such classes is suggested by one of the K3 models encountered in this note, realized as a chiral orbifold of a (2,2) LG theory with superpotential $W=W_1+W_2$, where
\begin{align}
W_1 &= \tfrac{1}{6}X_1^6 +\tfrac{1}{3}X_5^3~,&
W_2 &= \tfrac{1}{6}X_3^6 + \tfrac{1}{3}X_6^3~.
\end{align}
Here, $W_1$ and $W_2$  are the LG superpotentials of two (2,2) Gepner models with $c=3$, each of which, upon a $\Z_6$ non-chiral orbifold, yields a (2,2) sigma model on a two-torus at the $\su(3)$ point. This suggests that the Gepner model itself admits an alternative description as the heterotic string compactified on $T^4$ at the maximal enhancement point $\so(32)\oplus\su(3)\oplus\su(3)$, orbifolded by a finite group acting asymmetrically on the left- and right-moving degrees of freedom. It would be instructive to understand how the consistency conditions for asymmetric orbifolds translate into the Gepner description, and to extend this analysis to other Gepner models known to admit toroidal orbifold presentations~\cite{Eguchi:1988vra,Nahm:1999ps}.

\appendix
\section{Perturbative anomalies in K3 models}
\label{app:6danomalies}
In this appendix, we review the anomaly cancellation conditions for six-dimensional (1,0) supergravity and verify them for the three models obtained from (0,2) LGO models in section~\ref{subsec:stringyK3}. Consider a six-dimensional (1,0) supergravity theory with $N_{\text{t}}$ tensor multiplets and non-abelian gauge group
\begin{align}
G = G_1\times \dots\times G_N~,
\end{align}
where each $G_\alpha$ is a simple Lie group. We denote by $n^\alpha_\rep{r}$ the number of hypermultiplets transforming in the representation $\rep{r}$ of $G_\alpha$, and by $n^{\alpha,\beta}_{\rep{r},\rep{r}'}$ the number transforming in the representation $(\rep{r},\rep{r}')$ of $G_\alpha\times G_\beta$. The relevant trace indices are
\begin{align}
\operatorname{tr}_\rep{r} \mathscr{F}_\alpha^2 &= A_\rep{r}\operatorname{tr} \mathscr{F}_\alpha^2~,&
\operatorname{tr}_\rep{r} \mathscr{F}_\alpha^4 &= B_\rep{r}\operatorname{tr} \mathscr{F}_\alpha^4 + C_\rep{r}(\operatorname{tr} \mathscr{F}_\alpha^2)^2~,
\end{align}
where $\operatorname{tr}$ denotes the trace in the defining representation. These coefficients can be computed using the methods of~\cite{Okubo:1981td}, nicely reviewed in~\cite{Hamada:2023zol}.

Local anomalies can be canceled by the generalized Green--Schwarz mechanism~\cite{Green:1984sg,Green:1984bx,Sagnotti:1992qw} provided that the anomaly polynomial $I_8$ factorizes in terms of $N_{\text{t}}+1$ four-forms $X_4^\alpha$ as
\begin{align}
I_8 = X_4\cdot X_4~,
\end{align}
where $X\cdot Y=\Omega_{\alpha\beta}X^\alpha Y^\beta$ denotes the Lorentzian inner product on $\R^{1,N_{\text{t}}}$. This factorization imposes several consistency conditions on the massless spectrum. The vanishing of the irreducible gravitational and gauge anomalies, proportional to $\operatorname{tr}\mathscr{R}^4$ and $\operatorname{tr}\mathscr{F}_\alpha^4$, respectively, requires
\begin{align}
N_{\text{h}}-N_{\text{v}}+29N_{\text{t}} &=273~,&
B^\alpha_\rep{\mathrm{adj}} &= \sum_{\rep{r}}n^\alpha_\rep{r}B^\alpha_\rep{r}~,
\end{align}
where $N_{\text{h}}$ and $N_{\text{v}}$ denote the number of hypermultiplets and vector multiplets. The remaining anomaly coefficients are encoded in the symmetric intersection matrix $\mathbb{G}$, whose entries are
\begin{align}
\mathbb{G}_{00} &= 9-N_{\text{t}}~,
&
\mathbb{G}_{\alpha 0} &= \frac{\lambda_\alpha}{6}\left( \sum_{\rep{r}}n^\alpha_\rep{r}A^\alpha_\rep{r} - A^\alpha_\rep{\mathrm{adj}} \right)~,
\nonumber\\
\mathbb{G}_{\alpha \beta} &=\lambda_\alpha\lambda_\beta \sum_{\rep{r},\rep{r}'}n^{\alpha,\beta}_{\rep{r},\rep{r}'} A^\alpha_\rep{r} A^\beta_{\rep{r}'} ~,&
\mathbb{G}_{\alpha \alpha} &= \frac{\lambda_\alpha^2}{3}\left( \sum_{\rep{r}}n^\alpha_\rep{r}C^\alpha_\rep{r} - C^\alpha_\rep{\mathrm{adj}} \right)~,
\end{align}
with $\alpha,\beta=1,\dots, N$ distinct. The coefficients $\lambda_\alpha=\lambda(G_\alpha)$ are group-theoretic constants; their values can be found, for example, in~\cite{Hamada:2023zol}. The anomaly polynomial factorizes if there exist vectors  $b_0,b_\alpha$ in $\R^{1,N_{\text{t}}}$ satisfying
\begin{align}
b_0\cdot b_0 &= \mathbb{G}_{00}~,& 
b_0\cdot b_\alpha &= \mathbb{G}_{0\alpha}~,& 
b_\alpha\cdot b_\beta &= \mathbb{G}_{\alpha\beta}~.
\end{align}
Equivalently, these conditions can be written as
\begin{align}
\mathbb{G} = \mathbbm{b}^{\text{t}}\Omega\,\mathbbm{b}~,
\end{align}
where $\mathbbm{b}$ is an $(N_{\text{t}}+1)\times (N+1)$ matrix whose columns are the vectors $b_0,b_\alpha$.

If the gauge group contains an additional $\GU(1)$ factor, anomaly cancellation requires the existence of an additional vector $b_{\Lu(1)}$ satisfying~\cite{Erler:1993zy,Park:2011wv}
\begin{align}
b_0\cdot b_{\Lu(1)} &= \frac{1}{6}\sum_{q}n_q q^2~,&
b_{\Lu(1)}\cdot b_{\Lu(1)} &= \frac{1}{3}\sum_{q}n_q q^4~,&
b_\alpha\cdot b_{\Lu(1)} & =\lambda_\alpha\sum_{\rep{r},q}n_{\rep{r}_q}^\alpha A^\alpha_{\rep{r}}q^2~.
\end{align}
Here, the sums run over the $\GU(1)$ charges $q$ of the hypermultiplets, with $n_q$ denoting the number of hypermultiplets of charge $q$, and $n_{\rep{r}_q}^\alpha$ the number of hypermultiplets of charge $q$ transforming in the representation $\rep{r}$ of $G_\alpha$.

The massless spectra~\eqref{eq:stringyK3spectrum}, obtained in the main text from compactifications of the $\Spin(32)/\Z_2$ heterotic string on a string-size K3 surface $M$ with an abelian gauge bundle, define six-dimensional (1,0) supergravity theories with one tensor multiplet. The Lorentzian inner product on $\R^{1,1}$ is defined by
\begin{align}
\Omega = \begin{pmatrix}
0&1\\
1&0
\end{pmatrix}~.
\end{align}
For each model, we verify that the anomaly cancellation conditions are satisfied by exhibiting the corresponding Gram matrix $\mathbb{G}$ and anomaly coefficient matrix $\mathbbm{b}$. For the model obtained from compactification on $M_1$, with $\so(8)\oplus\su(12)\oplus\so(4)$ gauge symmetry, we find
\begin{align}
\mathbb{G} &= \begin{pmatrix}
8&-2&6&60&60\\
-2&-4&0&0&0\\
6&0&4&40&40\\
60&0&40&400&400\\
60&0&40&400&400\\
\end{pmatrix}~,&
\mathbbm{b} &=
\begin{pmatrix}
2&1&1&10&10\\
2&-2&2&20&20
\end{pmatrix}~.
\end{align}
For the model associated to compactification on $M_2$, with $\so(16)\oplus\su(8)\oplus\so(4)$ gauge symmetry, we find
\begin{align}
\mathbb{G} &= \begin{pmatrix}
8&-2&10&36&36\\
-2&-4&2&0&0\\
10&2&8&36&36\\
36&0&36&144&144\\
36&0&36&144&144\\
\end{pmatrix}~,&
\mathbbm{b} &=
\begin{pmatrix}
2&1&1&6&6\\
2&-2&4&12&12
\end{pmatrix}~.
\end{align}
Finally, for the model associated to compactification on $M_3$, with $\so(20)\oplus\su(6)\oplus\so(4)\oplus\Lu(1)$ gauge symmetry, we find
\begin{align}
\mathbb{G} &= \begin{pmatrix}
8&-2&14&24&24&84\\
-2&-4&4&0&0&24\\
14&4&12&32&32&72\\
24&0&32&64&64&192\\
24&0&32&64&64&192\\
84&24&72&192&192&432\\
\end{pmatrix}~,&
\mathbbm{b} &=
\begin{pmatrix}
2&1&1&4&4&6\\
2&-2&6&8&8&36\\
\end{pmatrix}~.
\end{align}

The family of (1,0) supergravities presented in section~\ref{subsec:K3spectrum}, with $\so(32-2\text{r})\oplus\su(\text{r})\oplus\Lu(1)$ gauge algebra and hypermultiplet content~\eqref{eq:hypersK3instanton}, does not have a factorized anomaly. Instead, the anomaly polynomial takes the more general form $I_8 = \tfrac{1}{2}X_4\cdot X_4 + F_{\Lu(1)}X_6$, where
\begin{align}
X_4 &= -\tfrac{1}{2}b_0 \mathrm{tr}\mathscr{R}^2+ b_{\so(32-2\text{r})}\mathrm{tr}\mathscr{F}_{\so(32-2\text{r})}^2+2b_{\su(\text{r})}\mathrm{tr}\mathscr{F}_{\su(\text{r})}^2+2b_{\Lu(1)}F_{\Lu(1)}^2~,
\nonumber\\
X_6 &=-16\left(\tr F_{\Lu(1)}\mathrm{tr}\mathscr{R}^2-16\mathrm{tr}\mathscr{F}_{\su(\text{r})}^3-48 F_{\Lu(1)}\mathrm{tr}\mathscr{F}_{\su(\text{r})}^2-16\tr F_{\Lu(1)}^3\right)~,
\end{align}
with anomaly coefficients
\begin{align}
b_0 &=
\begin{pmatrix}2\\2\end{pmatrix}~,&
b_{\so(32-2\tr)} &=
\begin{pmatrix}1\\-2\end{pmatrix}~,&
b_{\su(\tr)} &=
\begin{pmatrix}1\\ \frac{48}{\tr}-2\end{pmatrix}~,&
b_{\Lu(1)} &=
\begin{pmatrix}\tr\\ 48-2\tr\end{pmatrix}~.
\end{align}

\section{KM algebra details}

\subsection{Conformal embeddings}
\label{app:conformalembeddings}
For a simple Lie algebra $\Lg$, we denote by $\chi^{\Lg}_{\rep{r}}$ the characters of the $\Lg$ KM algebra at level one, where $\rep{r}$ labels a KM primary representation. These characters transform under $\SL(2,\Z)$ as a vector-valued modular form of weight zero. In particular,
\begin{align}
\chi^{\Lg}_{\rep{r}}(-1/\tau) = \sum_{\rep{r}'}\cS^{\Lg}_{\rep{r}\rep{r}'}\chi^{\Lg}_{\rep{r}'}(\tau)~,
\end{align}
where the modular matrix $\cS^{\Lg}$ is symmetric. The $\so(2\tr)$ current algebra at level $1$ has four primary representations, $\rep{1},\rep{v},\rep{s},\rep{s}$. The $\su(\tr)$ current algebra at level $1$ has $\tr$ primary representations $\rep{r}_m=\wedge^m\rep{f}$, with $m=0,\dots,\tr-1$, where $\rep{f}$ is the defining representation of $\su(\tr)$. For a $\Lu(1)$ current algebra, we  denote by $\Lu(1)_\tk$ the extension of the current algebra by two vertex operators of conformal weight $h_\sleft=\tk$, constructed as exponentials of the chiral boson that bosonizes the holomorphic current. The extended algebra has $2\tk$ primary representations, labeled by $m\in\Z_{2\tk}$, with characters
\begin{align}
\chi^{\Lu(1)_\tk}_m(\tau) = \frac{1}{\eta(\tau)}\sum_{l\in\Z}\text{q}^{\tk\left(l+\frac{m}{2\tk}\right)^2}~.
\end{align}
The corresponding modular matrices are
\begin{align}
\mathcal{S}^{\so(2\tr)} &=\frac{1}{2}
\begin{pmatrix}
1&1&1&1\\
1&1&-1&-1\\
1&-1&\ii^\tr&-\ii^\tr\\
1&-1&-\ii^\tr&\ii^\tr\\
\end{pmatrix}~,&
\mathcal{S}^{\su(\tr)}_{\rep{r}_m\rep{r}_{m'}} &= \frac{\ee^{2\pi\ii \frac{mm'}{\tr}}}{\sqrt{\tr}}~,&
\mathcal{S}^{\Lu(1)_\tk}_{mm'} &= \frac{\ee^{-2\pi\ii \frac{mm'}{2\tk}}}{\sqrt{2\tk}}~.
\end{align}
For even $\tr$, there is a conformal embedding $\su(\tr)\oplus\Lu(1)_{\frac{\tr}{2}}\subset \so(2\tr)$, with branching rules
\begin{align}
\chi^{\so(2\tr)}_{\rep{1}} &= \sum_{m=0,2,4,\dots,\tr-2} \chi^{\su(\tr)}_{\rep{r}_m}\chi^{\Lu(1)}_m~,&
\chi^{\so(2\tr)}_{\rep{v}} &= \sum_{m=1,3,5,\dots,\tr-1} \chi^{\su(\tr)}_{\rep{r}_m}\chi^{\Lu(1)}_m~,
\nonumber\\
\chi^{\so(2\tr)}_{\rep{s}} &= \sum_{m=0,2,4,\dots,\tr-2} \chi^{\su(\tr)}_{\rep{r}_m}\chi^{\Lu(1)}_{m+\frac{\tr}{2}}~,&
\chi^{\so(2\tr)}_{\rep{c}} &= \sum_{m=1,3,5,\dots,\tr-1} \chi^{\su(\tr)}_{\rep{r}_m}\chi^{\Lu(1)}_{m+\frac{\tr}{2}}~.
\end{align}
This embedding can be understood as a simple current extension: the $\su(\tr)\oplus\Lu(1)$ KM primary $(\rep{r}_2)_{+2}$ is a $\Z_{\tr/2}$ simple current of spin-$1$, whose extension yields the $\so(2\tr)$ current algebra, in agreement with the Lie algebra decomposition
\begin{align}
\text{adj}_{\so(2\tr)} = (\text{adj}_{\su(\tr)})_0\oplus\rep{1}_0\oplus(\Lambda^2\rep{\textbf{r}})_{+2}\oplus(\Lambda^2\brep{\textbf{r}})_{-2}~.
\end{align}

\subsubsection*{Decomposition of $\Lu(1)_\tk$ under shifts}
Consider a left-moving $\Lu(1)_\tk$ algebra, with holomorphic current $J_\sleft$ normalized as
\begin{align}
J_\sleft(z)J_\sleft(0)\sim \frac{2\tk}{z^2}~,
\end{align}
and the $\Z_\tN\subset\GU(1)$ symmetry generated by $g=\exp(2\pi\ii J_{\sleft 0}/\tN)$. The $\Lu(1)_\tk$ primaries need not transform homogeneously under this symmetry. It is therefore useful to decompose $\Lu(1)_\tk$ into a subalgebra $\Lu(1)_{\tK}$ whose primaries have definite $\Z_\tN$ charges. We have
\begin{align}
\Lu(1)_{\tk}&\supset\Lu(1)_{\tK}~,&
\text{K} = \tN^2\tk~.
\end{align}
This is a conformal embedding, since both factors have central charge $c_\sleft=1$, with branching rules\footnote{This decomposition can be understood as follows: the $\Lu(1)_\tK$ primary $M=2\tk \tN$ is a $\Z_\tN$ simple current of conformal weight $h_\sleft=\tk$. Extending $\Lu(1)_\tK$ by this simple current projects onto the $2\tk \tN$ primaries with $M=0\mod \tN$, which are grouped into $2\tk$ extended primaries under the $\Z_\tN$ action.}
\begin{align}
\chi^{\Lu(1)_\tk}_m = \sum_{\ell\in\Z_\tN} \chi^{\Lu(1)_{\tK}}_{\tN(2\tk\ell+m)}~.
\label{eq:conformalembeddingu1}
\end{align}
The $\Lu(1)_\tK$ primary representations are labeled by $M\in\Z_{2\tK}$. This decomposition refines the characters according to their $\Z_\tN$ charges: each component $\chi^{\Lu(1)_{\tK}}_{\tN(2\tk\ell+m)}$ transforms under $g$ with phase $\ee^{2\pi\ii(2\tk\ell+m)/\tN}$. Such decomposition is used in section~\ref{eq:smallvolumetransition} to decompose the level-$1$ $\su(2)$ current algebra (equivalently, the $\Lu(1)_1$ KM algebra) with respect to a $\Z_9$ action. Specifically, we use the $\Lu(1)_{81}$ subalgebra, with branching rules
\begin{align}
\chi^{\su(2)}_{\rep{1}} &= \sum_{\ell\in\Z_9} \chi^{\Lu(1)_{81}}_{18\ell}~,&
\chi^{\su(2)}_{\rep{2}} &= \sum_{\ell\in\Z_9} \chi^{\Lu(1)_{81}}_{18\ell+9}~.
\end{align}

\subsection{Orbifolds of free Weyl fermions}
\label{app:fermionorbifold}
Consider an even number $\tr$ of free left-moving Weyl fermions $\lambda_1,\dots,\lambda_\tr$, and a $\Z_\tN$ chiral symmetry generated by
\begin{align}
g &= \exp(2\pi\ii\nu J_{\sleft0})~,&
\nu&\in\tfrac{1}{\tN}\Z~,
\end{align}
where $J_\sleft=:\lambda_1\overline{\lambda}_1:+\dots+:\lambda_\tr\overline{\lambda}_\tr:$. The current algebra generated by the fermion bilinears admits the decomposition $\so(2\tr)_1\supset \su(\tr)_1\oplus\Lu(1)_{\frac{\tr}{2}}$. We further decompose $\Lu(1)_{\frac{\tr}{2}}\supset \Lu(1)_{\tR}$, with $\tR=\tr \tN^2/2$, so that the $\Lu(1)_\tR$ primaries carry definite $\Z_\tN$ charge. In this decomposition, the NS partition function of the free fermions---before orbifolding---takes the form
\begin{align}
Z^{(0,0)} = \sum_{m\in\Z_\tr} \sum_{\ell\in\Z_\tN} \chi^{\su(\tr)}_{\rep{r}_m} \chi^{\Lu(1)_\tR}_{\tN(m+\tr \ell)} ~.
\end{align}
The twisted partition functions are then obtained by modular orbits, yielding
\begin{align}
Z^{(a,b)} = \sum_{m\in\Z_\tr}\sum_{\ell\in\Z_\tN} \chi^{\su(\tr)}_{\rep{r}_m} \chi^{\Lu(1)_\tR}_{\tN(m+\tr\ell+\tr\nu a) } \,\ee^{2\pi\ii(m+\tr\ell+\tr\nu a)\nu b } ~,
\end{align}
and obey the modular transformation
\begin{align}
Z^{(a,b)}(-1/\tau) = \ee^{2\pi\ii \tr\nu^2 ab} Z^{(b,-a)}(\tau)~.
\end{align}
In the special case where $\tr$ is a multiple of $\tN$, the $\Lu(1)_{\tr/2}$ primaries themselves carry definite charges under $g$, and the twisted partition functions simplify to
\begin{align}
Z^{(a,b)} = \sum_{m\in\Z_\tr} \chi^{\su(\tr)}_{\rep{r}_m} \chi^{\Lu(1)_{\tr/2}}_{m+\tr\nu a} \,\ee^{2\pi\ii(m+\tr\nu a)\nu b } ~.
\end{align}

\subsection{Superconformal minimal models}
\label{app:minimalmodels}
The (2,2) LG model for one chiral multiplet $X$ with superpotential
\begin{align}
W = \tfrac{1}{k}X^k
\end{align}
flows in the IR to a (2,2) minimal model~\cite{Martinec:1988zu,Vafa:1988uu}, with central charge $c=3(1-2/k)$. This model can be represented by the coset
\begin{align}
\frac{\mathfrak{su}(2)_{k-2}\oplus\mathfrak{u}(1)_{2}}{\mathfrak{u}(1)_{k}}~,
\label{eq:minimalmodelcoset}
\end{align}
whose primary representations are labeled by three indices $(l,m,s)$ valued in the range
\begin{align}
(l,m,s)&\in\{0,1,\dots,k-2\}\times\Z_{2k}\times\Z_4~,&
l+m+s&=0\mod2~,
\end{align}
subject to the field identification $(l,m,s)\simeq(k-2-l,m+k,s+2)$. The conformal weight $h_\sleft$ and R-charge $q_\sleft$ of the left-moving primary representation $(l,m,s)$ satisfy
\begin{align}
h_\sleft &=\frac{l(l+2)-m^2}{4k}+\frac{s^2}{8}\mod 1~,&
q_\sleft &=-\frac{m}{k}+\frac{s}{2}\mod 2~.
\end{align}
We denote characters of the coset~\eqref{eq:minimalmodelcoset} by
$\chi^l_m{}^{(s)}$. They obey the modular transformation
\begin{align}
\chi^l_m{}^{(s)}(-1/\tau) = \frac{1}{2k} \sum_{l',m',s'} \sin\left(\frac{\pi (l+1)(l'+1)}{k}\right) \ee^{\ii\pi\frac{mm'}{k}-\ii\pi\frac{ss'}{2}}\chi^{l'}_{m'}{}^{(s')}(\tau)~.
\label{eq:Stransformationminmodel}
\end{align}
The primaries of the N=2 superconformal algebra itself are of the form $(l,m,0)\oplus(l,m,2)$ in the NS sector, and $(l,m,1)\oplus(l,m,3)$ in the Ramond sector. In particular, the characters of the SCA take the form
\begin{align}
\chi^l_m = \chi^l_m{}^{(0)}+\chi^l_m{}^{(2)}
\end{align}
in the NS sector. The NS partition function of the (2,2) SCFT obtained as the IR limit of the LG model associated with $X$ takes the form
\begin{align}
Z_X = \tfrac{1}{2}\sum_{l,m}\chi^l_m\overline{\chi}^l_m~,
\end{align}
with an overall factor that accounts for the field identification. It is convenient to split the left-moving characters into their two coset components as
\begin{align}
Z_X = \frac{1}{2}\sum_{l,m,\varepsilon}\chi^l_m{}^{(2\varepsilon)}\overline{\chi}^l_m~,
\end{align}
with an additional sum over $\varepsilon\in\Z_2$.

For a LG model composed of $N_\LG$ minimal factors of respective levels $k_1,\dots,k_{N_\LGsmall}$, the NS partition function is the product of the individual $Z_{X_i}$. We write it as
\begin{align}
Z_\LG = \frac{1}{2^{N_\LGsmall}}\sum_{\boldsymbol{l},\boldsymbol{m},\boldsymbol{\varepsilon}}\chi^{\boldsymbol{l}}_{\boldsymbol{m}}{}^{(2\boldsymbol{\varepsilon})}\overline{\chi}^{\boldsymbol{l}}_{\boldsymbol{m}}~,
\end{align}
with indices
\begin{align}
\boldsymbol{l} &=(l_1,\dots,l_{N_\LGsmall})~,&
\boldsymbol{m} &=(m_1,\dots,m_{N_\LGsmall})~,&
\boldsymbol{\varepsilon} &=(\varepsilon_1,\dots,\varepsilon_{N_\LGsmall})~,
\end{align}
valued in the appropriate range. For an orbifold of the type introduced in section~\ref{subsec:LGOconstruction}, with orbifold charges satisfying $\mu^\alpha_i\in\frac{1}{2}\Z$, the twisted Hilbert spaces have a particularly simple presentation, as the orbifold action preserves the coset. Using~\eqref{eq:Stransformationminmodel}, the twisted partition functions may be obtained via modular orbits, and take the form
\begin{align}
Z_{\LG}^{(a,b)} &=\frac{1}{2^{N_{\LGsmall}}}\sum_{\boldsymbol{l},\boldsymbol{m},\boldsymbol{\varepsilon}} 
\chi^{\boldsymbol{l}}_{\boldsymbol{m}+2a\cdot\boldsymbol{\mu}}{}^{(2\boldsymbol{\varepsilon}+2a\cdot\boldsymbol{\mu})} \overline{\chi}^{\boldsymbol{l}}_{\boldsymbol{m}}\,\ee^{2\pi\ii a\cdot A_\LGsmall\cdot b - 2\pi\ii\sum_{i=1}^{N_\LGsmall} b\cdot\boldsymbol{\mu}_i m_i/k_i}~.
\end{align}

\subsection{Chiral orbifold of \texorpdfstring{$W=\frac{1}{3}X^3$}{W=X3/3}}
\label{app:chiralbosonorbifold}
The (2,2) LG model of a chiral superfield $X=(\Phi,\Gamma)$ with superpotential $W=\frac{1}{3}X^3$ admits a $\Z_9$ symmetry,
\begin{align}
\Phi&~\to~\ee^{\frac{2\pi\ii}{9}}\Phi~,&
\Gamma&\to\ee^{-\frac{4\pi\ii}{9}}\Gamma~.
\end{align}
This action is a $\Z_9$ subgroup of the $\GU(1)_\sleft$ symmetry of the LG model, under which the (0,2) components $\Phi$ and $\Gamma$ have charges $q_\sleft=\tfrac{1}{3}$ and $q_\sleft=-\frac{2}{3}$, respectively. In the IR, it can therefore be expressed as $g=\exp(\frac{2\pi\ii}{3} J_{\sleft0})$, where $J_\sleft$ is the left-moving R-current of the (2,2) SCFT. Since the $\Z_9$ action is asymmetric between left- and right-movers, determining the orbifold theory requires decomposing the N=2 characters according to their $\Z_9$ charges. This decomposition is particularly simple because the $c=1$ N=2 minimal model admits a compact boson representation. In particular, the coset characters $\chi^l_m{}^{(s)}$ can be expressed in terms of $\Lu(1)_6$ characters as
\begin{align}
\chi^l_m{}^{(s)} = \chi^{\Lu(1)_6}_{3s-2m}~.
\end{align}
The (NS,NS) partition function of the LG model then takes the form
\begin{align}
Z = \sum_{m\in\Z_6}(\chi^{\Lu(1)_6}_{-2m}+\chi^{\Lu(1)_6}_{-2m+6})\overline{\chi}^0_m~.
\end{align}
To resolve the $\Z_9$ charges, we further decompose $\Lu(1)_6\supset\Lu(1)_{54}$, following~\eqref{eq:conformalembeddingu1}. The resulting twisted partition functions are
\begin{align}
Z^{(a,b)} = \sum_{m\in\Z_6}\sum_{\ell\in\Z_6}\chi^{\Lu(1)_{54}}_{18\ell-6m+2a}\overline{\chi}^0_m\,\ee^{2\pi\ii\left(\frac{\ell}{3}-\frac{m}{9}+\frac{a}{27}\right)b}~.
\end{align}

\section{Additional examples}
\label{app:additionalexamples}
In this appendix we present further (0,2) GLSMs whose large radius phases realize heterotic CY compactifications with abelian gauge bundles.

\subsection{A single \texorpdfstring{$\Lu(1)$}{u(1)} instanton on K3}
\label{app:generalU1instanton}
In section~\ref{subsec:K3GLSMs} we presented three GLSMs describing K3 surfaces $M$ whose Picard groups are generated by two classes $\eta_1,\eta_2$, with abelian bundles fibered over the class $\upomega=\eta_1-\eta_2$. More generally, we can consider classes $\upomega=w^1\eta_1+w^2\eta_2$, where $w^1$ and $w^2$ are integers whose product is fixed by the heterotic Bianchi identity~\eqref{eq:heteroticBianchiK3} for a given abelian configuration $\bL$. Without loss of generality, we take $w^1\geq-w^2>0$.

Beyond the three models discussed in the main text, with $w=(1,-1)$, we find seven additional possibilities. For each, we list the complete superfield content except for the fermi multiplets $\Lambda_A$ encoding the abelian configuration $\bL$. The listed fields produce a gauge anomaly precisely canceled by the contribution $\cA_\cE=(\bL\cdot\bL)ww^{\text{t}}$ from these additional fermi multiplets. Each model admits an LGO phase with a locus of enhanced (2,2) supersymmetry. We give the corresponding (2,2) superpotential $W$ and orbifold group $\Gamma$.
\begin{center}
$\GLSMblock{T_M}{
\begin{array}{cccccc}
\Phi_{1,2} & \Phi_{3,4} & \Phi_5 & \Gamma_5 \\
1 & 0 & 2 & -4 \\
0 & 1 & 2 & -4 \\
\end{array}
}$
\quad
$\GLSMblock{\cO}{
\begin{array}{ccc}
\Phi_6 & \Gamma_1 & \Gamma_2 \\
-18 & 1 & 17 \\
0 & 0 & 0 \\
\end{array}
}$
\quad
$\GLSMblock{\cO}{
\begin{array}{ccc}
\Phi_7 & \Gamma_3 & \Gamma_4 \\
0 & 0 & 0 \\
-9 & 1 & 8 \\
\end{array}
}$
\qquad 
$\begin{array}{rl}
w &=(2,-1) \qquad \bL\cdot\bL=6\\[8pt]
W&=\tfrac{1}{18}X_1^{18}+\tfrac{1}{9}X_3^9+\tfrac{1}{3}X_5^3\\
\Gamma&=\Z_{18}\times\Z_9
\end{array}$

\textcolor{pink}{\hrulefill}

$\GLSMblock{T_M}{
\begin{array}{cccccc}
\Phi_{1,2} & \Phi_{3,4} & \Phi_5 & \Gamma_5 \\
1 & 0 & 2 & -4 \\
0 & 1 & 2 & -4 \\
\end{array}
}$
\quad
$\GLSMblock{\cO}{
\begin{array}{ccc}
\Phi_6 & \Gamma_1 & \Gamma_2 \\
-24 & 1 & 23 \\
0 & 0 & 0 \\
\end{array}
}$
\quad
$\GLSMblock{\cO}{
\begin{array}{ccc}
\Phi_7 & \Gamma_3 & \Gamma_4 \\
0 & 0 & 0 \\
-8 & 1 & 7 \\
\end{array}
}$
\qquad 
$\begin{array}{rl}
w &=(3,-1) \qquad \bL\cdot\bL=4\\[8pt]
W&=\tfrac{1}{24}X_1^{24}+\tfrac{1}{8}X_3^8+\tfrac{1}{3}X_5^3\\
\Gamma&=\Z_{24}\times\Z_8
\end{array}$

\textcolor{pink}{\hrulefill}

$\GLSMblock{T_M}{
\begin{array}{cccccc}
\Phi_{1,2} & \Phi_{3,4} & \Phi_5 & \Gamma_5 \\
1 & 0 & 2 & -4 \\
0 & 1 & 2 & -4 \\
\end{array}
}$
\quad
$\GLSMblock{\cO}{
\begin{array}{ccc}
\Phi_6 & \Gamma_1 & \Gamma_2 \\
-42 & 1 & 42 \\
0 & 0 & 0 \\
\end{array}
}$
\quad
$\GLSMblock{\cO}{
\begin{array}{ccc}
\Phi_7 & \Gamma_3 & \Gamma_4 \\
0 & 0 & 0 \\
-7 & 1 & 6 \\
\end{array}
}$
\qquad 
$\begin{array}{rl}
w &=(6,-1) \qquad \bL\cdot\bL=2\\[8pt]
W&=\tfrac{1}{42}X_1^{42}+\tfrac{1}{7}X_3^7+\tfrac{1}{3}X_5^3\\
\Gamma&=\Z_{42}\times\Z_7
\end{array}$

\textcolor{pink}{\hrulefill}

$\GLSMblock{T_M}{
\begin{array}{cccccc}
\Phi_{1,2} & \Phi_{3,4} & \Phi_5 & \Gamma_5 \\
1 & 0 & 2 & -4 \\
0 & 1 & 2 & -4 \\
\end{array}
}$
\quad
$\GLSMblock{\cO}{
\begin{array}{ccc}
\Phi_6 & \Gamma_1 & \Gamma_2 \\
-15 & 1 & 14 \\
0 & 0 & 0 \\
\end{array}
}$
\quad
$\GLSMblock{\cO}{
\begin{array}{ccc}
\Phi_7 & \Gamma_3 & \Gamma_4 \\
0 & 0 & 0 \\
-10 & 1 & 9 \\
\end{array}
}$
\qquad
$\begin{array}{rl}
w &=(3,-2) \qquad \bL\cdot\bL=2\\[8pt]
W&=\tfrac{1}{15}X_1^{15}+\tfrac{1}{10}X_3^{10}+\tfrac{1}{3}X_5^3\\
\Gamma&=\Z_{15}\times\Z_{10}
\end{array}$

\textcolor{pink}{\hrulefill}

$\GLSMblock{T_M}{
\begin{array}{cccccc}
\Phi_{1,2} & \Phi_{3,4} & \Phi_5 & \Gamma_5 \\
1 & 0 & 1 & -3 \\
0 & 1 & 1 & -3 \\
\end{array}
}$
\quad
$\GLSMblock{\cO}{
\begin{array}{ccc}
\Phi_6 & \Gamma_1 & \Gamma_2 \\
-12 & 1 & 11 \\
0 & 0 & 0 \\
\end{array}
}$
\quad
$\GLSMblock{\cO}{
\begin{array}{ccc}
\Phi_7 & \Gamma_3 & \Gamma_4 \\
0 & 0 & 0 \\
-6 & 1 & 5 \\
\end{array}
}$
\qquad
$\begin{array}{rl}
w &=(2,-1) \qquad \bL\cdot\bL=4\\[8pt]
W&=\tfrac{1}{12}X_1^{12}+\tfrac{1}{6}X_3^6+\tfrac{1}{4}X_5^4\\
\Gamma&=\Z_{12}\times\Z_6
\end{array}$

\textcolor{pink}{\hrulefill}

$\GLSMblock{T_M}{
\begin{array}{cccccc}
\Phi_{1,2} & \Phi_{3,4} & \Phi_5 & \Gamma_5 \\
1 & 0 & 1 & -3 \\
0 & 1 & 1 & -3 \\
\end{array}
}$
\quad
$\GLSMblock{\cO}{
\begin{array}{ccc}
\Phi_6 & \Gamma_1 & \Gamma_2 \\
-20 & 1 & 19 \\
0 & 0 & 0 \\
\end{array}
}$
\quad
$\GLSMblock{\cO}{
\begin{array}{ccc}
\Phi_7 & \Gamma_3 & \Gamma_4 \\
0 & 0 & 0 \\
-5 & 1 & 4 \\
\end{array}
}$
\qquad
$\begin{array}{rl}
w &=(4,-1) \qquad \bL\cdot\bL=2\\[8pt]
W&=\tfrac{1}{20}X_1^{20}+\tfrac{1}{5}X_3^5+\tfrac{1}{4}X_5^4\\
\Gamma&=\Z_{20}\times\Z_5
\end{array}$

\textcolor{pink}{\hrulefill}

$\GLSMblock{T_M}{
\begin{array}{cccccc}
\Phi_{1,2} & \Phi_{3,4} & \Phi_{5,6} & \Gamma_{5,6} \\
1 & 0 & 1 & -2 \\
0 & 1 & 1 & -2 \\
\end{array}
}$
\quad
$\GLSMblock{\cO}{
\begin{array}{ccc}
\Phi_7 & \Gamma_1 & \Gamma_2 \\
-12 & 1 & 11 \\
0 & 0 & 0 \\
\end{array}
}$
\quad
$\GLSMblock{\cO}{
\begin{array}{ccc}
\Phi_8 & \Gamma_3 & \Gamma_4 \\
0 & 0 & 0 \\
-4 & 1 & 3 \\
\end{array}
}$
\qquad
$\begin{array}{rl}
w &=(3,-1) \qquad \bL\cdot\bL=2\\[8pt]
W&=\tfrac{1}{12}X_1^{12}+\tfrac{1}{4}X_3^4+\tfrac{1}{3}X_5^3+\tfrac{1}{3}X_6^3\\
\Gamma&=\Z_{12}\times\Z_4
\end{array}$

\end{center}

\subsection{Branched covers of Hirzebruch surfaces}
\label{app:Hirzebruchcover}
A generalization of the K3 geometry $M_1$ introduced in section~\ref{subsec:K3GLSMs} is obtained by realizing a K3 surface $M$ as a double covering of the Hirzebruch surface $\mathbb{F}_p$. Let $\eta_1,\eta_2$ denote the classes in the Picard lattice of $M$ pulled back from $\mathbb{F}_p$.  Their intersection pairing is twice that of the corresponding classes on $\mathbb{F}_p$:
\begin{align}
D = 2\begin{pmatrix}
-p&1\\1&0\\
\end{pmatrix}~.
\end{align}
The class $\upomega=\eta_1-\eta_2$ is anti-self-dual along the locus $\rho^2=(p+1)\rho^1$ in the FI parameter space, and satisfies $\upomega.\upomega = -2(p+2)$. Compatibility with the Bianchi identity $(\bL\cdot\bL)(\upomega.\upomega) = -48$ requires $12/(p+2)\in\Z$, since $\bL\cdot\bL$ is even. Thus, the allowed values are $p\in\{0,1,2,4,10\}$. This geometry is realized by a (0,2) GLSM with gauge group $\GU(1)^2$ and the following field content:
\begin{center}
\arraycolsep=4pt
$\GLSMblock{T_M}{
\begin{array}{cccccc}
\Phi_1 & \Phi_2 & \Phi_3 & \Phi_4 & \Phi_5 & \Gamma_5 \\
1 & 1 & 0 & 0 & 2 & -4 \\
p & 0 & 1 & 1 & p+2 & -2(p+2)\\
\end{array}
}$
\qquad
$\GLSMblock{\cO}{
\begin{array}{ccc}
\Phi_6 & \Gamma_1 & \Gamma_2 \\
-\text{m}_p-1  &  1 & \text{m}_p \\
-p & p & 0 \\
\end{array}
}$
\qquad
$\GLSMblock{\cO}{
\begin{array}{ccc}
\Phi_7 & \Gamma_3 & \Gamma_4 \\
0 & 0 & 0 \\
-\text{n}_p-1 & 1 & \text{n}_p \\
\end{array}
}$
\end{center}
where $\text{m}_p$ and $\text{n}_p$ are the positive integers
\begin{align}
\text{m}_p & = 5+\frac{12}{p+2}~,&
\text{n}_p & = (p+1)(p+5)+\frac{12}{p+2}~.
\end{align}
The large radius phase corresponds to $\rho^2>p\rho^1>0$. The resulting K3 surface need not be smooth; smoothness depends on the hypersurface equation defining the branch locus in $\mathbb{F}_p$. In particular, for $p=10$, the surface is necessarily singular, whereas for the other allowed values of $p$, smooth choices exist---in accordance with the classification of~\cite{Hayashi:2022}. The anomaly matrix associated with this field content is
\begin{align}
\cA &= -\frac{24}{p+2}\begin{pmatrix}
1&-1\\
-1&1
\end{pmatrix}~,
\end{align}
and can be canceled by introducing additional Fermi multiplets $\Lambda_1,\dots,\Lambda_{\tr}$ with suitable charges, encoding a rank-$\tr$ abelian bundle over $M$ in the large radius phase. For $p>0$, negative FI parameters instead lead to a hybrid phase.

\subsection{Calabi--Yau hypersurfaces in \texorpdfstring{$(\P^1)^{d+1}$}{(P1)d+1}}
\label{app:CYhypersurfaces}
A class of GLSMs realizing CY hypersurfaces in $(\P^1)^{d+1}$ may be constructed as follows. The gauge theory, with $\GU(1)^{d+1}$ gauge group, has $d+1$ pairs of chiral superfields, each pair carrying unit charge under one gauge factor, and one fermi superfield of charge $-2$ under every factor. These fields define a hypersurface in $(\P^1)^{d+1}$ specified by a polynomial of degree two in each pair of chiral fields. This matter content is gauge anomalous---we supplement it by additional fermi multiplets $\Lambda_1,\dots,\Lambda_{\tr}$ that cancel the mixed gauge anomalies, encoding a rank-$\tr$ abelian bundle, together with extra chiral and fermi multiplets encoding trivial bundle factors. The CY hypersurface $Y$ inherits $d+1$ integral classes $\eta_\alpha$ from $(\P^1)^{d+1}$, satisfying $(\eta_\alpha)^2=0$ and $\eta_{\alpha_1}\eta_{\alpha_2}\dots\eta_{\alpha_d}=2$ (for $\alpha_1,\dots,\alpha_d$ distinct). In particular, when all FI parameters are equal and strictly positive, the classes $\eta_\alpha-\eta_\beta$ are primitive (1,1) forms and may be used to construct abelian HYM bundles.

We illustrate this construction in low dimensions. For $d=1$, the following GLSM describes an elliptic curve $C$ with a rank-$6$ trivial bundle.
\begin{center}
$\GLSMblock{T_C}{
\begin{array}{ccc}
\Phi_{1,2}&\Phi_{3,4}&\Gamma_0 \\
1&0&-2\\
0&1&-2\\
\end{array}
}$
\qquad
$\GLSMblock{\cO}{
\begin{array}{ccc}
\Phi_5&\Gamma_1 &\Gamma_2 \\
-4&1&3\\
0&0&0\\
\end{array}
}$
\qquad
$\GLSMblock{\cO}{
\begin{array}{ccc}
\Phi_6&\Gamma_3 &\Gamma_4 \\
0&0&0\\
-4&1&3\\
\end{array}
}$
\qquad
$\GLSMblock{\cE}{
\begin{array}{cc}
\Lambda_{1,2} &\Lambda_{3,4} \\
1&-1\\
-1&1\\
\end{array}
}$
\end{center}
For $d=2$, we obtain a GLSM for a K3 surface $M$ equipped with the bundle $\cO^{\oplus 3}\oplus\cE$, where $\cE$ is a rank-$3$ abelian bundle.
\begin{center}
$\GLSMblock{T_M}{
\begin{array}{cccc}
\Phi_{1,2} & \Phi_{3,4} & \Phi_{5,6} & \Gamma_0 \\
1 & 0 & 0 & -2 \\
0 & 1 & 0 & -2 \\
0 & 0 & 1 & -2 \\
\end{array}
}$
\qquad
$\GLSMblock{\cO}{
\begin{array}{ccc}
\Phi_7 & \Gamma_1 & \Gamma_2 \\
-6 & 1 & 5\\
0 & 0 & 0 \\
0 & 0 & 0 \\
\end{array}
}$
\qquad
$\GLSMblock{\cO}{
\begin{array}{ccc}
\Phi_8 & \Gamma_3 & \Gamma_4 \\
0 & 0 & 0 \\
-6 & 1 & 5\\
0 & 0 & 0 \\
\end{array}
}$
\qquad
$\GLSMblock{\cO}{
\begin{array}{ccc}
\Phi_9 & \Gamma_5 & \Gamma_6 \\
0 & 0 & 0 \\
0 & 0 & 0 \\
-6 & 1 & 5\\
\end{array}
}$
\qquad
$\GLSMblock{\cE}{
\begin{array}{ccc}
\Lambda_1 & \Lambda_2 & \Lambda_3 \\
2 & 0 & -2 \\
-2 & 2 & 0 \\
0 & -2 & 2
\end{array}
}$
\end{center}
For $d=3$, we obtain a GLSM for a CY threefold $Y$, with $h^{1,1}(Y)=4$ and $h^{2,1}(Y)=68$, and a bundle $\cO^{\oplus 4}\oplus\cE$, where $\cE$ is a rank-$4$ abelian bundle.
\begin{center}
$\GLSMblock{T_Y}{
\begin{array}{ccccc}
\Phi_{1,2} & \Phi_{3,4} & \Phi_{5,6} & \Phi_{7,8} & \Gamma_0 \\
1 & 0 & 0 & 0 & -2 \\
0 & 1 & 0 & 0 & -2 \\
0 & 0 & 1 & 0 & -2 \\
0 & 0 & 0 & 1 & -2 \\
\end{array}
}$
\;
$\GLSMblock{\cO}{
\begin{array}{ccc}
\Phi_9 & \Gamma_1 & \Gamma_2 \\
-8 & 1 & 7 \\
0 & 0 & 0 \\
0 & 0 & 0 \\
0 & 0 & 0 \\
\end{array}
}$
\;
$\GLSMblock{\cO}{
\begin{array}{ccc}
\Phi_{10} & \Gamma_3 & \Gamma_4 \\
0 & 0 & 0 \\
-8 & 1 & 7 \\
0 & 0 & 0 \\
0 & 0 & 0 \\
\end{array}
}$
\;
$\GLSMblock{\cO}{
\begin{array}{ccc}
\Phi_{11} & \Gamma_5 & \Gamma_6 \\
0 & 0 & 0 \\
0 & 0 & 0 \\
-8 & 1 & 7 \\
0 & 0 & 0 \\
\end{array}
}$
\;
$\GLSMblock{\cO}{
\begin{array}{ccc}
\Phi_{12} & \Gamma_7 & \Gamma_8 \\
0 & 0 & 0 \\
0 & 0 & 0 \\
0 & 0 & 0 \\
-8 & 1 & 7 \\
\end{array}
}$
\;
$\GLSMblock{\cE}{
\begin{array}{cccc}
\Lambda_1 & \Lambda_2 & \Lambda_3 & \Lambda_4 \\
3 & -1 & -1 & -1 \\
-1 & 3 & -1 & -1 \\
-1 & -1 & 3 & -1 \\
-1 & -1 & -1 & 3
\end{array}
}$
\end{center}
Each GLSM also admits a LGO phase, with a locus where $\Gamma_0$ decouples from the superpotential and the remaining superfields $\Phi_i,\Gamma_i$ assemble into (2,2) chiral multiplets, half of which become massive. At this locus, the theory reduces to a chiral orbifold of a (2,2) LG model with $c_\sleft=c_\sright=3d$, together with free left-moving fermions, by a finite group $\Gamma$. Omitting the massive fields, the corresponding (2,2) superpotentials and orbifold groups are
\begin{align}
d&=1~:&
W &= \tfrac{1}{4}(X_1^4+X_3^4)~,&
\Gamma &= (\Z_4)^2~,
\nonumber\\
d&=2~:&
W &= \tfrac{1}{6}(X_1^6+X_3^6+X_5^6)~,&
\Gamma &= (\Z_6)^3~,
\nonumber\\
d&=3~:&
W &= \tfrac{1}{8}(X_1^8+X_3^8+X_5^8+X_7^8)~,&
\Gamma &= (\Z_8)^4~.
\end{align}

The K3 GLSM can also be used to construct a heterotic flux vacuum via the TLSM construction~\cite{Adams:2006kb}. In this case, the Fermi multiplets $\Lambda_{1,2,3}$ are replaced by shift multiplets $\Omega^{I=1,2}$, with metric $\cG_{IJ}$ and shift charges $w^I$ given by
\begin{align}
\mathcal{G}&=
\begin{pmatrix}
4 & -2 \\
-2 & 4 \\
\end{pmatrix}~,&
w^1&=(1,0,-1)~.&
w^2&=(0,1,-1)~,
\end{align}
in the notation of~\cite{Israel:2026bco}. The large radius phase of the TLSM then describes a six-dimensional non-K\"ahler space, obtained as a principal $T^2$ fibration over the K3 surface $M$, with the two circle fibers associated to the linearly independent K3 classes $\upomega^1=\eta_1-\eta_3$ and $\upomega^2=\eta_2-\eta_3$.

\section{Line bundle cohomology}
\label{app:linebundlecohomology}
In a heterotic compactification on a CY $d$-fold $Y$ with an abelian gauge bundle $\cE\to Y$, the charged massless spectrum is determined by the cohomology dimensions $h^1(Y,\cL)$ of the line bundles $\cL$ in which the left-moving fermions transform. While the Hirzebruch--Riemann--Roch theorem determines the holomorphic Euler characteristic,
\begin{align}
\chi(Y,\cL)=\sum_{r=0}^d(-1)^r h^r(Y,\cL) = \int_Y \mathrm{ch}(\cL)\mathrm{td}(T_Y)~,
\end{align}
the individual cohomology dimensions require a separate computation. For the geometries considered in the main text, obtained as branched coverings of product of projective lines, this can be reduced to line bundle cohomology on $(\P^1)^d$.

More generally, consider an abelian branched cover $\pi:Y\to S$ of a K\"ahler $d$-fold $S$, and a holomorphic vector bundle $\cE\to Y$. Since $\pi$ is finite, the Leray spectral sequence gives
\begin{align}
H^r(Y,\cE)\simeq H^r(S,\pi_\ast \cE)~.
\end{align}
If $\cE$ is obtained as the pullback $\cE=\pi^\ast \cF$ of a vector bundle $\cF\to S$, the projection formula\footnote{Namely, $\pi_\ast(\pi^\ast\cF)\simeq \cF\otimes\pi_\ast\cO_Y$---see, for example, chapter II.5 of~\cite{Hartshorne:1977}.} yields 
\begin{align}
H^r(Y,\cE)\simeq H^r(S,\cF\otimes\pi_\ast\cO_Y)~.
\end{align}
The pushforward $\pi_\ast\cO_Y$ admits an explicit description in terms of the covering data~\cite{Pardini:1991}. In the cases of interest, the cover is specified by a collection of line bundles $\cT_i\to S$, $i=1,\dots,N$. The space $Y$ is defined by 
\begin{align}
(\xi_1^{k_1},\dots,\xi_N^{k_N}) = (s_1,\dots,s_N)~,
\end{align}
where $(\xi_1,\dots,\xi_N)$ are the fiber coordinates on $\oplus_i\cT_i$, and $(s_1,\dots,s_N)$ is a section of $\oplus_i\cT_i^{\otimes k_i}$. The covering group $\Gamma=\Z_{k_1}\times\dots\times\Z_{k_N}$ acts by multiplying each $\xi_i$ by a $k_i$-th root of unity. Consequently, the pushforward $\pi_\ast\cO_Y$ decomposes as\footnote{Here, for $\boldsymbol{\ell}=(\ell_1,\ldots,\ell_N)\in\Gamma$, each $\ell_i$ is taken in the range $\{0,\dots,k_i-1\}$.}
\begin{align}
\pi_\ast \cO_X &= \mathop{\oplus}_{\boldsymbol{\ell}\in\Gamma} \cT_{\boldsymbol{\ell}}^\ast ~,&
\cT_{\boldsymbol{\ell}} &=\cT_1^{\otimes \ell_1}\otimes\dots\otimes\cT_N^{\otimes\ell_N}~,
\end{align}
and hence
\begin{align}
h^r(Y,\cE)=\sum_{\boldsymbol{\ell}\in\Gamma} h^r(S, \cF\otimes \cT_{\boldsymbol{\ell}}^\ast )~.
\end{align}
For line bundles on $S\simeq(\P^1)^d$, the resulting cohomology groups follow from the K\"unneth formula,
\begin{align}
h^r((\P^1)^d,\cO(l_1,\dots,l_d) ) = \sum_{r_1+\dots+r_d=r}h^{r_1}(\P^1,\cO(l_1))\times\dots\times h^{r_d}(\P^1,\cO(l_2))~,
\end{align}
and the standard results
\begin{align}
h^0(\P^1,\mathcal{O}(l)) &=
\begin{cases}
l+1&\text{if } l\geq 0~,\\
0&\text{if } l< 0~,\\
\end{cases}&
h^1(\P^1,\mathcal{O}(l)) &=
\begin{cases}
0&\text{if } l\geq -1~,\\
-l-1&\text{if } l< -1~.\\
\end{cases}
\end{align}

The Calabi--Yau $d$-folds $Y_1$, $Y_2$, and $Y_3$ introduced in the main text arise as double, triple, and biquadratic covers of $(\P^1)^d$, respectively. Their structure-sheaf pushforwards are
\begin{align}
\pi_\ast \cO_{Y_1} &\simeq \cO\oplus\cO(-2,\dots,-2)~,\nonumber\\
\pi_\ast \cO_{Y_2} &\simeq \cO\oplus\cO(-1,\dots,-1)\oplus\cO(-2,\dots,-2)~,\nonumber\\
\pi_\ast \cO_{Y_3} &\simeq \cO\oplus\cO(-1,\dots,-1)^{\oplus 2}\oplus\cO(-2,\dots,-2)~.
\end{align}

\subsubsection*{Charged matter on $Y_3$}
We now consider the CY threefold $Y_3$, a $\Z_2\times\Z_2$ branched covering of $(\P^1)^3$. The rank-$6$ abelian bundle is $\cE=(\cL_1\oplus\cL_2\oplus\cL_3)^{\oplus 2}$, with constituent line bundles $\cL_{i=1,2,3}$ obtained by pulling back line bundles on $(\P^1)^3$,
\begin{align}
\cL_1 &= \pi^\ast\cO(2,-1,-1)~,&
\cL_2 &= \pi^\ast\cO(-1,2,-1)~,&
\cL_3 &= \pi^\ast\cO(-1,-1,2)~,
\end{align}
with first Chern classes denoted by $c_1(\cL_i)=\upomega_i$. Using the above procedure, we compute the relevant cohomology dimensions, listed in table~\ref{tab:linecohomology}. They satisfy the expected Serre duality relations $h^r(Y_3,\cL)=h^{3-r}(Y_3,\cL^\ast)$.
\begin{table}[h!]
\centering
\arraycolsep=8pt
$\begin{array}{ccccc}
c_1(\cL) & h^0(Y_3,\cL) & h^1(Y_3,\cL) & h^2(Y_3,\cL) & h^3(Y_3,\cL) \\ \hline
0 &1&0&0&1\\ 
\upomega_i &0&0&8&0\\ 
- \upomega_i &0&8&0&0\\ 
2 \upomega_i &0&0&64&0\\
-2\upomega_i  &0&64&0&0\\
\upomega_i-\upomega_j &0&8&8&0\\
\upomega_i+\upomega_j &0&8&0&0\\
-\upomega_i-\upomega_j &0&0&8&0\\
\end{array}$
\caption{Line bundle cohomology for the CY threefold $Y_3$.}
\label{tab:linecohomology}
\end{table}

The charged chiral spectrum of the heterotic compactification follows from the branching $\so(32) \supset \so(20)\oplus\su(2)^{\oplus 3}\oplus\Lu(1)^{\oplus 3}$. The $496$-dimensional adjoint decomposes into:
\begin{align}
\rep{R}_{0,0,0} &=  (\rep{190},\rep{1},\rep{1},\rep{1})\oplus(\rep{1},\rep{3},\rep{1},\rep{1})\oplus(\rep{1},\rep{1},\rep{3},\rep{1})\oplus(\rep{1},\rep{1},\rep{1},\rep{3})\oplus(\rep{1},\rep{1},\rep{1},\rep{1})^{\oplus3}~,
\nonumber\\
\rep{R}_{+1,0,0} &= \colunderline{\rep{R}_{-1,0,0}} = (\rep{20},\rep{2},\rep{1},\rep{1})~,
\nonumber\\
\rep{R}_{0,+1,0} &= \colunderline{\rep{R}_{0,-1,0}} = (\rep{20},\rep{1},\rep{2},\rep{1})~,
\nonumber\\
\rep{R}_{0,0,+1} &= \colunderline{\rep{R}_{0,0,-1}} = (\rep{20},\rep{1},\rep{1},\rep{2})~,
\nonumber\\
\colunderline{\rep{R}_{0,+1,+1}} &= \rep{R}_{0,-1,-1} = \colunderline{\rep{R}_{0,+1,-1}} = \colunderline{\rep{R}_{0,-1,+1}} = (\rep{1},\rep{1},\rep{2},\rep{2})~,
\nonumber\\
\colunderline{\rep{R}_{+1,0,+1}} &= \rep{R}_{-1,0,-1} = \colunderline{\rep{R}_{+1,0,-1}} = \colunderline{\rep{R}_{-1,0,+1}} = (\rep{1},\rep{2},\rep{1},\rep{2})~,
\nonumber\\
\colunderline{\rep{R}_{+1,+1,0}} &= \rep{R}_{-1,-1,0} = \colunderline{\rep{R}_{+1,-1,0}} = \colunderline{\rep{R}_{-1,+1,0}} = (\rep{1},\rep{2},\rep{2},\rep{1})~,
\nonumber\\
\rep{R}_{+2,0,0} &= \colunderline{\rep{R}_{-2,0,0}} = \rep{R}_{0,+2,0} = \colunderline{\rep{R}_{0,-2,0}} = \rep{R}_{0,0,+2} = \colunderline{\rep{R}_{0,0,-2}} = (\rep{1},\rep{1},\rep{1},\rep{1})~.
\end{align}
The representations underlined in color correspond to charged chiral multiplets, whose multiplicities are determined by the cohomology dimensions $h^1(Y_3,\cL)$ in table~\ref{tab:linecohomology}.

\bibliographystyle{./utphys}
\bibliography{abelianbundle}

\end{document}